\documentclass[fleqn,usenatbib]{mnras}

\usepackage[T1]{fontenc}

\DeclareRobustCommand{\VAN}[3]{#2}
\let\VANthebibliography\thebibliography
\def\thebibliography{\DeclareRobustCommand{\VAN}[3]{##3}\VANthebibliography}

\usepackage{graphicx}	% Including figure files
\usepackage{amsmath}	% Advanced maths commands
\usepackage{pdflscape}
\usepackage{adjustbox}
\usepackage{caption}
\usepackage[skip=0.5ex]{subcaption}
\usepackage{float}
\usepackage{rotating,threeparttable}
\usepackage{afterpage}
\usepackage{newtxtext,newtxmath}
\defcitealias{irabor2018}{Paper~IV}
\defcitealias{kalprep}{Paper~III}
\defcitealias{cornissh2013}{Paper~II}
\defcitealias{cornish2012}{Paper~I}
\title[Data processing and catalogue.]{The Co-Ordinated Radio and Infrared Survey for High-Mass Star Formation. V. The CORNISH-South Survey and Catalogue.}

\author[T. Irabor et al.]{
T. Irabor,$^{1}$\thanks{E-mail: T.Irabor@leeds.ac.uk (TI)}
M.G. Hoare,$^{1}$ 
M. Burton,$^{13}$
W.D. Cotton,$^{3}$
P. Diamond,$^{2}$
S. Dougherty,$^{21}$
\newauthor S.P. Ellingsen,$^{15}$
R. Fender,$^{14}$
G.A. Fuller,$^{2,20}$ 
S. Garrington, $^{2}$
P.F. Goldsmith,$^{5}$
J. Green,$^{12}$
A.G. Gunn$^{2}$
\newauthor J. Jackson,$^{7}$
S. Kurtz,$^{4}$
S.L. Lumsden,$^{1}$
J. Marti,$^{11}$
I. McDonald,$^{2,22}$
S. Molinari,$^{16}$
T.J. Moore,$^{8}$
\newauthor M. Mutale,$^{1}$
T. Muxlow, $^{2}$
T. O'Brien, $^{2}$
R.D. Oudmaijer,$^{1}$
R. Paladini,$^{19}$
J.D. Pandian,$^{6}$
J.M. Paredes,$^{10}$
\newauthor A.M.S. Richards, $^{2}$
A. Sanchez-Monge, $^{20}$
R. Spencer, $^{2}$
M.A. Thompson,$^{1,9}$ 
G. Umana,$^{18}$
J.S. Urquhart,$^{17}$
\newauthor M. Wieringa,$^{12}$ 
and  A. Zijlstra, $^{2}$
\\
$^{1}$Physics and Astronomy, University of Leeds, LS2 9JT, UK\\
$^{2}$Jodrell Bank Centre for Astrophysics, Department of Physics and Astronomy, University of Manchester, Manchester, M13 9PL, UK\\
$^{3}$The National Radio Astronomy Observatory, Charlottesville, VA 22903, USA\\
$^{4}$Institute of Radio Astronomy and Astrophysics,
National Autonomous University of Mexico, 58089 Morelia, Michoacán, México\\
$^{5}$Jet Propulsion Laboratory California Institute fo Technology, Pasadena CA, 91109\\
$^{6}$Department of Earth \& Space Sciences, Indian Institute of Space Science and Technology, Trivandrum 695547, India\\
$^{7}$Green Bank Observatory, 155 Observatory Rd, P.O. Box 2, Green Bank, WV 24944, USA\\
$^{8}$Astrophysics Research Institute, Liverpool John Moores University, Twelve Quays House, Egerton Wharf, CH41 1LD, UK\\
 $^{9}$Centre for Astrophysics Research, University of Hertfordshire, College Lane, Hatfield AL10 9AB, UK\\
 $^{10}$Cosmos Science Institute, University of Barcelona	, ICCUB, Martí i Franqués, 1, 08028 Barcelona, Spain\\
$^{11}$Departamento de F\'isica (EPSJ), Universidad de Ja\'en, Campus Las Lagunillas s/n, A3, E-23071  Ja\'en, Spain\\
$^{12}$CSIRO Space and Astronomy, PO Box 1130, Bentley WA 6102, Australia\\
$^{13}$Armagh Observatory and Planetarium,College Hill, BT61 9DB, Northern Ireland\\
$^{14}$Department of Physics, University of Oxford, Keble Road, Oxford, OX1 3RH, UK\\
$^{15}$School of natural Sciences, College of Sciences and Engineering, University of Tasmania, Hobart 7001, TAS, Australia\\
$^{16}$Istituto Nazionale di Astrofisica - IAPS, Via Fosso del Cavaliere 100, I-00133 Roma, Italy\\
$^{17}$Centre for Astrophysics and Planetary Science, University of Kent, Canterbury, CT2\,7NH, UK\\
$^{18}$INAF - Osservatorio Astrofisico di Catania, Via S. Sofia 78, I-95123, Catania, Italy\\
$^{19}$Infrared Processing Center, California Institute of Technology, Pasadena, CA 91125, USA\\
$^{20}$Physikalisches Institut, University of Cologne, Z\"ulpicher Str. 77, 50937 K\"oln, Germany\\
$^{21}$The ALMA headquarters, Santiago, Alonso de Córdova 3107, Chile\\
$^{22}$Department of Physics and Astronomy, Open University, Walton Hall, Milton Keynes, MK7 6AA, UK\\
}

\date{Accepted XXX. Received YYY; in original form ZZZ}

\pubyear{2022}

\begin{document}
\label{firstpage}
\pagerange{\pageref{firstpage}--\pageref{lastpage}}
\maketitle

\begin{abstract}
We present the first high spatial resolution radio continuum survey of the southern Galactic plane. The CORNISH project has mapped the region defined by $\mathrm{295\degr<l<350\degr}$; $\mathrm{|b|<1\degr}$ at 5.5-GHz, with a resolution of 2.5$\arcsec$ (FWHM). As with the CORNISH-North survey, this is designed to primarily provide matching radio data to the \textit{Spitzer} GLIMPSE survey region. The CORNISH-South survey achieved a root mean square noise level of $\mathrm{\sim}$ 0.11 mJy beam$\mathrm{^{-1}}$, using the 6A configuration of the Australia Telescope Compact Array (ATCA). In this paper, we discuss the observations, data processing and measurements of the source properties. Above a 7$\mathrm{\sigma}$ detection limit, 4701 sources were detected, and their ensemble properties show similar distributions with their northern counterparts. The catalogue is highly reliable and is complete to 90 per cent at a flux density level of 1.1 mJy. We developed a new way of measuring the integrated flux densities and angular sizes of non-Gaussian sources. The catalogue primarily provides positions, flux density measurements and angular sizes. All sources with IR counterparts at 8$\mathrm{\mu m}$ have been visually classified, utilizing additional imaging data from optical, near-IR, mid-IR, far-IR and sub-millimetre galactic plane surveys. This has resulted in the detection of 524 H II regions of which 255 are ultra-compact H II regions, 287 planetary nebulae, 79 radio stars and 6 massive young stellar objects. The rest of the sources are likely to be extra-galactic. These data are particularly important in the characterization and population studies of compact ionized sources such as UCHII regions and PNe towards the Galactic mid-plane.  
\end{abstract}


%%%%%%%%%%%%%%%%%%%%%%%%%%%%%%%%%%%%%%%%%%%%%%%%%%

%%%%%%%%%%%%%%%%% BODY OF PAPER %%%%%%%%%%%%%%%%%%

\section{Introduction}
Understanding the formation and evolution of the content of our Galaxy requires studying an unbiased population of objects covering different evolutionary stages utilizing a wide range of wavebands. Cm-wave radio continuum surveys are useful for probing the ionized gas components such as H II regions and planetary nebulae.

The ultra-compact (UC) HII population provides a means to probe the early phases of massive star formation, where the young stars are still deeply embedded in their natal molecular clouds. They are characterized by physical sizes $\mathrm{\leq}$ 0.1 pc, high emission measures $\mathrm{\geq 10^7\ pc\ cm^{-6}}$, high electron densities $\mathrm{\geq 10^4\ cm^{-3}}$ and lifetimes typical of $\mathrm{10^5}$ yrs  \citep[]{wood21989,come1996,kurt2000,kurtz2002,church_2002_new}.  Given that these populations often form in clusters, radio and infrared observations at arcsecond resolution are required to resolve their morphologies and provide insight into their immediate environment \citep[]{ChurchwellEd1990UHrt,hoa2007}. An unbiased survey of the population of UCHII regions will allow us to test evolutionary models of massive star formation, H II region dynamics, galactic structure and massive star formation rate in our Galaxy \citep{church_2002_new,hoa2007,davies2011,urq_2013,harry}.  Radio observations of UCHII regions are also more useful when carried out at a frequency where thermal free-free emission is optically thin \citep[$\geq$5-GHz:][]{ChurchwellEd1990UHrt,wood21989}.

IR surveys with high sensitivity and arcsecond resolution of the Galactic plane have made possible studies of an unbiased and statistically representative population of Galactic objects, thereby aiding the studies of massive star formation and stellar evolution. In the northern Galactic plane, these surveys include the  the mid-infrared Galactic Legacy Infrared Mid-Plane Survey Extraordinaire (GLIMPSE)  by the \textit{Spitzer} satellite \citep{churchwell2009}, the mid-infrared inner Galactic plane survey using the Multiband Infrared Photometer for \textit{Spitzer} (MIPSGAL\footnote{\url{http://mipsgal.ipac.caltech.edu/}}: \citealt{carey2009}), the far-infrared Herschel Infrared Galactic Plane (Hi-GAL) survey \citep{molinari2010}, the near-infrared Galactic plane survey (GPS) of the United Kingdom Infrared Deep Sky Survey (UKIDSS) project \citep{lucas2008}, sub-millimetre APEX Telescope Large Area Survey of the Galaxy (ATLASGAL: \citealt{schu2009}) and the H$\mathrm{\alpha}$ Isaac Newton Telescope Photometric Survey (IPHAS: \citealt{drew2005}).

To complement these surveys, the CORNISH project delivered a uniform and high resolution (1.5$\mathrm{\arcsec}$) radio continuum dataset of the northern Galactic plane at 5-GHz. It achieved a sensitivity $\mathrm{\sim 0.43}$ mJy beam$\mathrm{^{-1}}$ using the VLA in B and BnA configurations (\citealt{cornish2012,cornissh2013}: hereafter \citetalias{cornish2012} and \citetalias{cornissh2013}, respectively). These data made possible an unbiased census of UCHII regions that is the largest selection to date in the northern Galactic plane (\citealt{kalprep}:  hereafter \citetalias{kalprep}). \citet{kal2018PhD} found that 70 per cent of UCHII regions had a cometary morphology using the CORNISH data and follow-up higher resolution radio data. Through a multi-wavelength analysis, \citet{irabor2018} (hereafter \citetalias{irabor2018}) uncovered an unbiased population of compact planetary nebulae (PNe), of which 7 per cent were newly discovered PNe.  A subset of the PNe population has properties that are typical of young sources (\citetalias{irabor2018} and \citealt{frag2016}). The study of such objects is critical, given that the transition window from the post-AGB to the PN phase is small (usually < 1000 yrs). Other studies that used the CORNISH data in massive star formation studies include \citet{urq2013,cesa2015,trem2015,yang2019} and \citet{djor2019}. Note that the Global view on Star formation in the Milky Way (GLOSTAR) Galactic plane survey \citep{Brunthaler2021}, is using the JVLA to go deeper than the CORNISH survey of the northern Galactic plane.

In the southern Galactic plane, the VST Photometric H$\mathrm{\alpha}$ Survey  \citep[VPHAS+:][]{drew2014} and the Vista Variables in the Via Lactea (VVV) survey \citep{min2011} delivered high resolution H$\mathrm{\alpha}$ and near-infrared data respectively, in addition to existing infrared and sub-millimetre data (see Table \ref{table1}). Surveys of masers at radio wavelengths such as the Methanol Multibeam Survey (MMB; \citealt{green_MMB_2012,green_MMB_2017}) and the $\mathrm{H_{2}O}$ Southern Galactic Plane Survey (HOPS; \citealt{walsh2011}) are useful tracers of massive star formation. Existing radio continuum surveys include the Molongolo Galactic Plane Survey (MGPS) \citep{murphy2007,green1999} that surveyed the $\mathrm{245\degr <l< 365\degr}$ and $\mathrm{|b|\leq 10\degr}$ region at 843 MHz with a resolution of 45$\arcsec$, the 1.4-GHz Southern Galactic Plane Survey (SGPS) that mapped the $\mathrm{253\degr <l< 358\degr}$; $\mathrm{|b|<1\degr.5}$ region, with a resolution of 100\arcsec\ and sensitivity of 1 mJy beam$\mathrm{^{-1}}$ \citep{mcclure2005,hav2006}, the GaLactic and Extragalactic All-sky Murchison Widefield Array (GLEAM) survey at 72–231 MHz, with a resolution of 4$\arcmin$–2$\arcmin$ that covers all the southern plane \citep{hurley-walker2019} and the TIFR GMRT Sky Survey (TGSS) survey at 150 MHz and 25\arcsec\ resolution covering the sky above declination $\mathrm{-53\degr}$ \citep{intema2017}. These surveys are too low a resolution to resolve UCHII regions and are at frequencies where these objects are optically thick. 

In this paper we present a new radio continuum survey from the CORNISH project of the southern Galactic Plane. The observations are described in Section \ref{observation}. The calibration and imaging are discussed in Sections \ref{calibrations} and \ref{imaging}.  Data quality and measurement of source properties are presented in Sections \ref{data_quality} and \ref{catalog} along with a discussion of the catalogue and its reliability. Furthermore, example sources  and a comparison of the CORNISH catalogue with other surveys are presented in Section \ref{catalog}.

\begin{table*}
  \begin{center}
    \caption{Multi-wavelength high resolution and sensitivity surveys in the southern Galactic plane.}
    \label{table1}
    \begin{tabular}{|l|c|c|r|r|} 
    \hline
      Survey &Bands& Resolution and Sensitivity & Coverage&Reference\\
      \hline
      GLIMPSE & 3.6$\mathrm{\mu}$m, 4.5$\mathrm{\mu}$m, 5.8$\mathrm{\mu}$m, and 8.0$\mathrm{\mu}$m & $\mathrm{<}$2$\mathrm{\arcsec}$; 0.4 mJy at 8$\mathrm{\mu m}$ ($\mathrm{5\sigma}$)& $\mathrm{295\degr <l< 360\degr}$; $\mathrm{|b|\leq 1\degr}$&\citet{churchwell2009} \\
      MIPSGAL & 24$\mathrm{\mu}$m & $\mathrm{\sim 10\arcsec}$; 1.3 mJy (5$\mathrm{\sigma}$)& $\mathrm{295\degr <l< 350\degr}$; $\mathrm{|b|\leq 1\degr}$& \citet{carey2009}\\
      VVV  & J, H, Ks & $\mathrm{\sim 1\arcsec}$; 16.5 mag in K ($\mathrm{5\sigma}$)&$\mathrm{295\degr <l< 350\degr}$; $\mathrm{|b|\leq 2\degr}$& \citet{min2011}\\
      VPHAS+ & H$\mathrm{\alpha}$, r, i &$\mathrm{\sim}$1$\mathrm{\arcsec}$; 20.56 mag in H$\mathrm{\alpha}$ (5$\mathrm{\sigma}$) & $\mathrm{+210\degr \lesssim l\lesssim +40\degr}$; $\mathrm{|b|\leq 5\degr}$& \citet{drew2014}\\
      Hi-Gal & 70, 160, 250, 350 and 500$\mathrm{\mu}$m  & 6 $-$  40$\mathrm{\arcsec}$; $\mathrm{\sim}$ 13-27 mJy (1$\mathrm{\sigma}$) &$\mathrm{-70\degr \lesssim l \lesssim 68\degr}$; $\mathrm{|b|\leq 1\degr}$&\citet{molinari2010}\\
      ATLASGAL & 870$\mathrm{\mu}$m & 18$\mathrm{\arcsec}$, $\mathrm{\sim}$50 - 70 mJy beam$\mathrm{^{-1}}$ (1$\mathrm{\sigma}$)&  $\mathrm{280\degr <l< 60\degr}$; $\mathrm{|b|<1.5\degr}$&\citet{schu2009}\\
      \hline
    \end{tabular}
  \end{center}
\end{table*}

\section{Observations}\label{observation}
The CORNISH program observed the southern Galactic plane with the Australia Telescope Compact Array (ATCA\footnote{\url{https:/www.narrabri.atnf.csiro.au/}}) using the 2-GHz bandwidth of the Compact Array Broadband Backend (CABB) correlator \citep{wilson2011}. Observations were carried out with the 6A\footnote{\url{https://www.narrabri.atnf.csiro.au/cgi-bin/obstools/baselines.cgi?array=6a}} array configuration for about 400 hours at two different frequency bands (4.5 - 6.5-GHz and 8 - 10-GHz) in full-polarization, centred on 5.5-GHz and 9-GHz, simultaneously. This paper focuses on the 4.5 - 6.5-GHz band. Data reduction of the 8 - 10-GHz data is still on-going and will be presented in a future paper (T. Irabor et al., In preparation). Observations were between 2010 and 2012 and the observation parameters are summarized in Table ~\ref{oberve_param}.

\begin{table}
\caption{Summary of observation parameters for the CORNISH-South survey.}\label{oberve_param}
\begin{adjustbox}{width=7cm,center}
\begin{tabular}{l|r}\hline 
\multicolumn{1}{l|}{Parameters}
& \multicolumn{1}{l}{Value} \\ \hline 
Observation region&$\mathrm{295\degr < l < 350\degr}$; $|b|\leq 1\degr$ \\
Total Time&$\mathrm{\sim}$ 400 hours\\
Number of antennas&6\\
Number of baselines&15\\
Observation period& 2010 to 2012\\
Observation frequency&5.5-GHz\\
Bandwidth&2-GHz\\
Longest baseline&6 km\\
Size of single dish&22 m\\
Field of view/ Primary beam (FWHM) &$\mathrm{\sim 10\arcmin}$\\
Synthesized beam (FWHM)& 2.5$\mathrm{\arcsec}$\\
Root mean square (rms) noise level&$\mathrm{\sim}$ 0.11 mJy beam$\mathrm{^{-1}}$\\

\hline
\end{tabular}
\end{adjustbox}
\end{table}

The survey utilized on-the-fly mapping, such that the antennas were scanning continuously, while the phase centres were sequentially moved in a traditional mosaic pattern. This resulted in the doubling of the uv-coverage in a single run and a primary beam that is elongated in the scanning direction\footnote{\url{http://www.narrabri.atnf.csiro.au/observing/users_guide/html/atug.html}}. The phase centres were spaced at $\mathrm{7.4'}$ in a mosaic pattern that is a scaled version of the hexagonal mosaic implemented in the NVSS survey \citep{condon21998}. This spacing delivers a sensitivity that is uniform to 10 per cent at 5.5-GHz.

The survey region was divided into 33 blocks (33 days of observations) covering 110 deg$\mathrm{^2}$ of the Galactic plane, defined by $\mathrm{295\degr < l < 350\degr}$ and $\mathrm{|b|\leq 1\degr}$. The mosaic was scanned in Galactic latitude and required $\sim$18 pointings to cover the $\mathrm{2\degr}$ of a row (i.e $\mathrm{-}$1$\mathrm{\degr}$  to +1$\mathrm{\degr}$).  At a scan rate of 7$\mathrm{\farcm}$4\,/\,10\,s, each row was completed in 3.2 minutes, including the turnaround time. A secondary calibrator was observed for 2 mins after observations of eight rows. A further eight rows were then observed with a secondary calibrator to complete one \textit{uv}-cut of a block of observation, summing up to 54 mins. To achieve an optimum \textit{uv}-coverage, this was repeated eleven times at different LSTs, resulting in 1.8 minutes of on-source time. Twelve hours were spent on each block, including flux calibration and set up time. Sixteen rows corresponds to 1.7$\mathrm{\degr}$ in longitude, hence 33 of these 1.7$\mathrm{\degr \times 2\degr}$ blocks were required to cover the survey area. PKS B1934-638 was observed at the end of each block of observations as the primary flux calibrator. PKS B0823-500 was also observed at the beginning of each block of observations as a backup flux calibrator. The secondary calibrators and corresponding days of observation are presented in Table \ref{phase_cal}.

\begin{figure}
\centering
	\includegraphics[height=6.5cm, width=\columnwidth]{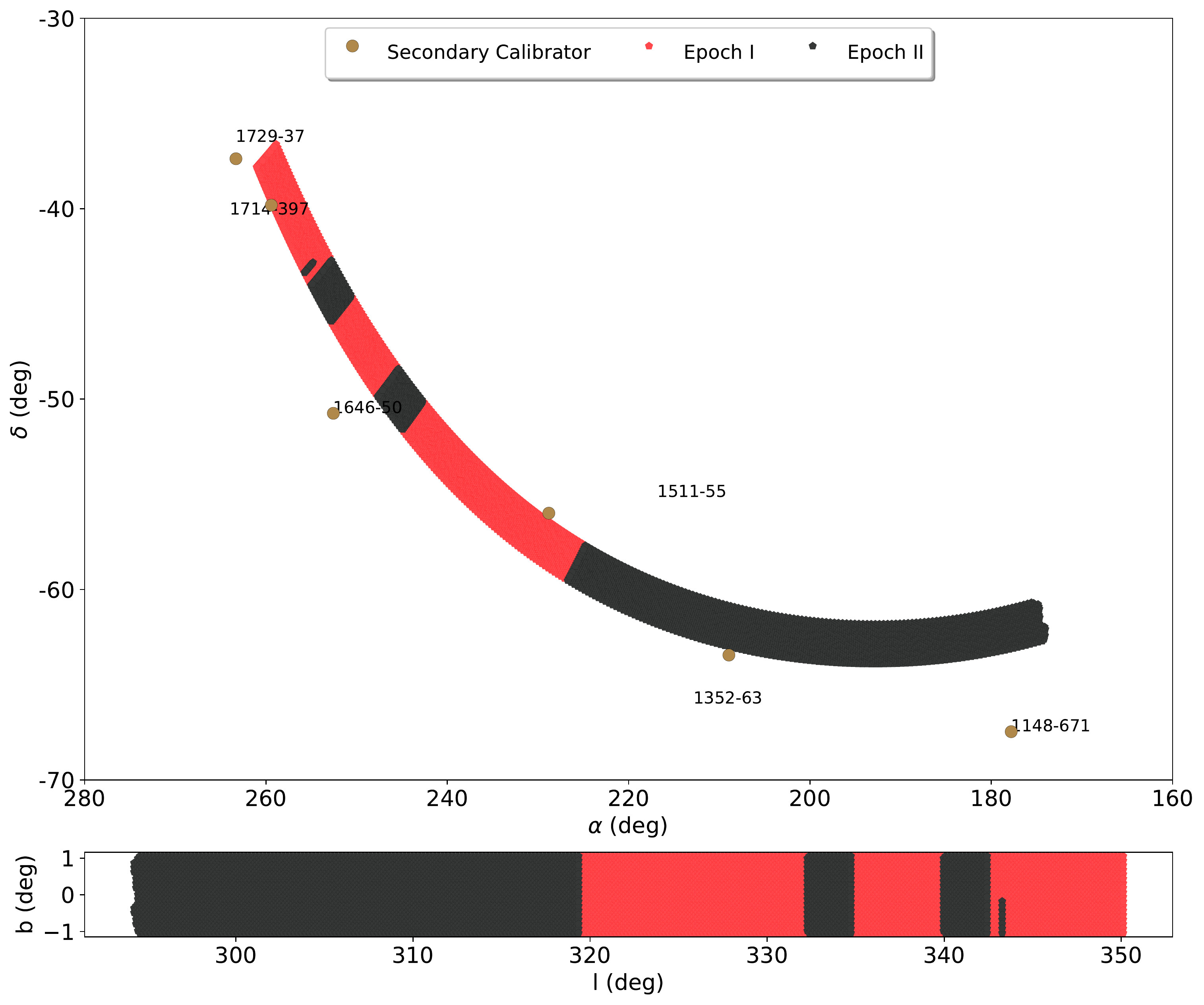}
	\caption{Coverage of the CORNISH-South data and positions of the six secondary calibrators used for calibration. Epoch 1 (red shaded regions) is defined by block 2010-12-21 to 2011-01-07 and epoch II (black shaded regions) is defined by 2011-12-20 to 2012-01-07.}
\label{region_cab}
\end{figure}

The observations fall naturally into two epochs (see Figure \ref{region_cab}), based on the observation period (between 2010 and 2012). Fields that were missed or days with observation difficulties due to bad weather or correlator problems were repeated to achieve as full a coverage as possible within the allocated time.

\begin{table*}
\caption{Secondary calibrators for each block of observations and the corresponding longitude range.}\label{phase_cal}

\begin{tabular}{l|l|c|c|l|c}\hline 
\multicolumn{1}{c}{Date}
& \multicolumn{1}{l}{Calibrator (s)} 
& \multicolumn{1}{c}{Longitude Range}
& \multicolumn{1}{c}{Date}
& \multicolumn{1}{l}{Calibrator (s)}
& \multicolumn{1}{c}{Longitude Range}\\ \hline 

2010$-$12$-$21 &1714$-$397&348.5 - 350.0 &2011$-$12$-$21&1511$-$55& 317.8 - 319.4\\
2010$-$12$-$22 &1729$-$37&346.8 - 348.5&2011$-$12$-$22&1352$-$63&304.1 - 305.7\\
2010$-$12$-$23&1729$-$37&345.1 - 346.8&2011$-$12$-$23&1511$-$55&  316.0 - 317.6\\
2010$-$12$-$24&1646$-$50&343.4 - 345.1&2011$-$12$-$24&1352$-$63, 1511$-$55& 314.3 - 315.9\\
2010$-$12$-$25&1646$-$50&341.7 - 343.4 &2011$-$12$-$25&1352$-$63&310.9 - 312.5\\ 
2010$-$12$-$26&1646$-$50&340.0 - 341.7&2011$-$12$-$26&1352$-$63&309.2 - 310.8 \\
2010$-$12$-$27&1646$-$50&338.3 - 340.0 &2011$-$12$-$27&1148$-$671&297.2 - 298.8\\
2010$-$12$-$28&1646$-$50&336.5 - 338.3&2011$-$12$-$28&1148$-$671, 1511$-$55&295.5 - 297.1\\
2010$-$12$-$29&1646$-$50&334.8 - 336.5 &2011$-$12$-$29&1148$-$671&   299.0 - 304.0\\
2010$-$12$-$30&1646$-$50&331.4 - 333.0&2011$-$12$-$30&1352$-$63&305.8 - 307.5 \\
2010$-$12$-$31&1646$-$50&  329.7 - 331.4&2011$-$12$-$31&1148$-$671& 302.4 - 304.0  \\
2011$-$01$-$01&1511$-$55, 1646$-$50&328.0 - 329.7&2012$-$01$-$01&1148$-$671&300.7 - 302.3\\
2011$-$01$-$02&1511$-$55, 1646$-$50&326.3 - 328.0 &2012$-$01$-$02&1352$-$63, 1148$-$671&307.5 - 309.1\\
2011$-$01$-$04&1511$-$55&324.6 - 326.3&2012$-$01$-$03&1148$-$671, 1352$-$63& 294.3 - 295.4\\
2011$-$01$-$05&1511$-$55&  322.9 - 324.6&2012$-$01$-$04&1352$-$63&312.6 - 314.2\\
2011$-$01$-$06&1511$-$55&321.2 - 322.9&2012$-$01$-$05&1148$-$671, 1352$-$63&301.5 - 326.2\\
2011$-$01$-$07&1511$-$55&319.5 - 321.2  &2012$-$01$-$07&1352$-$63&310.1 - 310.9\\
2011$-$12$-$20&1646$-$50, 1511$-$55&333.1 - 334.7&\\
\hline
\end{tabular}
\end{table*}

\section{Data Reduction Pipeline}\label{calibrations}
 
For us to achieve a similar uniform processing as in the CORNISH-North survey, we implemented a similar semi-automated calibration and imaging pipeline (see Figure 1 in \citetalias{cornissh2013}). The pipeline was implemented in \textit{python} language using \textit{mirpy}\footnote{\url{https://pypi.org/project/mirpy/}} to directly interface with the Multichannel Image Reconstruction, Image Analysis and Display (MIRIAD) software packages \citep{sault1995}.

\subsection{Flagging and Calibration}

The raw data were converted to a MIRIAD \textit{uv}-file format using the \textit{atlod} task. Known bad channels and radio-frequency interference (RFI) at the time of observations were flagged. The system variables such as \textit{uv}-coverage were inspected for each 12-hour block of observations. The system variables allowed quick identification of blocks with poor \textit{uv}-coverage or times with bad visibilities. Additionally, the amplitude and phase variations with time were inspected visually before any flagging. This was to ensure that no more data were flagged than necessary, to minimize any impact on image fidelity.

The epochs (I and II) presented different flagging demands. All XX polarization data of all baselines to antenna ca01\footnote{\url{https://www.narrabri.atnf.csiro.au/observing/users_guide/html/chunked/aph.html}} were flagged for epoch I. For epoch II, the YY polarization data of all baselines to antenna ca01 were flagged instead. This was due to a ripple in the bandpass at the time of observation, which would lead to false structures in the final images\footnote{\url{https://atcaforum.atnf.csiro.au/viewtopic.php?f=11&t=184}}. We performed both automated (\textit{pgflag}) and manual flagging (\textit{uvflag}) on the flux calibrator, secondary calibrators and science data.

Flagging parameters were determined manually for each block of observations and written to a master configuration file that was applied automatically when the pipeline was run.  After an initial flagging of the primary and secondary calibrators, bandpass and flux calibrations were performed. For blocks with two secondary calibrators, the solutions in phase and amplitude were estimated for each calibrator and then combined to produce a global calibration table. A second pass of flagging was then performed on the calibrated data before performing a second and final calibration. This was to ensure that the final calibration was performed on properly flagged data. The final global flagging and calibration tables produced were then applied to the science data and split into individual pointings.

\section{Imaging}\label{imaging}

The calibrated visibilities of individual fields were imaged by iteratively cleaning down to the maximum residual peak flux (MRF: see Section \ref{decon}) using multi-frequency synthesis (mfs; \citealt{sault1999}). This is implemented in MIRIAD using \textit{invert} with the \textit{'mfs'}, \textit{'sdb'} options and the multi-frequency CLEAN \textit{mfclean} tasks. Multi-frequency imaging accounts for the spectral variation across the observation bandwidth of 2-GHz. Thus, the resulting image consists of the normal flux component and the flux times the spectral component ($\mathrm{I\alpha}$), where $S_{\nu}\propto \nu^\alpha$.

The dirty images were created using a robust weighting scheme \citep{briggs1995} of 0.5 robustness and gridded to an image pixel size of 0.6$\mathrm{\arcsec}$, $\mathrm{\sim}$, which is about one third of the minor axis of the synthesized beam. The robust weighting scheme provides a trade-off between uniform and natural weighting, which is a trade-off between resolution and sensitivity. The choice of 0.5 allows an improved sensitivity without sacrificing the resolution. To ensure uniform resolution, we have forced a restoring circular Gaussian beam with an FWHM of 2.5$\mathrm{\arcsec}$ using \textit{restor} (see Section \ref{syn_beam}). The residuals were not corrected for the change in resolution, so care should be taken when summing values below the maximum residual flux (see \ref{decon}).

\subsection{Maximum Residual Flux}\label{decon}
The maximum residual flux (MRF) is an important parameter in the deconvolution process and should be carefully chosen. CLEANing deeper than necessary would result in many artefacts in the form of faint spurious sources and under-estimated peak fluxes and flux densities of real sources. This is due to the CLEAN algorithm cleaning noise spikes and re-distributing flux from point sources to noise peaks and is known as the clean bias effect \citep{white_1997,condon21998}. This effect is more pronounced for sources with lower signal-to-noise ratio (SNR), given that the amount of distributed flux is independent of the flux density of the source. According to \cite{pran2000}, the clean bias effect can be reduced by setting the MRF above the noise level. However, a shallow CLEAN, where the MRF is too far above the noise level, will create residual images dominated by sidelobes. Therefore, it is important that MRF be carefully chosen, especially for a blind survey such as CORNISH. Unlike for the CORNISH-North survey the \textit{mfclean} task does not employ windowing. Equation (\ref{eq1}) is an expression for the MRF, where $\mathrm{f_{mrf}}$ is a constant multiplicative factor and $\mathrm{rms}$ is the root mean square noise level.

\begin{equation}
\mathrm{MRF \simeq f_{mrf} \times rms}
\label{eq1}
\end{equation}
	
Given that the clean bias is also dependent on the rms noise level, we have determined the $\mathrm{rms}$ in Equation (\ref{eq1}) by imaging the central portion of each field's dirty map in Stokes V. Stokes V maps usually have very few sources or no sources, since there are few circularly polarized sources at 5.5-GHz \citep{rob1975,ho2006}. Hence, the Stokes V maps are dominated by thermal noise.

\begin{figure}
\centering
	\includegraphics[height=6.5cm, width=\columnwidth]{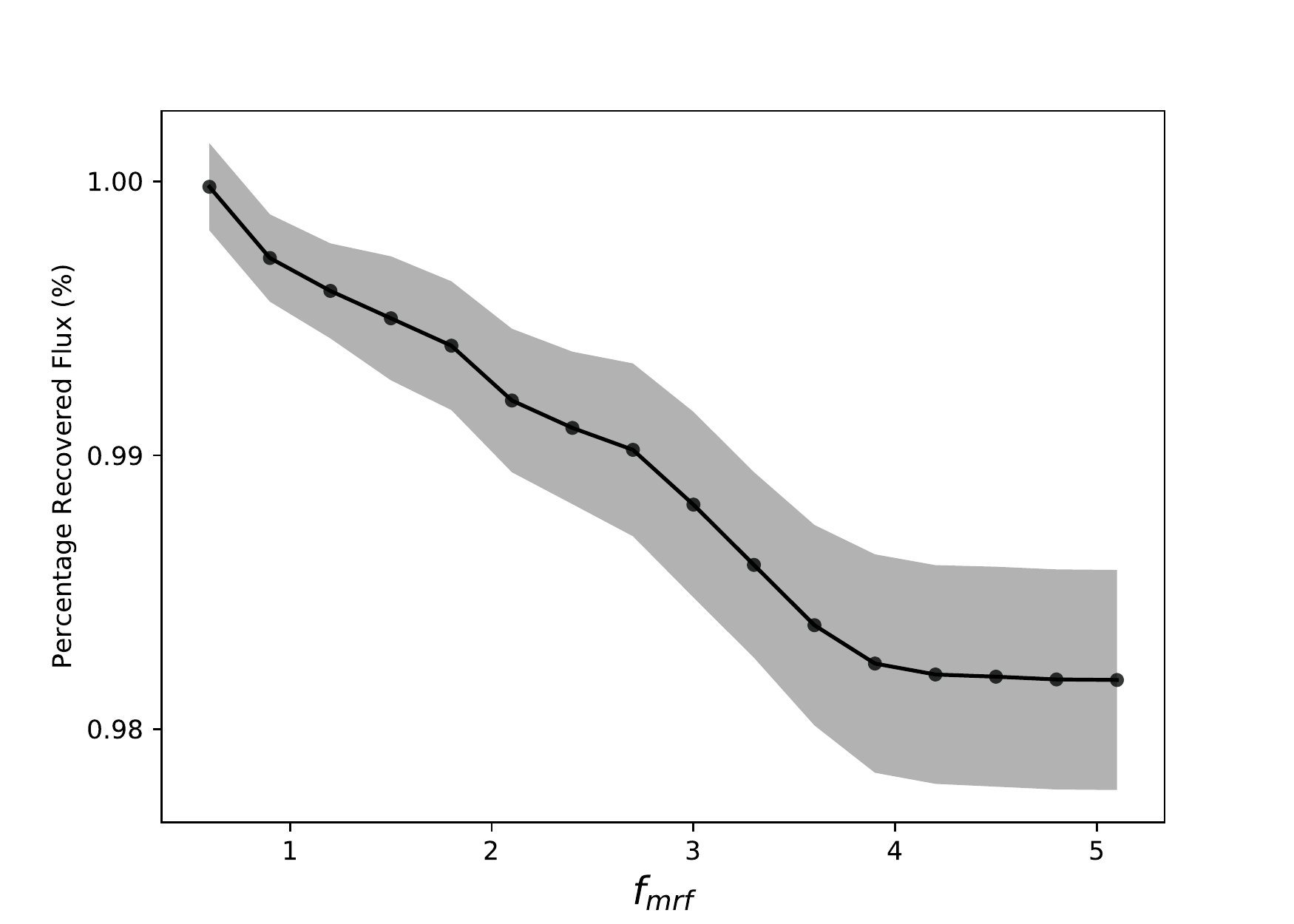}
	\caption{Fraction of recovered flux density vs. $\mathrm{f_{mrf}}$ for an artificial source of 50 mJy that is introduced into an empty field (115050.00-612353.14). Shaded grey region represents the ratio of the rms noise level to the recovered flux density.}
\label{uvdata1}
\end{figure}

In order to determine an average and appropriate $\mathrm{f_{mrf}}$ value for the CORNISH-South data, we followed the same procedure as in \citetalias{cornissh2013}. Artificial point sources and sources with simple morphologies were introduced into the \textit{uv} data of an empty field  and the field was imaged with $\mathrm{f_{mrf}}$ values from 0.5 to 5, using our imaging pipeline. The flux densities of these artificial point sources were then measured and compared with the original flux densities. Figure \ref{uvdata1} shows the resulting plot of the fraction of recovered flux density against the $\mathrm{f_{mrf}}$ values. The range of $\mathrm{f_{mrf}}$ values resulted in a $\mathrm{>}$ 98 per cent recovered flux density. Below $\mathrm{f_{mrf}}$=1.5 the field is over CLEANed with a $\mathrm{>}$40 per cent decrease in the rms noise level, compared to the Stokes V map. 

To further constrain the choice of an average $\mathrm{f_{mrf}}$ value, for each iteration, the residual images of the sources with simple morphologies were inspected for sidelobe structures. At $\mathrm{f_{mrf}}$=1.5, the Stokes I map is CLEANed deeply enough to remove sidelobes and recover $\mathrm{> 99}$ per cent of the flux density. Based on these tests, an average value of $\mathrm{f_{mrf}=1.5}$ was used in the imaging process. For fields with imaging artefacts as a result of very bright sources ($\mathrm{>1 Jy}$) and/or extended sources, we have re-imaged with higher values of $\mathrm{f_{mrf}}$ and manually CLEANed, where necessary. We discuss the procedure for estimating the effect of the clean bias on the CORNISH-South data in Section \ref{cb}.

\subsection{Imaging Extended Sources}\label{ext_art}
The shortest baseline of the ATCA 6A configuration is $\mathrm{\sim}$337 m, which corresponds to a spatial scale of $\mathrm{\sim}$40$\mathrm{\arcsec}$.  Nevertheless, the \textit{uv}-plane at these short baselines is sparsely sampled and the deconvolution algorithm models extended emission as a series of delta functions. Thus, the algorithm struggles to re-construct extended emission due to the difficulty in interpolating the sparsely sampled plane. This is reflected in the image as scattered flux, which results in high rms noise and imaging artefacts, especially in regions with bright and large structures. 

\begin{figure}
\centering
	\includegraphics[height=16cm, width=\columnwidth]{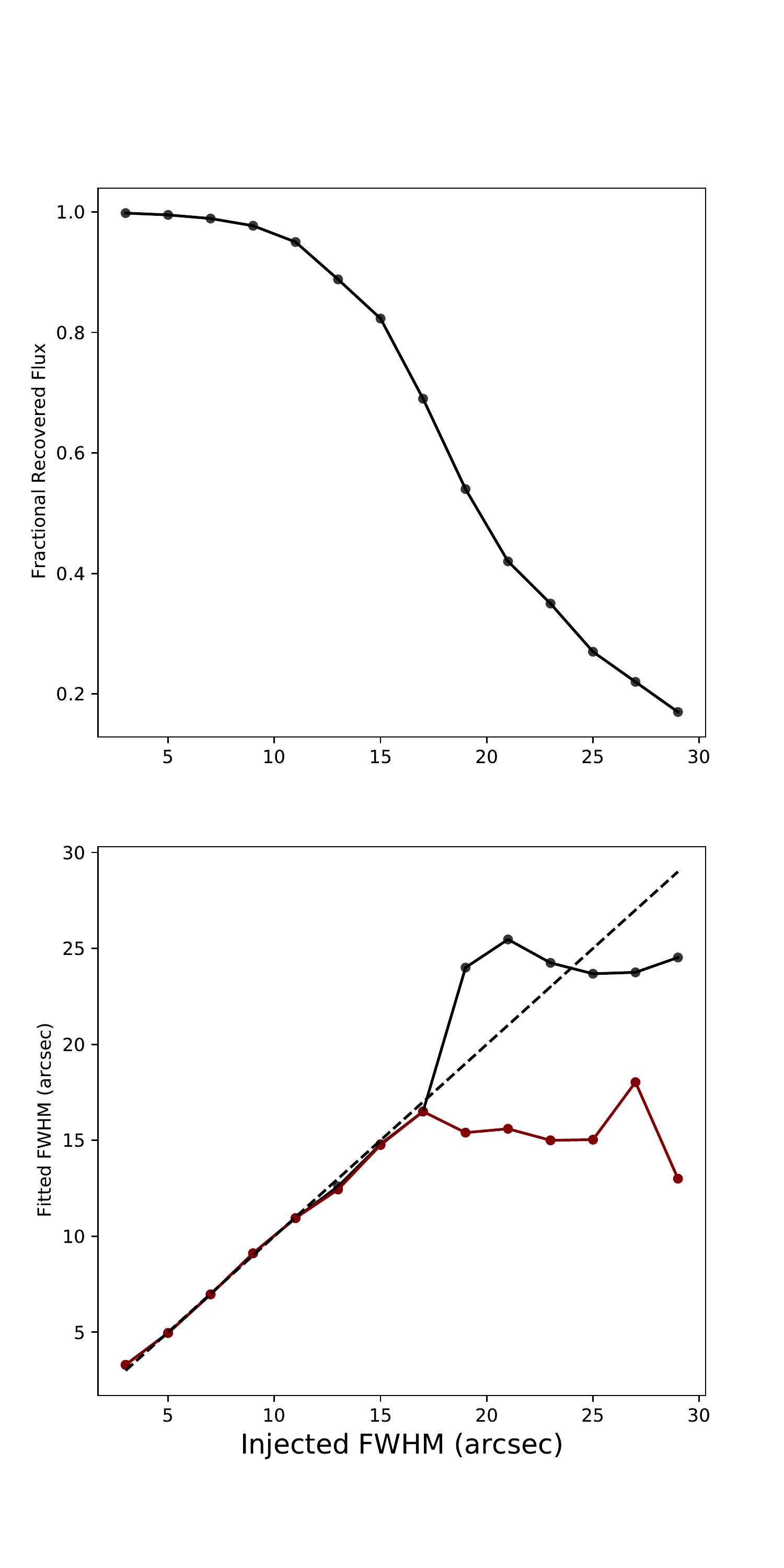}
	\vspace{-1cm}
	\caption{Top panel: Fractional recovered flux density as a function of FWHM for an artificial Gaussian source of 1 Jy. Bottom panel: Recovered sizes as a function of the injected FWHM. The red line traces the Gaussian fits while the black line after 17$\mathrm{\arcsec}$ traces the polygon fit. The black dashed line is a fit to the Gaussian measurements. The transition from  Gaussian fitted sizes to polygon sizes happens between 17$\mathrm{\arcsec}$ and 20$\mathrm{\arcsec}$ (see text for details). Both plots share same x-axis.}
\label{gauss_im}
\end{figure}

To estimate the maximum recoverable angular size above which the deconvolution begins to fail or struggles to recover the flux density of extended sources, we injected a 1 Jy artificial Gaussian source into the \textit{uv} data of an empty field. Holding the flux density constant, the FWHM was increased from 3 to 30$\mathrm{\arcsec}$ in  2$\mathrm{\arcsec}$ steps. The data were imaged with our imaging pipeline and then the injected Gaussian properties were measured for each iteration using the AEGEAN source fitting algorithm (see Section \ref{sourcefind}). Figure \ref{gauss_im} shows the fraction of recovered flux density as a function of the injected Gaussian FWHM (top panel) and a plot of the measured sizes against the injected FWHM (bottom panel).

For sizes above 17$\mathrm{\arcsec}$, the source was fitted with two Gaussians by AEGEAN; the FWHM plotted in Figure \ref{gauss_im} (bottom panel) is the Gaussian with the larger $\mathrm{\theta_{maj}}$ (red line). Our pipeline combines the multiple Gaussians into a polygon and measures the geometric mean as the angular size (see Section \ref{sourcefind}), which is traced by the black line. The maximum FWHM of the fitted Gaussian is $\mathrm{\sim 17\arcsec}$ (red line) as illustrated in Figure \ref{gauss_im} (bottom panel). However, by switching to a polygon that encloses the emission, we can recover up to $\mathrm{\sim 24\arcsec}$ (black line). Above $\mathrm{17\arcsec}$, the corresponding flux density is reduced by $\mathrm{\geq 40}$ per cent as the flux density steeply drops off. For the real data, we expect that the reduction in flux density and limited size of an extended source should be influenced by morphology as well as rms noise level. Compared to the northern counterpart with an estimated maximum size of $\sim \mathrm{14\arcsec}$ \citepalias{cornissh2013}, we expect to see more extended sources in the CORNISH-South catalogue. Most structures that are less than 15\arcsec should give reasonable estimates of the angular size and the integrated flux density is expected to be within a factor of 2 of the true value.

\subsection{Mosaicking}

Individually imaged fields were linearly mosaicked onto $\mathrm{20\arcmin \times 20\arcmin}$ grid tiles, using \textit{linmos}, a MIRIAD task\footnote[13]{\url{https://www.atnf.csiro.au/computing/software/miriad/doc/linmos.html}} that overlap by 1$\mathrm{\arcmin}$). The tiles are arranged in equatorial coordinates (J2000) and 1825 tiles were needed to cover the survey region, extending to the edges of the survey. Tiles covering the edges for $\mathrm{|b|>1^{\circ}}$ may not be full; i.e less than $\mathrm{20\arcmin}$, and some sources at the edges of the survey region may have poor image fidelity due to poor \textit{uv}-coverage. Wide-band primary beam correction, which is the inverse of the primary beam response as a function of the radius and frequency, was performed by \textit{linmos} for each field before linearly mosaicking them to form a tile. \textit{Linmos} uses the $\mathrm{\alpha I}$ plane  and takes into account the OTF scanning during the primary beam correction. In order to improve the accuracy of the primary beam correction, the \textit{`bw'} option in \textit{linmos} was used to specify the bandwidth of the images (2-GHz). 

Linear mosaicking is performed in \textit{linmos} using the standard mosaic equation by minimizing the rms noise \citep{salt1996}. In order to properly account for the geometry and avoid interpolation problems during mosaicking, the overlapping fields were put on the same pixel grid using the `\textit{offset}' key in \textit{invert}. If this is not taken into account, the position of sources in overlapping tiles will be altered because \textit{linmos} does not automatically account for the geometric correction during linear mosaicking.

\section{Data Quality}\label{data_quality}
\subsection{Calibrators}\label{calib_posit}

\begin{figure*}
\centering
	\includegraphics[scale=0.2]{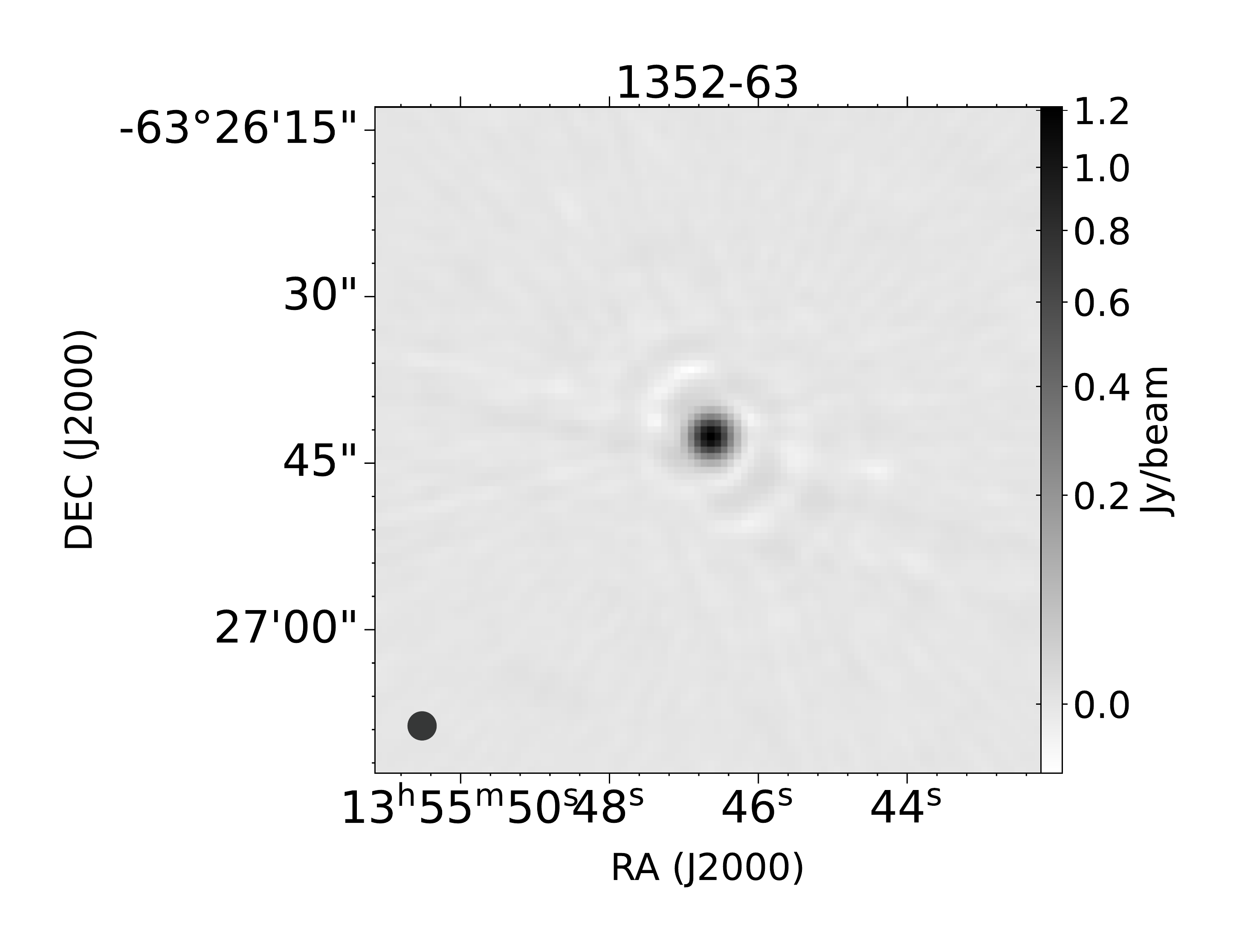}
	\includegraphics[scale=0.2]{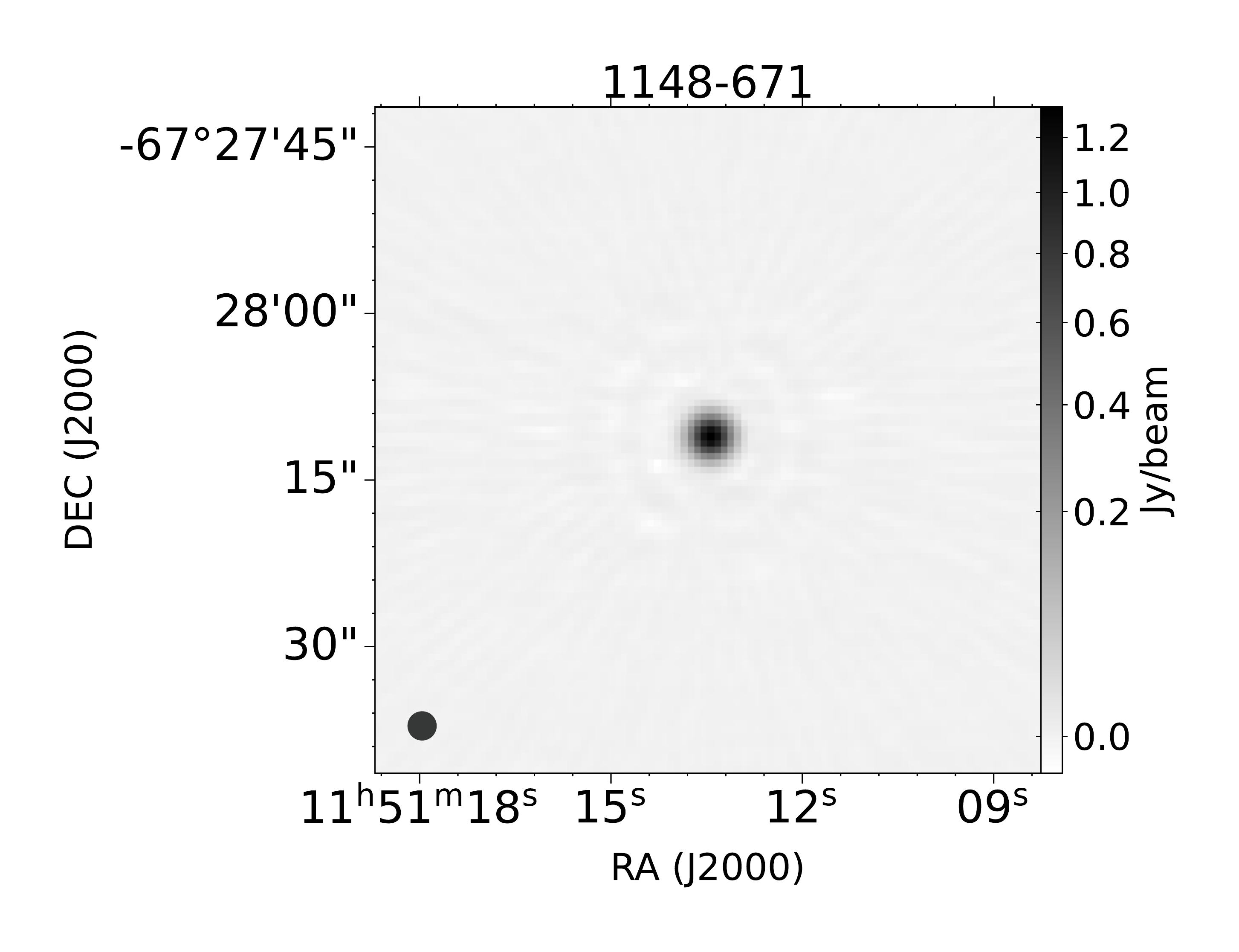}
	\includegraphics[scale=0.2]{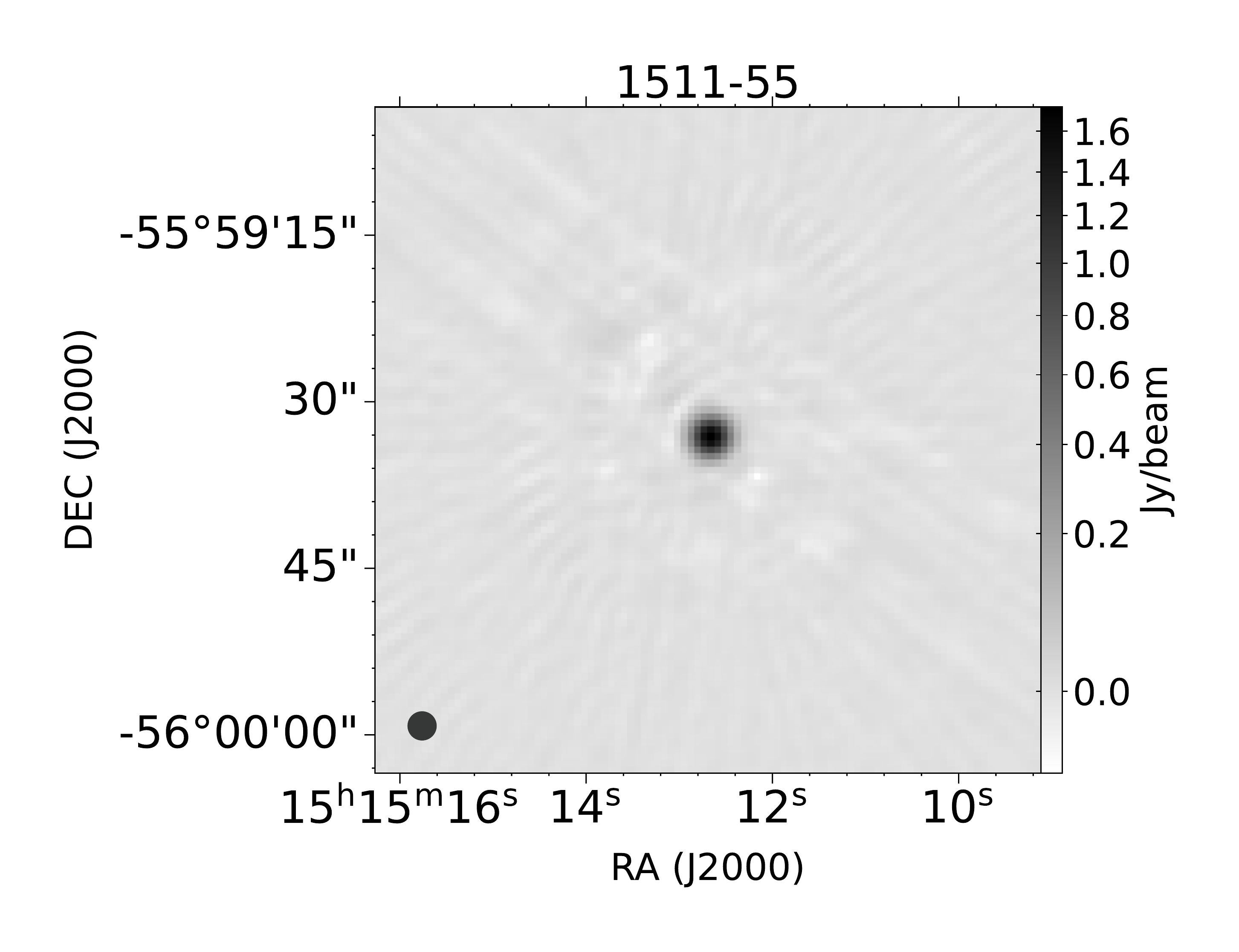}
    \includegraphics[scale=0.2]{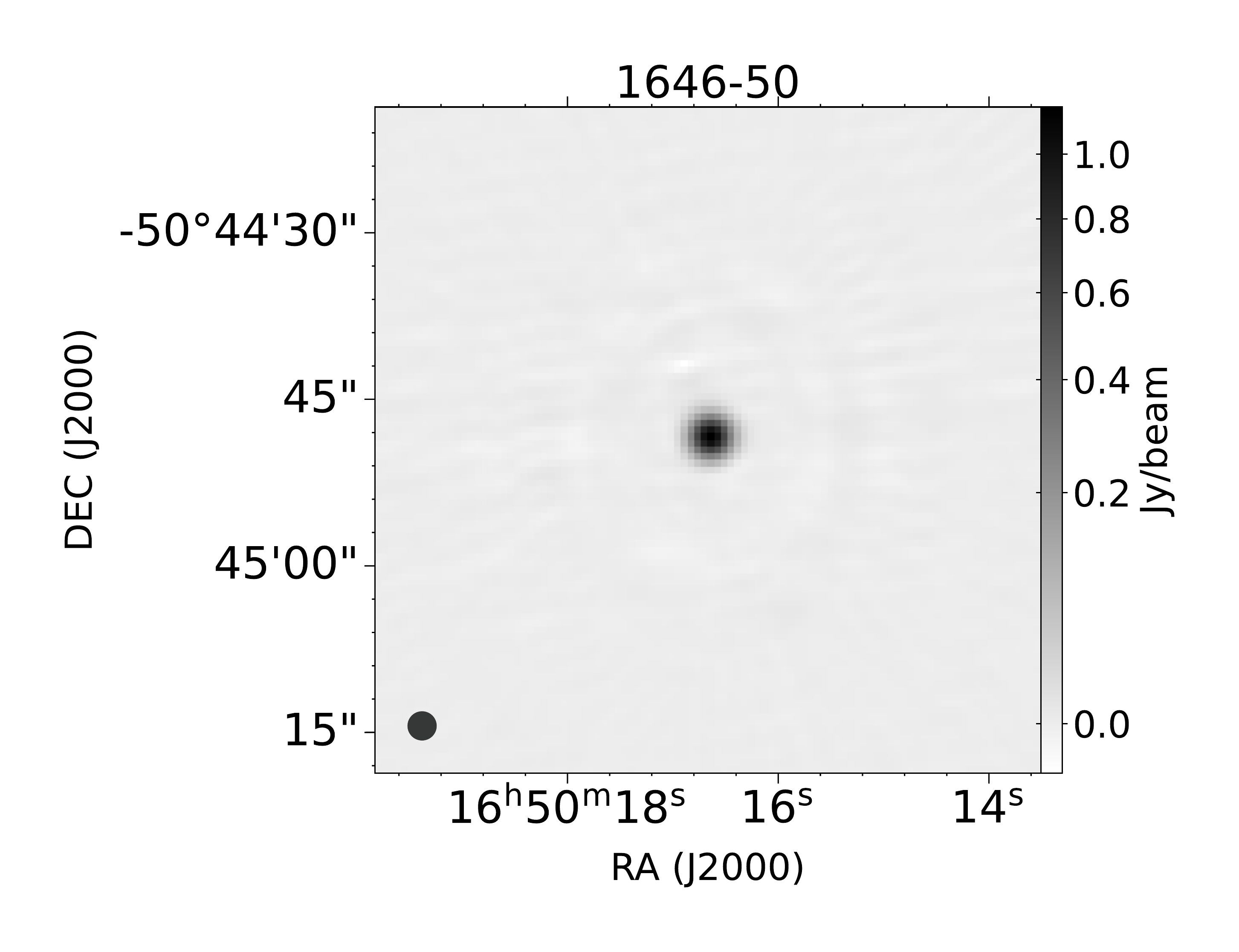}
  	\includegraphics[scale=0.2]{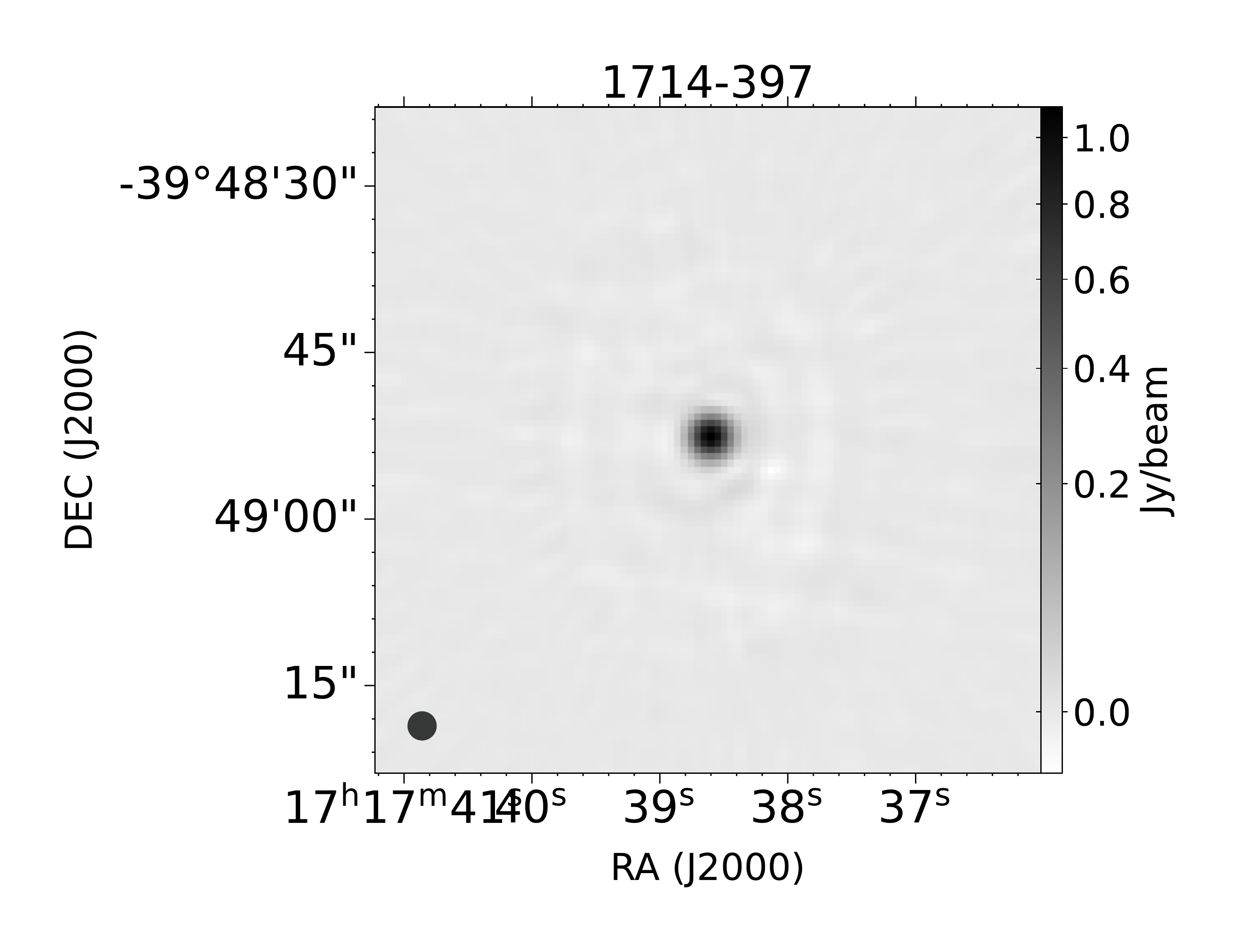}
 	\includegraphics[scale=0.2]{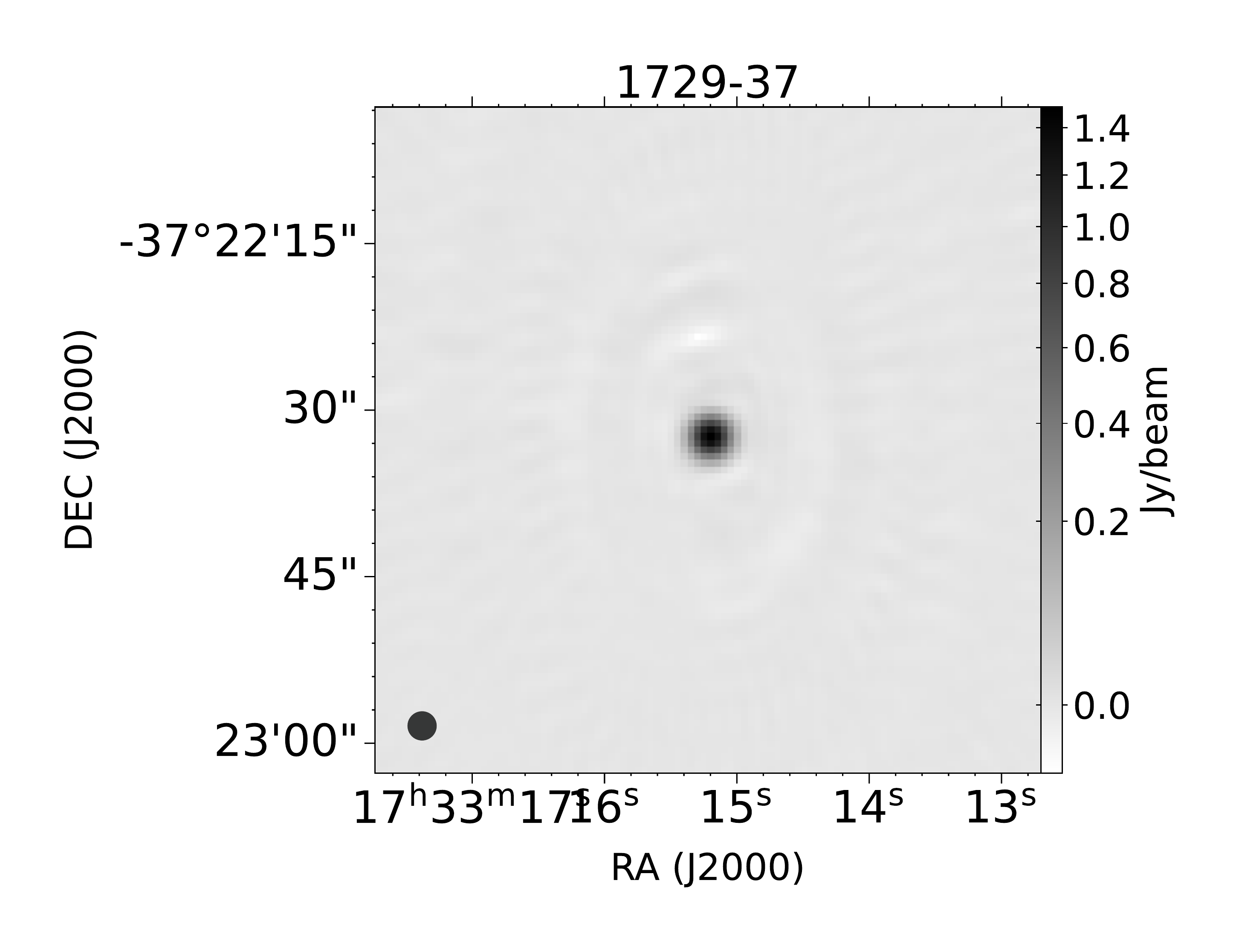}
	\caption{Images of the six secondary calibrators 1352-63 ($\mathrm{1.064 \pm 0.007}$ Jy), 1148-671 ($\mathrm{1.557 \pm 0.007}$ Jy), 1511-55 ($\mathrm{2.338 \pm 0.009}$ Jy), 1646-50 ($\mathrm{2.063 \pm 0.018}$ Jy), 1714-397 ($\mathrm{1.172 \pm 0.006}$ Jy), 1729-37 ($\mathrm{1.73 \pm 0.01}$ Jy). The quoted flux densities are from the ATCA calibrator manual. The images are all 1$\mathrm{^\prime}$ by 1$\mathrm{^\prime}$ with the size of the beam shown at the bottom left.}
\label{calimage}
\end{figure*}

The six secondary calibrators along the Galactic plane are shown in Figure \ref{region_cab}. For observation days with two secondary calibrators, calibration was done separately and then the solutions were combined. Before imaging the science data, the calibrators were imaged. This was to inspect the images for artefacts, jets or anything else that could affect the data. All the secondary calibrators as shown in Figure \ref{calimage} are point sources with no jets or any other structure $\mathrm{\geq 5\sigma}$ within the field.

B1934-638 was the preferred primary flux calibrator with a flux density of 4.95 Jy at 5.5-GHz. An additional backup flux calibrator (B0823-500) with a flux density of 2.93 Jy at 5.5-GHz was also observed at the beginning of each day's observation. B0823-500 was used for flux calibration when the B1934-638 data were bad or when it could not be observed due to time constraints e.g. for blocks 28 and 35 (2011-12-30 and 2012-01-07). The flux densities of the secondary calibrators from the ATCA calibrator manual\footnote{\url{https://www.narrabri.atnf.csiro.au/calibrators/calibrator_database.html}} at 5.5-GHz are  $\mathrm{1.064 \pm 0.007}$ Jy (1352-63), $\mathrm{1.557 \pm 0.007}$ Jy (1148-671), $\mathrm{2.338 \pm 0.009}$ Jy (1511-55), $\mathrm{2.063 \pm 0.018}$ Jy (1646-50), $\mathrm{1.172 \pm 0.006}$ Jy (1714-397) and $\mathrm{1.73 \pm 0.01}$ Jy (1729-37). Because the calibrators are standard calibrators, \textit{mfcal} uses the appropriate flux density variation with frequency during calibration. 

\begin{figure}
\centering
	\includegraphics[height=4.0cm, width=\columnwidth]{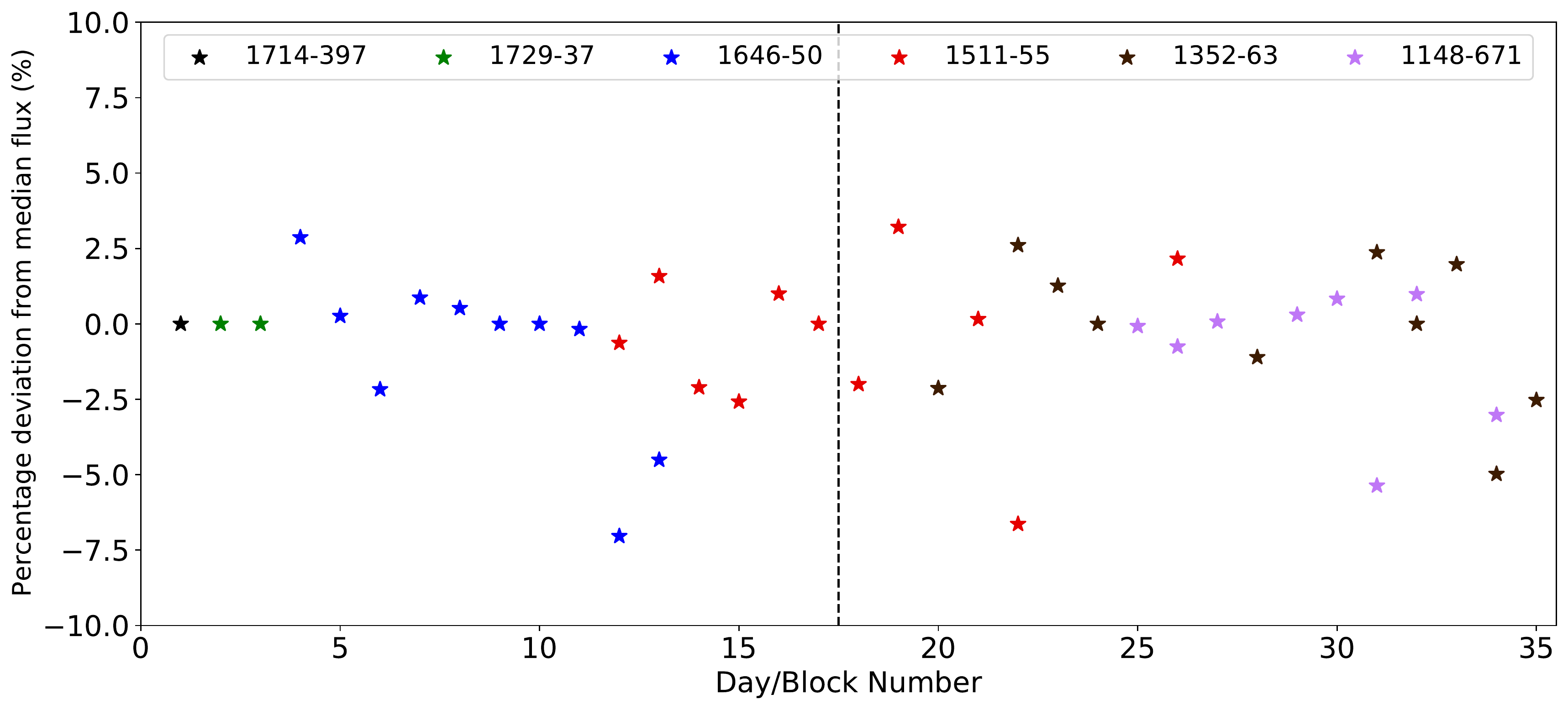}
	\caption{Percentage deviation from the median flux density vs. block/day number for the six secondary calibrators. Each point represents measurements from a block's observations.}
\label{calibta}
\end{figure}

With the available data, the flux densities of the secondary calibrators were measured after flagging and calibration. Figure \ref{calibta} shows the deviation in percentage from the median flux densities of the six secondary calibrators over the 35 days of observations. Each point in Figure \ref{calibta} represents a single day's observations. Different calibrators were used for the different epochs with the exception of 1511-55 that overlaps both epochs. The secondary calibrators show a percentage flux density deviation from the mean flux density that is less than 10 per cent and a standard deviation of 4.0 per cent. Based on the scatter in the flux density deviation of the secondary calibrators in Figure \ref{calibta}, we have adopted a 10 per cent calibration error for the CORNISH-South data. The mean positional accuracy of all six calibrators\footnote{https://www.narrabri.atnf.csiro.au/calibrators/calupdate.html} is 0.1$\mathrm{\arcsec}$.

\subsection{Synthesized Beam}\label{syn_beam}
\begin{figure*}
\centering
	\includegraphics[width=18cm]{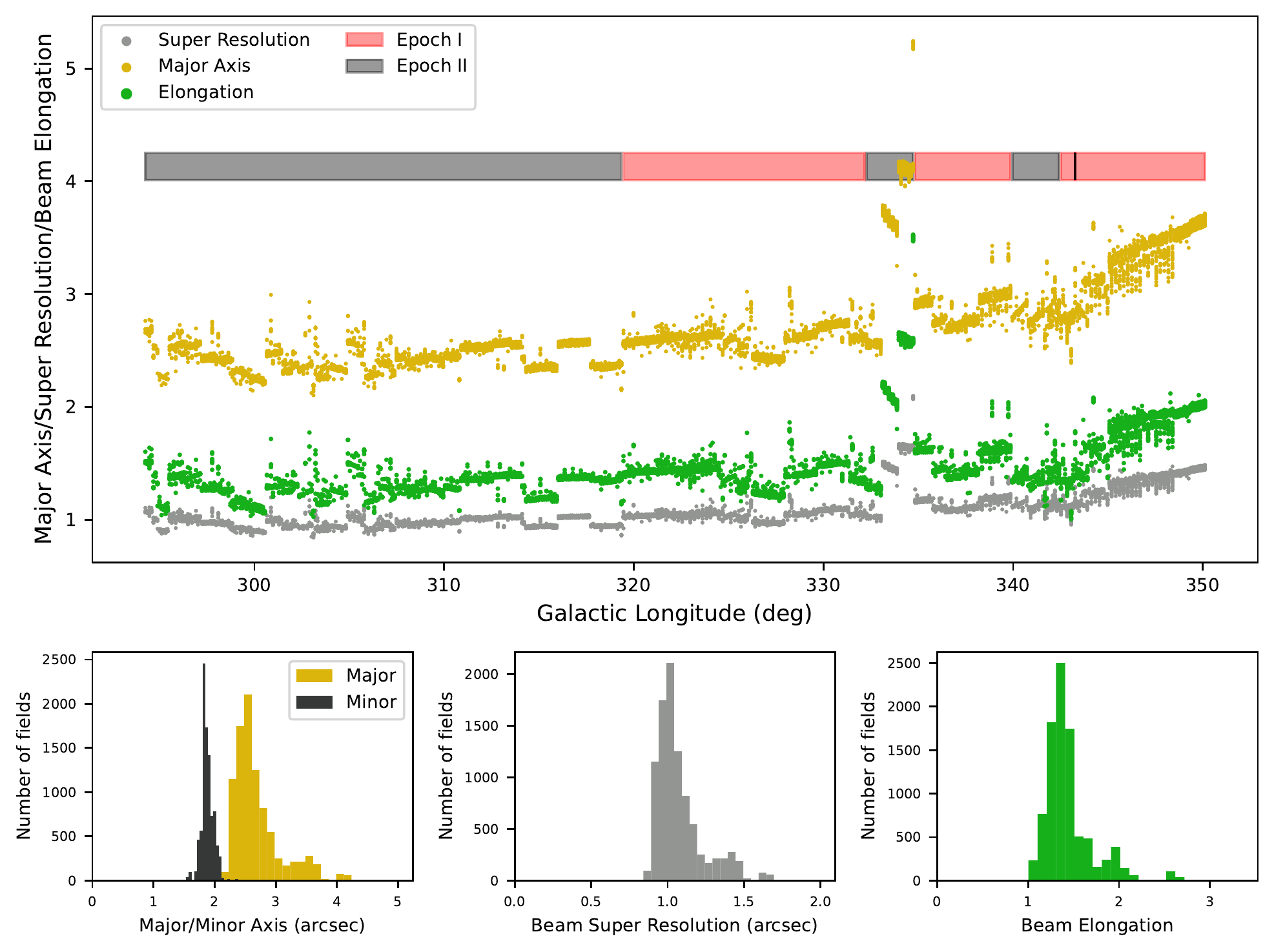}
	\caption{Top panel: Galactic longitude distribution of the major axis and beam elongation before super-resolution is shown. The synthesized beam of an East-West array like ATCA is a function of the declination of the field, hence the greater major axis for l > 344$\mathrm{^{\circ}}$. Additionally, the super-resolution distribution is presented. Bottom panel: Major and minor axes distribution of the imaged fields with a median major ($\mathrm{\theta_{maj}}$) of 2.5$\mathrm{\arcsec}$ and median minor axis ($\mathrm{\theta_{min}}$) of 1.8$\mathrm{\arcsec}$ (bottom left). Forcing a restoring beam size of 2.5$\mathrm{\arcsec}$ (FWHM) will result in the distribution of the beam super-resolution shown in the middle panel (bottom). 66 per cent of the fields have super-resolution that is greater than 1. The beam elongation is the ratio of the major to the minor axis (bottom right) . 96 per cent of the fields have elongation less than 2.}
\label{major}
\end{figure*}

Figure \ref{major} (bottom right) shows the distribution of the unconstrained major and minor axes of the dirty beam. The major axis has a median of $\sim$2.5$\mathrm{\arcsec}$ and extends up to 5.5$\mathrm{\arcsec}$. The minor axis distribution shows a narrower distribution with a peak about 1.8$\mathrm{\arcsec}$. For uniformity across the survey region, the median value of the major axis distribution (2.5$\mathrm{\arcsec}$) was chosen as the size of a circular restoring beam. This results in a super-resolution (major axis/restoring beam of 2.5\arcsec) distribution presented in Figure \ref{major} (bottom middle). Ninety percent (90 per cent) of the fields have super-resolution that is less than 1.3. We also plan to release the calibrated uv dataset so that users can re-image fields with whichever beam they require, as well as convolving the residuals with their chosen beam.

The resulting elongation of the synthesized beam before super-resolution, i.e. the ratio of the major to the minor axis, is shown in Figure \ref{major} (bottom right). Ninety-six percent (96 per cent) of the fields have elongation less than 2 with a peak at $\mathrm{\sim}$1.4. This means that there are a few fields ($\mathrm{<}$4 per cent) where the major axis is 2 or more times greater than the minor axis. The variation of the beam's major axis across the survey region is presented in Figure \ref{major}. The elongation of the synthesized beam of an East-West array like ATCA is a function of the declination of the field, hence the greater elongation for l > 344$\mathrm{^{\circ}}$ for more equatorial declinations. This will also explain the fields with major axis $>3\arcsec$ ($ \mathrm{4\arcsec > \theta_{maj} > 3\arcsec}$) seen within the longitude region >344$\mathrm{^{\circ}}$ (Figure \ref{major} : Top panel). Fields with higher major axes ($\mathrm{> 3.5\arcsec}$) in Figure \ref{major} fall within the longitude 333$\mathrm{^{\circ}}$ to 335$\mathrm{^{\circ}}$ region. Figure \ref{major} (top panel) also highlights the epochs. Epoch II shows a less elongated beam, with major axes lower than 3$\mathrm{\arcsec}$ for $\mathrm{\sim}$93 per cent of the fields. The fields with higher major axes within the longitude 333$\mathrm{^{\circ}}$ to 335$\mathrm{^{\circ}}$ region are seen to come from epoch II. This is due to the poor \textit{uv}-coverage resulting from less than the typical 12 scans for the block observed on 2011-12-20. However, only a few fields were affected, making up $\mathrm{\sim}$6 per cent of the epoch II data.

\subsection{Sensitivity/Root Mean Square (rms) noise Level}

\begin{figure}
	\includegraphics[width=\columnwidth]{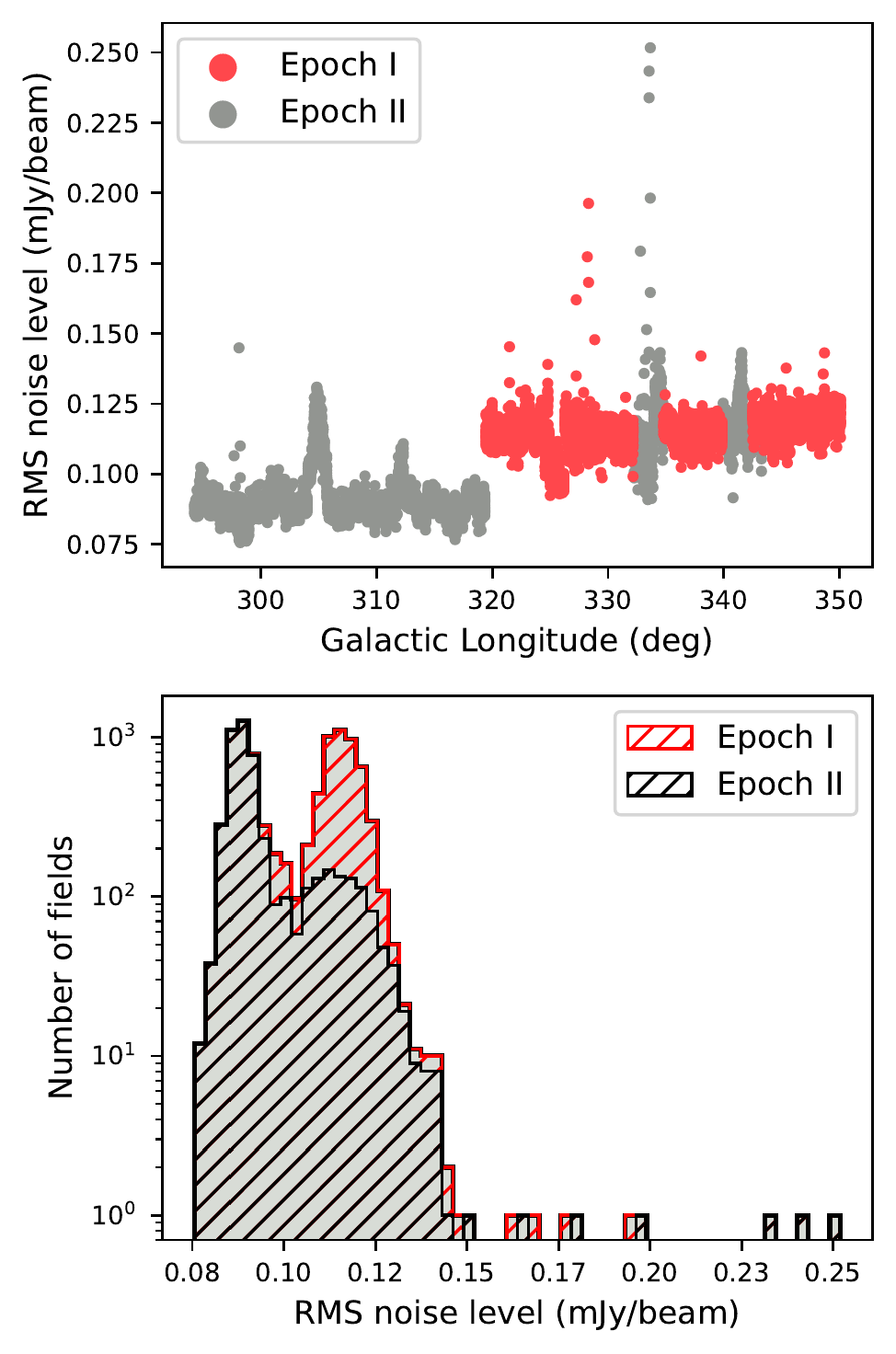}
	\caption{Top panel: The variation of the rms noise in Stokes I maps across Galactic longitude. Bottom panel:  Distribution of the rms noise in Stokes I maps (grey region) measured within an aperture size of $\mathrm{3\arcmin}$. The hatched regions represent the rms noise from the epoch I (red hatched region) and II (black hatched region. The mean rms noise is $\mathrm{\sim 0.11}$ mJy beam$\mathrm{^{-1}}$. The SIQR of epoch I is  0.003 and that of epoch II is 0.004.}
\label{rms_reg}
\end{figure}

The rms noise level achieved across the survey region in CLEANed Stokes I maps is shown in Figure \ref{rms_reg}. The noise level is fairly uniform, having a mean of 0.11 mJy beam$\mathrm{^{-1}}$. The rms noise level around a few very bright source clusters is particularly high as expected. The noisy region between longitude 333$\mathrm{^{\circ}}$ and 335$\mathrm{^{\circ}}$ is seen to reflect the poor uv-coverage seen in Figure \ref{major}. Figure \ref{rms_reg} (bottom panel) further shows a distribution of the rms noise with an elongated tail that corresponds to regions with poor uv-coverages. The two peaks correspond to the different epochs, where the second peak is dominated by epoch I data. The noise level of epoch II data (hatched grey shaded region) is better than that of epoch I (hatched red shaded region). Given that the two epochs are separated by eleven months, the intervening series of system maintenance and tests \footnote{\url{https://www.narrabri.atnf.csiro.au/observing/schedules/2011OctSem/CA.pdf}} could have improved the efficiency and overall performance of the array. The region of epoch II data with rms noise > 0.11 mJy beam$\mathrm{^{-1}}$ corresponds to fields where there were fewer than the average twelve scans (see Section \ref{observation}).

\subsection{Clean Bias}\label{cb}

To estimate the effect of the clean bias by CLEANing down to the MRF (see Equation \ref{eq1}), point sources of random flux densities between 1 and 15 mJy were injected into the \textit{uv} data of six empty tiles chosen from the two epochs. The positions of the sources were chosen to fall about the centres of individual fields, away from any source. The fields were imaged and mosaicked using the imaging pipeline. The flux densities and sizes were then measured from the mosaicked tiles at the injected positions, using our aperture photometry pipeline. This was to avoid flux density bias caused by thermal flux fluctuations \citep{fran2015}. This procedure was repeated ten times to get a better average estimate of the clean bias.

Measured flux densities were subtracted from the injected flux densities to get an estimate of the clean bias effect. The median clean bias from averaging the measurements is estimated to be 0.14 mJy and was not different across the epochs. This value is about half the clean bias estimated for the CORNISH-North catalogue of 0.33 mJy \citepalias{cornissh2013}. For $\mathrm{\geq}$7$\mathrm{\sigma}$ sources, this effect is $\mathrm{<}$ 13 per cent. Thus, we conclude that the clean bias will not significantly degrade the quality of the flux densities compared to the statistical and absolute flux density calibration uncertainty of 10 per cent.

\section{Catalogue}\label{catalog}

\subsection{Source Finding and Characterization}\label{sourcefind}
The automated source finding and source characterization part of our pipeline utilizes the AEGEAN software package\footnote{\url{https://github.com/PaulHancock/Aegean/wiki/Quick-Start-Guide}} \citep{Hancock_Aegean_2012,Hancock_Aegean2_2018}. AEGEAN uses the flood-fill algorithm, where two thresholds ($\mathrm{\sigma_s}$: seeding threshold, $\mathrm{\sigma_f}$: flooding threshold) are defined, such that $\mathrm{\sigma_s\geqslant \sigma_f}$. The seeding threshold is used to seed an island, while the flooding threshold is used to grow the island (see \citealt{Hancock_Aegean_2012}). Detected pixels are then grouped into contiguous islands and characterized by fitting with one or more overlapping 2D Gaussians.

We have used the background and noise estimation function, \textit{Bane} (\citealt{Hancock_Aegean_2012}) to compute the background and rms noise for each tile. \textit{Bane} uses a grid algorithm that estimates the rms noise ($\mathrm{\sigma_{BANE}}$) and background level within a sliding box of a defined size that is centered on a grid point. The pixel values within the box are then subjected to sigma clipping ($\mathrm{3\sigma}$), which reduces any effect source pixels may introduce (\citealt{Hancock_Aegean_2012}; see also \citealt{bertin1996}).  Given that radio images do not have very complicated backgrounds, we do not expect the noise properties across a $\mathrm{20\arcmin\times 20\arcmin}$ tile to change much. Therefore, we used the default boxcar size of $\mathrm{30\theta_{bm}}$ ($\mathrm{\theta_{bm}}$: synthesized beam size) for the CORNISH-South data, which is about 75\arcsec. This size has been demonstrated to be optimized for the completeness and reliability of compact sources \citep{huy2012}. For source finding we defined a 4.5$\sigma_s$ seeding threshold and 4.0$\mathrm{\sigma_f}$ flooding threshold to create an initial catalogue of the CORNISH-South data. 

\subsection{Quality Control}
\subsubsection{Elimination of Duplicate Sources}

With a $\mathrm{60\arcsec}$ overlap of the tiles, sources closer to the edges of the tiles were detected more than once. However, because overlapping regions are formed from the same fields, the difference in position would be a fraction of the synthesized beam. This will also affect the peak flux, depending on the local rms noise. An extended source fitted with multiple Gaussians could have different parameters from one tile to another because it is closer to the edge on one tile and fully imaged on another tile.

To eliminate such duplicated sources, we searched for sources with similar positions, $\mathrm{<2.0\arcsec}$ and similar peak flux, $\mathrm{Peak_{min}/Peak_{max}>0.7}$. In addition to both conditions, the distance of each duplicated source, relative to the centre of the tile, was calculated the source closer to the centre of a tile was retained, over the ones closer to the edges.

\subsubsection{Elimination of Spurious Sources}\label{spurious_sources}
The choice of a cut-off threshold for a catalogue, in terms of the SNR, is a trade-off between completeness and reliability. A low threshold catalogue, e.g. 3$\mathrm{\sigma}$, will result in more real sources but with many unreal sources as well, while a high threshold will result in a highly reliable catalogue but miss real sources with low SNR. In order to determine an appropriate cut-off threshold for the highly reliable CORNISH-South catalogue, we attempted to estimate the number of spurious sources at a given threshold. Based on the analysis in \citetalias{cornissh2013} (see \citealt{hopkins2002}), 15 tiles were selected to represent all 1825 tiles. These tiles were chosen such that there were no sources with very bright side lobes and contained only point sources and fairly extended sources. 

\begin{figure}
	\includegraphics[height=7.5cm, width=\columnwidth]{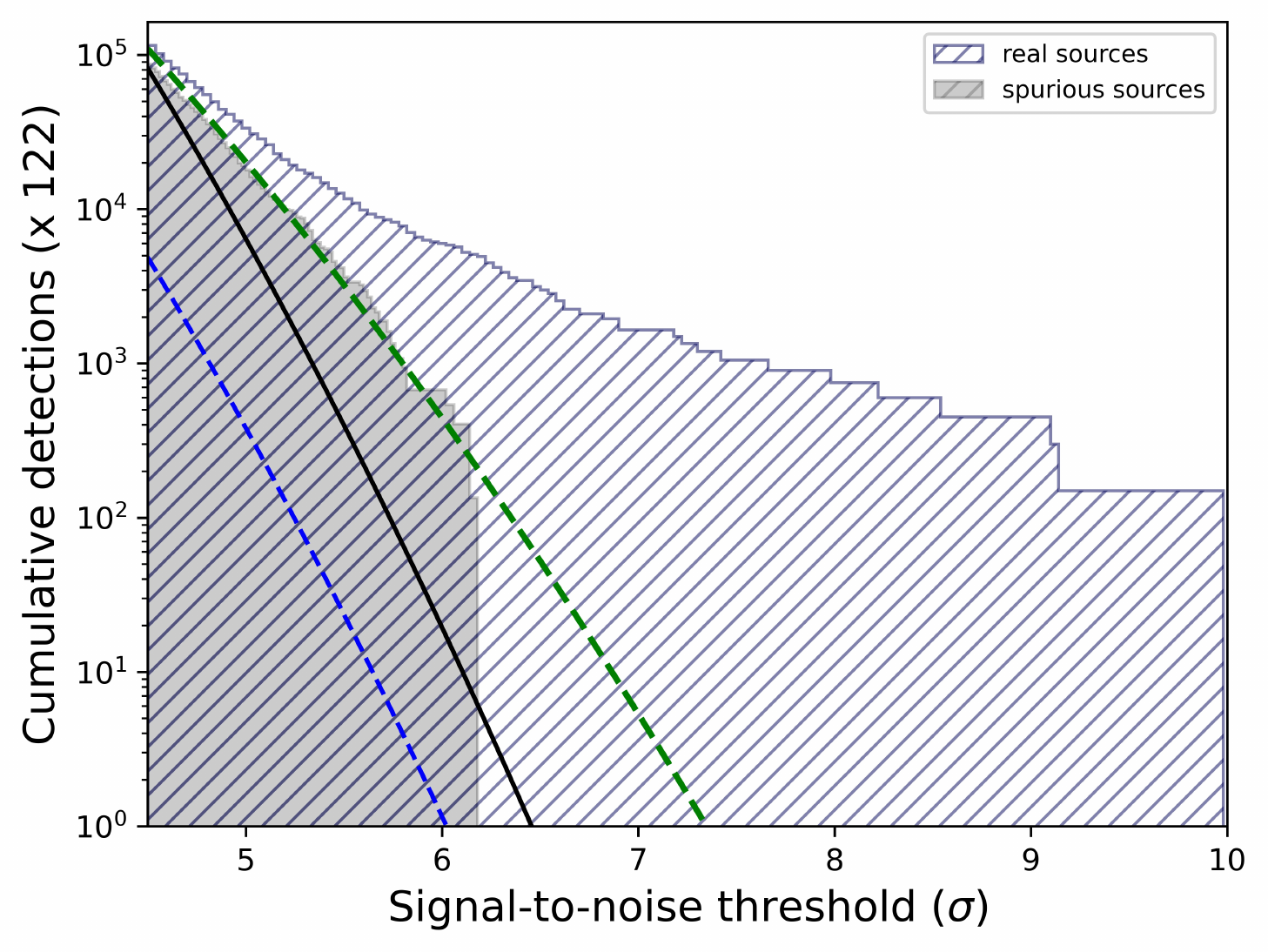}
	\caption{Number of sources as a function of signal-to-noise ratio for 15 real and inverted tiles. The hatched grey region represents spurious sources, while the hatched blue region represents real sources. The blue line is a fit to the spurious sources assuming the total number of possible detections equals the number of beams within the CORNISH survey region ($\mathrm{2.02\times 10^8 \ beams}$).  The black line is the fit to the number of spurious sources by allowing it to be a free parameter and adjusting the curve to fit. The green line represents a fit to bins higher than 5$\mathrm{\sigma_{a}}$ and adjusting the width of the Gaussian ($\mathrm{\sigma=0.8\sigma_{gauss}}$).}
\label{found_sources}
\end{figure}

To estimate the number of spurious sources as a function of SNR, i.e. the ratio of peak flux to rms noise level, the tiles were inverted by multiplying the pixel values in the tiles by -1. A seeding threshold of 4.5$\mathrm{\sigma_s}$ was then used to search for sources on both sets of tiles (normal and inverted tiles).  To account for an average estimate of the number of sources across the survey region, the detections from the 15 tiles were multiplied by 122 (1825/15).  Figure \ref{found_sources} shows a cumulative histogram of the detected sources before and after inversion (the later being obviously all spurious), as a function of the SNR. The rms noise level used for the SNR is measured in an annulus with a 5\arcsec \ gap from the source aperture ($\mathrm{\sigma_a}$: see Section \ref{aperturephot}). Detections below 5$\sigma_{a}$ are dominated by spurious sources ($\mathrm{> 90}$ per cent) as indicated by the grey shaded region. Spurious detections fall off steeply compared to real sources above 5$\mathrm{\sigma_{a}}$ and then fall off to 1 in the 15 tiles between 6.0$\mathrm{\sigma_{a}}$ and 6.5$\mathrm{\sigma_{a}}$.

Following the analysis in \citetalias{cornissh2013}, if the population of detected sources is assumed to be governed by Gaussian statistics, then the fraction, f($\mathrm{\sigma}$), of the population that falls within a given detection threshold can be expressed as 
 
\begin{equation}
\mathrm{f(\sigma)=1-errf(\sigma/\sqrt{2})} \ ,
\label{doi}
\end{equation}

\noindent where errf($\mathrm{\sigma}$) is the Gaussian error function, given by 

\begin{equation}
\mathrm{errf(\sigma)=\frac{1}{\sqrt{\pi}}\int^\sigma_{-\sigma}e^{-t^2}dt\\
=\frac{2}{\sqrt{\pi}}\int^\sigma_{0}e^{-t^2}dt\\}.
\label{errfunct}
\end{equation}

A plot of f($\mathrm{\sigma}$) is presented in Figure \ref{found_sources}, assuming the total number of possible detections equals the number of beams within the CORNISH survey region ($\mathrm{2.02\times 10^8 \ beams}$). With this assumption, the total number of spurious sources is underestimated (blue dashed line). However, the number of sources can be allowed to be a free parameter, resulting in a fit represented by the black line. The black line appears to predict the number of spurious sources at $\mathrm{4.5 \sigma_{a}}$ but falls off rather too steeply, compared to the number of spurious sources. Fitting f($\mathrm{\sigma}$) to bins higher than 5$\mathrm{\sigma_{a}}$ and adjusting the width of the Gaussian ($\mathrm{\sigma=0.8\sigma_{gauss}}$) results in a better fit (green dashed line) and predicts the number of spurious sources to be less than 10 at 7$\mathrm{\sigma_{a}}$. Figure \ref{found_sources} shows that the fraction of spurious sources decreases from 25 per cent to 5 per cent to 1 per cent at 5$\sigma$, 5.5$\sigma$ and 6$\sigma$, respectively. Based on this analysis, the cut-off threshold for a reliable CORNISH-South catalogue to be accepted is set at 7$\mathrm{\sigma_{a}}$.

\subsubsection{Gaussian Sources}\label{aperturephot}

For compact sources fitted with a single Gaussian by AEGEAN, the integrated flux densities we adopt those given by AEGEAN. However, for the rms noise we re-measure this in an annulus around an  aperture ($\mathrm{\sigma_{a}}$). The source aperture was defined by an elliptical aperture which extends to 3$\mathrm{\sigma}$ of the Gaussian major ($\mathrm{\theta_{maj}}$) and minor ($\mathrm{\theta_{min}}$) axes. An annulus with the same shape as the source aperture of width 15$\mathrm{\arcsec}$, offset at 5$\mathrm{\arcsec}$ from the source aperture, was then defined to measure the rms noise and background level. The choice of an annulus offset of 5$\mathrm{\arcsec}$ allows an estimation of the background around the immediate locale of the source. An annulus of width $\mathrm{15\arcsec}$ provides a statistically large area in pixels over which to compute the background and rms noise level, compared to the synthesized beam area of 19.7 pixels. This is more local than $\mathrm{\sigma_{BANE}}$ and is consistent with what was used for the CORNISH-North survey.    \citetalias{cornissh2013} provides equations and further details on the aperture photometry (see also \citetalias{irabor2018}).

\subsubsection{Extended Non-Gaussian Sources}\label{extend_poly}

Extended non-Gaussian sources were automatically detected by searching for contiguous islands with more than one overlapping Gaussian. A single optimal 2D polygon was then defined using the \textit{convex hull} algorithm to trace the outer outline of the Gaussians, enclosing the emission. The assumption is that overlapping Gaussians trace a single extended source. The extent of the generated 2D polygon is strongly affected by the extent of the individual Gaussians. Thus, before generating the polygons, there was the need to make sure the catalogue is clean (i.e. free from sidelobes), otherwise the generated polygon may be over-estimated, stretched by the sidelobes. Additionally, some extended sources that were not properly imaged may appear as individual Gaussians, spreading over an area. In such cases, manual intervention was needed to trace the outline of the real emission. Such cases account for $\mathrm{\sim}$ 10 per cent of non-Gaussian sources. 

\begin{figure*}
\begin{center}
	\includegraphics[height=6cm, width=7.0cm]{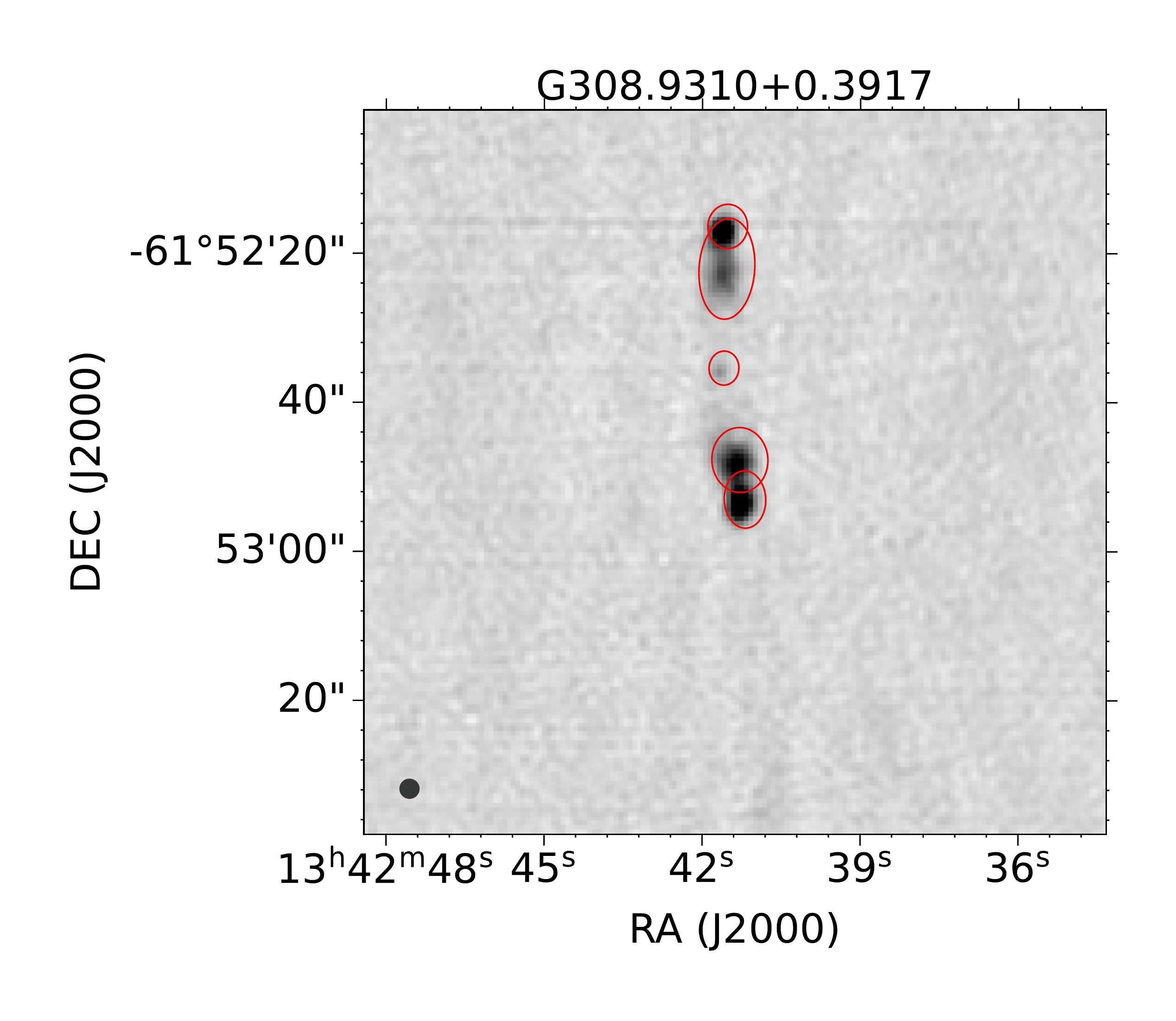}
    \includegraphics[height=6cm, width=7.0cm]{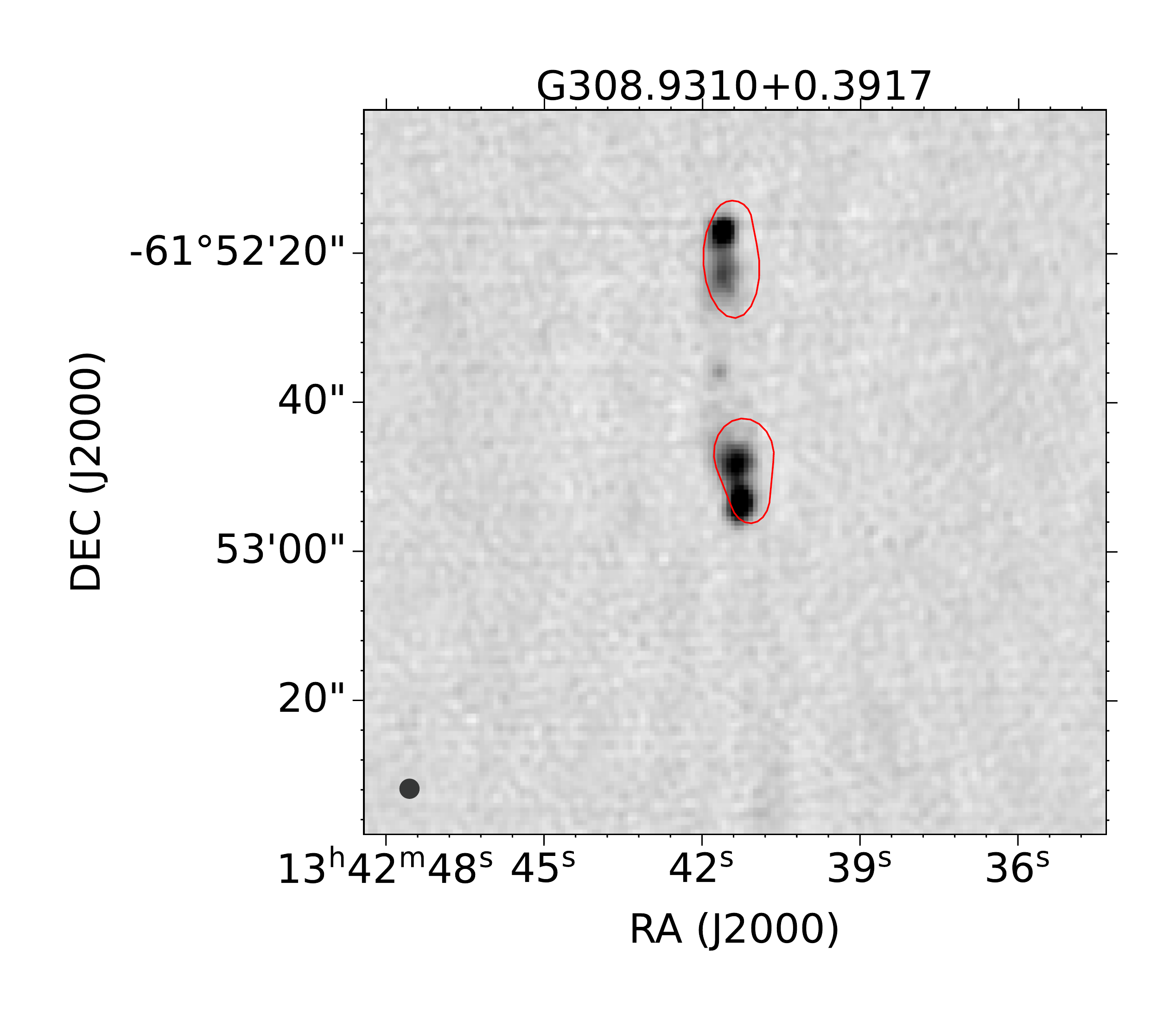}
    \includegraphics[height=6cm, width=7.0cm]{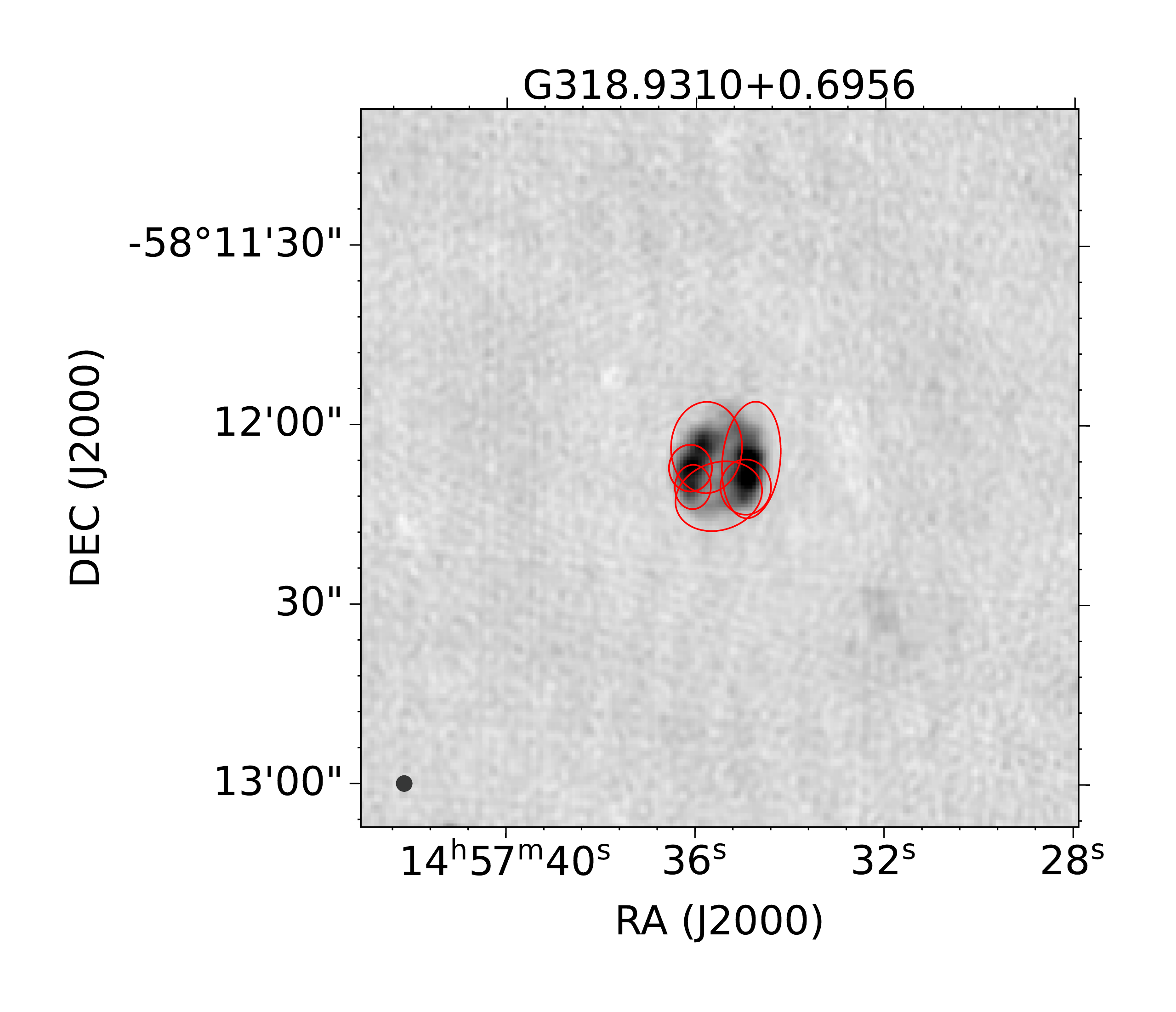}
    \includegraphics[height=6.0cm, width=7.0cm]{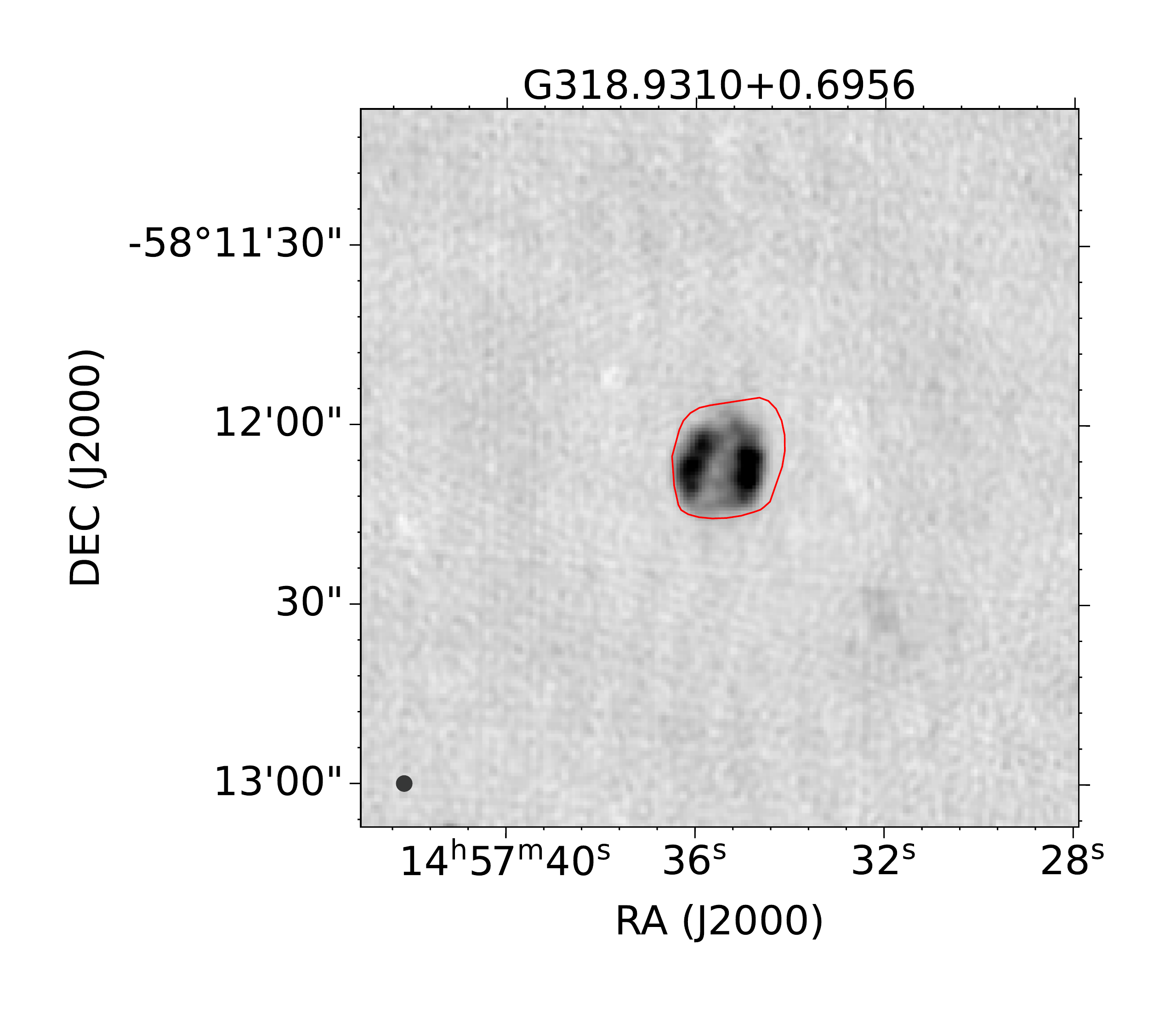}
	\caption{Examples of generated polygons for extended sources fitted with multiple Gaussian by the AEGEAN source finder. The fitted Gaussians are overlaid (left panel), while the defined polygon is shown (right panel). The top source (radio galaxy) is an example of a source that shows the centre source as a Gaussian and the lobes as non-Gaussian sources. The source at the bottom is an example of a double lobe HII region.}
\label{ext_sim}
\end{center}
\end{figure*}

Given the defined polygons, new intensity weighted centres ($\mathrm{\alpha_0}$, $\mathrm{\delta_0}$) and diameters were then determined. The diameter of the 2D polygon, defined by n-sides and n-vertices, was calculated by determining the radius of each vertex to the intensity weighted centre and then estimating the geometric mean of the radii and multiplying by a factor of 2. Figure \ref{ext_sim} shows an example of two sources that were first fitted by multiple Gaussians and the subsequently generated polygons. For these extended sources, aperture photometry was used to measure the source properties within the defined polygon that encloses the emission and the rms noise level within the annulus (see \citetalias{cornissh2013} and \citetalias{irabor2018}). The geometric mean of the 2D polygon is $\mathrm{\large{\theta_E=2\left(  \sqrt[n]{r_1 r_2 ... r_n} \right)}}$, where $\mathrm{r_1, r_2 ...r_n }$ are the radii of the vertices.

Having removed duplicates, spurious sources, and sources $\mathrm{< 7\sigma_a}$ (see Section \ref{spurious_sources}), we then used visual inspection to further eliminate artefacts due to side-lobes that are close to very bright sources. This was aided by comparison with the GLIMPSE \citep{churchwell2009} data. As we are primarily interested in a complete sample of UCHII regions, radio sources in areas with artefacts that have clear IR counterparts were retained. Often these artefacts are caused by bright H II regions themselves and we expect other H II regions to be present in such clustered star forming regions. After these eliminations, the final CORNISH-South catalogue has a total of 4701 high quality sources above 7$\mathrm{\sigma_a}$.

\subsection{Measurements and Uncertainties}\label{meas_uncern}

In order to create a uniform catalogue that is similar to the CORNISH-North catalogue, we have used the same sets of equations given in \citetalias{cornissh2013} to estimate the properties and associated errors of the sources (also see \citealt{con1997}). For the well-defined and unresolved sources, defined by a single Gaussian fit, the AEGEAN Gaussian fit measurements are the catalogued properties. However, for the catalogued rms noise level, we have re-measured within an annulus around the source for both extended and non-extended sources. This was to create a CORNISH (North and South) catalogue with uniform noise measurements, given that we have implemented sigma clipping to remove sources within the annulus (see \citetalias{cornissh2013}).  

The integrated flux density and associated error are given by 
\begin{equation}
\mathrm{S =\frac{A\pi }{4 ln(2)}  \frac{\theta_{maj} \theta_{min}}{\theta^2_{bm}}}
\label{gauss1}
\end{equation}

and  

\begin{equation}
\mathrm{\frac{\sigma^2_S}{S^2} \approx \frac{\sigma^2_A}{A^2}+\frac{\theta^2_{bm}}{\theta_{maj} \theta_{min}}  \left[ \frac{\sigma^2(\theta_{maj})}{\theta^2_{maj}}+\frac{\sigma^2(\theta_{min})}{\theta^2_{min}} \right]} ,
\label{gauss2}
\end{equation}
where $\mathrm{A}$ is the peak amplitude, $\mathrm{\theta_{min}}$ is the minor axis, $\mathrm{\theta_{maj}}$ is the major axis and $\mathrm{\theta_{bm}}$ is the synthesized beam size. $\mathrm{\sigma\theta_{maj}}$ and   $\mathrm{\sigma\theta_{min}}$ are the errors on the Gaussian fits. The catalogued angular size and associated error is the geometric mean of the major and minor axes, which can be calculated from 

\begin{equation}
\mathrm{\theta_{mean}=\sqrt{\theta_{maj}\theta_{min}}}
\label{mean_size1}
\end{equation}

and 
\begin{equation}
\mathrm{\sigma (\theta_{mean})=\frac{\theta_{mean}}{2}\sqrt{\frac{\sigma^2 (\theta_{maj})}{\theta_{maj}^2}+\frac{\sigma^2 (\theta_{min})}{\theta_{min}^2}}}
\label{mean_size_err}
\end{equation}

\begin{figure}
\centering
	\includegraphics[width=\columnwidth]{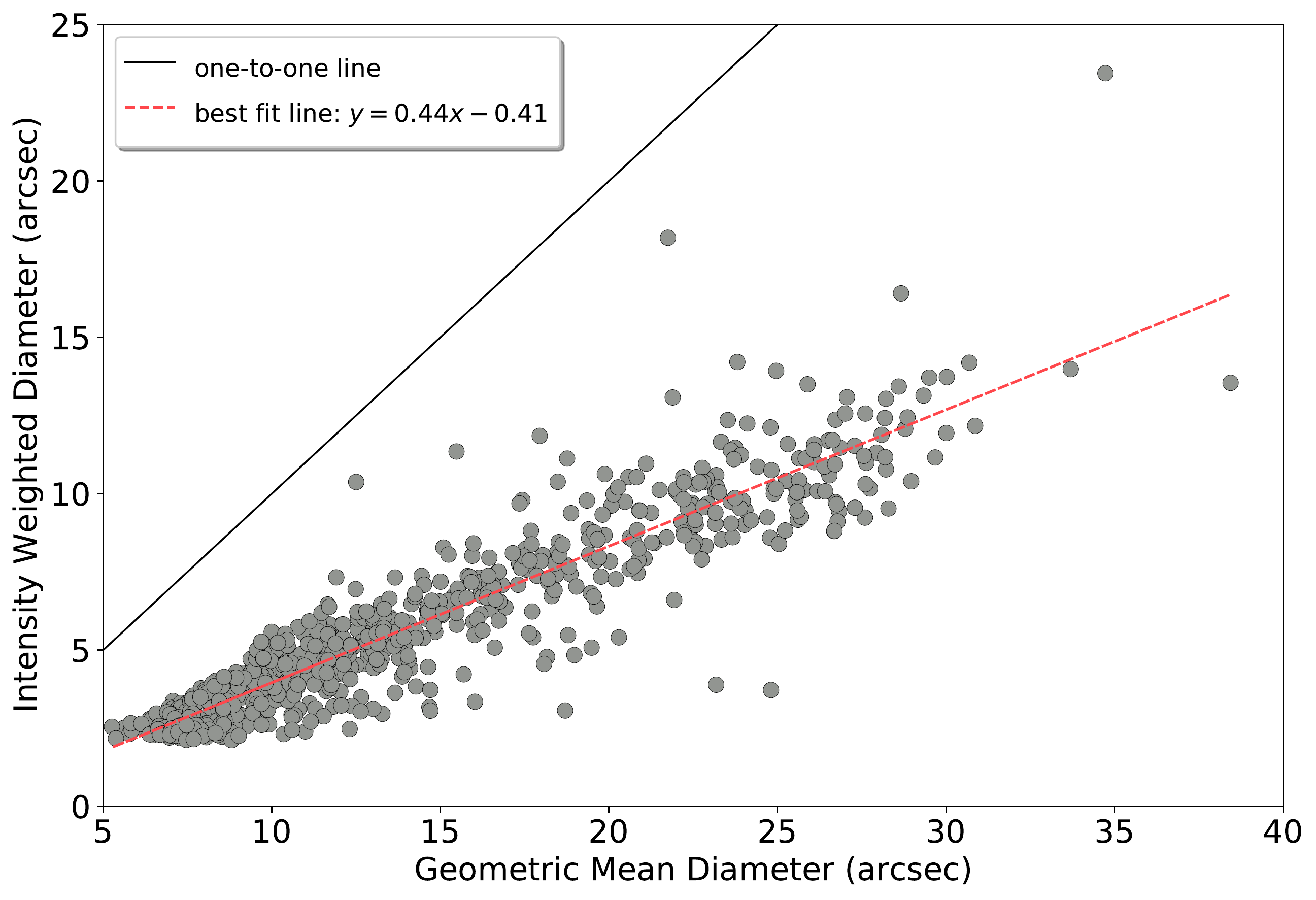}
	\caption{A plot of the intensity-weighted diameter against the geometric mean diameter for the polygonal sources. The one-to-one line (black) shows that the geometric mean diameter is consistently larger than the intensity weighted diameter. The best-fit line (green) is given by $\mathrm{y=0.44x - 0.41}$.}
\label{mea_ext}
\end{figure}

The integrated flux density of the polygonal sources, measured using aperture photometry, is given by  
\begin{equation}
\mathrm{S= {\left( \sum^{N_{src}}_{i=1} A_i - N_{src}\overline{B}    \right)   \bigg/a_{bm}}}\ ,
\label{phot_poly1}
\end{equation}
where $\mathrm{\sum^{N_{src}}_{i=1} A_i}$ is the total flux density in a given aperture over $\mathrm{N_{src}}$ pixels, $\mathrm{\overline{B}}$ is the median background flux and $\mathrm{a_{bm}}$ is the beam area (19.66 pixels). The associated error is given by 

\begin{equation}
\mathrm{\sigma^2_S=\left(  \sigma (\sum A_i)^2 +\frac{\pi N^2_{src} \sigma^2_g}{2N_{sky}} \right) \bigg /a^2_{bm}}\ ,
\label{phot_poly2}
\end{equation}
where $\mathrm{\sigma^2_g}$ is the variance and $N_{sky}$ is the number of pixels in the annulus.  For CORNISH-North, the angular sizes for the polygonal sources were intensity-weighted diameters and were given by 
\begin{equation}
\mathrm{d_{w} = \sum^{N_{src}}_{i=1} {r_i A_i } \bigg /\sum^{N_{src}}_{i=1} A_i} \ ,
\label{singing}
\end{equation}
where $\mathrm{d_w}$ is the intensity-weighted diameter and $\mathrm{\sum^{Nsrc}_{i=1}A_i}$ is the sum of the flux within the defined source aperture. This works well for simple extended sources.  However, the sizes of very extended and double-lobed sources could be under-estimated by $\mathrm{\geq}$50 per cent, in some cases. Figure \ref{mea_ext} shows a comparison of the intensity-weighted diameters and the geometric mean diameters for the polygonal sources (see Section \ref{extend_poly}). The intensity-weighted diameters are consistently smaller compared to the geometric mean diameters. The best-fit line to the scatter is given by: $\mathrm{y=0.44x-0.41}$. Based on this, the catalogued size for a polygonal source is the geometric mean. This can be useful in interpreting and re-scaling the CORNISH-North sizes for extended sources.

\subsection{Completeness}\label{incomp}
It is important to demonstrate the completeness of our 5.5-GHz catalogue as a function of flux density. In order to quantify this, artificial point sources were injected into the calibrated \textit{uv}-data of nine fairly empty tiles having no imaging artefacts, fairly homogenous noise distribution and from both epoch I and epoch II. The flux densities of the artificial point sources were chosen to be in the range of 0.2 mJy ($\mathrm{\sim 2\sigma}$) to 4 mJy ($\mathrm{\sim 40\sigma}$). The positions and flux densities were randomly assigned, while avoiding positions of real sources and making sure that the artificial sources do not overlap. In total, 5000 sources were injected into the tiles. The tiles were imaged, and the properties of the injected sources were measured with the same pipeline that produced the catalogue. This procedure was repeated 10 times and then the average values of the measured properties over the 10 iterations were compared to the injected properties.

\begin{figure}

\includegraphics[height=7cm, width=\columnwidth]{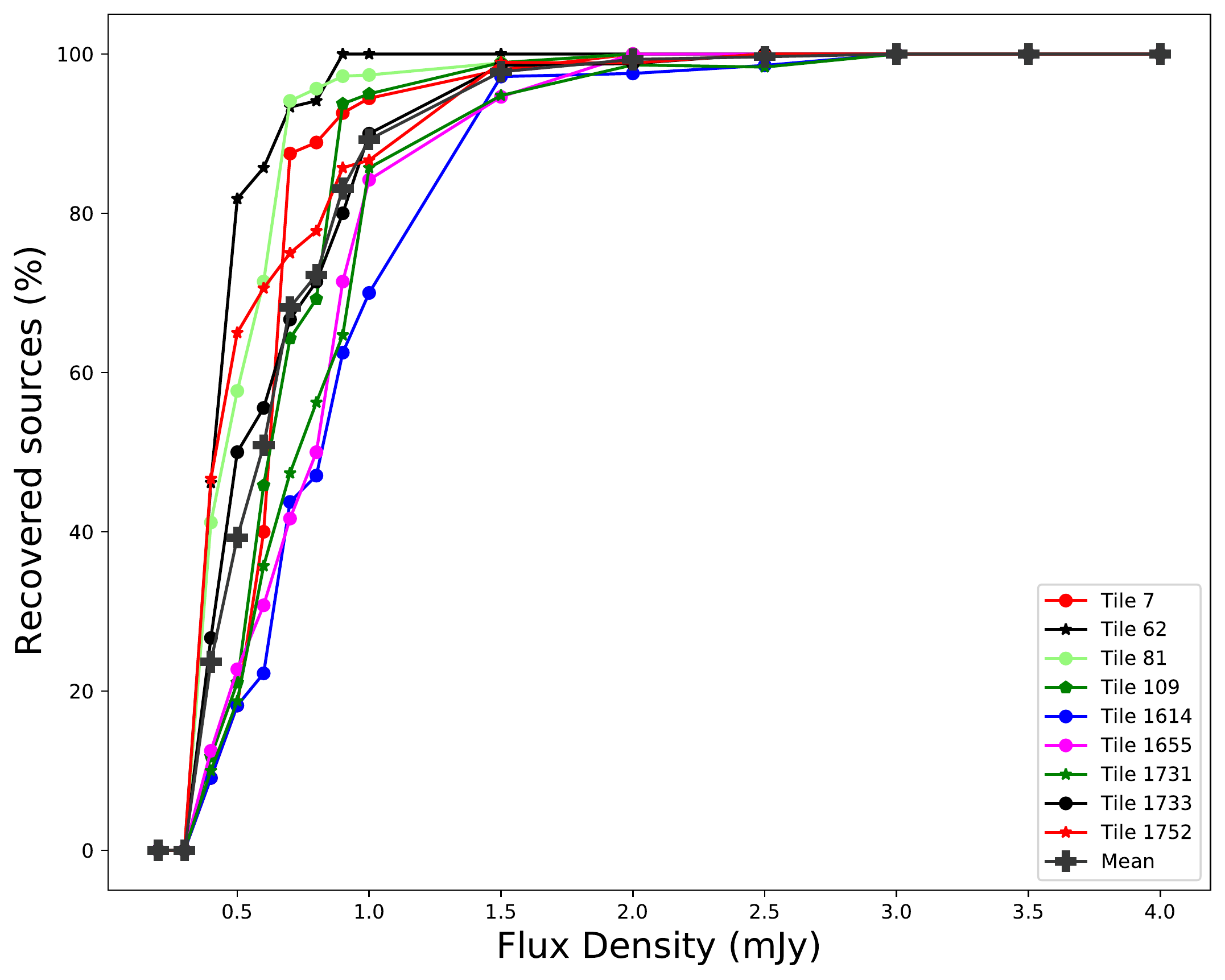}
\caption{Percentage completeness as a function of flux density for artificial point sources injected into 9 representative tiles. Completeness is the number of extracted sources divided by number of injected sources at binned flux densities. The black line with the '+' symbol shows the mean completeness.}
\label{comp_level}
\end{figure}

Figure \ref{comp_level} shows the completeness level measured by the percentage of detected sources as a function of their injected flux densities. The mean completeness level (percentage) is also shown as a black line with the '+' symbol. The percentage completeness from the graph shows > 90 per cent for 1.5 mJy and essentially 100 per cent for > 3 mJy.  The completeness for <0.3 mJy is 0 per cent because it is below the seeding threshold of 4.5$\mathrm{\sigma_s}$. Table \ref{comp_tab} shows the noise level and completeness for the individual tiles at 50 per cent and 90 per cent. Tiles from epoch I with higher noise levels show lower completeness level compared to tiles from epoch II. The stated rms noise level is an average across each tile and so the completeness will be affected by local rms noise surrounding a given source. At 1.4 mJy ($\mathrm{\sim}$7$\mathrm{\sigma}$), the percentage completeness is about 90 per cent for the worst case (Tile 1614).  Based on the mean completeness level, for point sources, the CORNISH-South data is 90 per cent complete at 1.1 mJy. The completeness will be worse around very bright sources, but as can be seen from Figure \ref{rms_reg}, this represents less than 0.3 per cent of the total area of the survey.  

\begin{table}
\begin{center}
\caption{Completeness level of the 5.5-GHz CORNISH-South data across 9 representative tiles at 50 per cent and 90 per cent.}\label{complete_cal}

\begin{tabular}{l|c|c|c|l|c}\hline 
\multicolumn{1}{c}{\textbf{Tile}}
& \multicolumn{1}{c}{\textbf{Epoch}}
& \multicolumn{1}{c}{\textbf{rms}}
& \multicolumn{1}{c}{\textbf{50 per cent}}
& \multicolumn{1}{l}{\textbf{90 per cent}}\\  
&&(mJy beam$\mathrm{^{-1}}$)&mJy&mJy\\

\hline

7&II &0.09&0.63&0.85\\
62&II &0.09&0.43&0.64\\
81&II &0.09&0.45&0.68\\
109&II &0.10&0.63&0.88\\
1614&I &0.19&0.81&1.38\\
1655&I &0.14&0.80&1.30\\
1731&I &0.12&0.72&1.27\\
1733&I &0.12&0.51&1.02\\
1752&I &0.16&0.42&1.14\\
Mean& I \& II&0.12&0.60&1.09\\

\hline
\end{tabular}
\label{comp_tab}
\end{center}
\end{table}

\subsection{Catalogue Ensemble Properties}
Figures \ref{popper6} to \ref{supwlat} present the distribution of the ensemble physical properties of the CORNISH-South sources. We identified 4701 sources above the 7$\mathrm{\sigma}$ limit, of which the properties of 608 are measured with a polygon. 

\subsubsection{Angular Size}
\begin{figure}
\centering
	
	\includegraphics[height=6.5cm, width=\columnwidth]{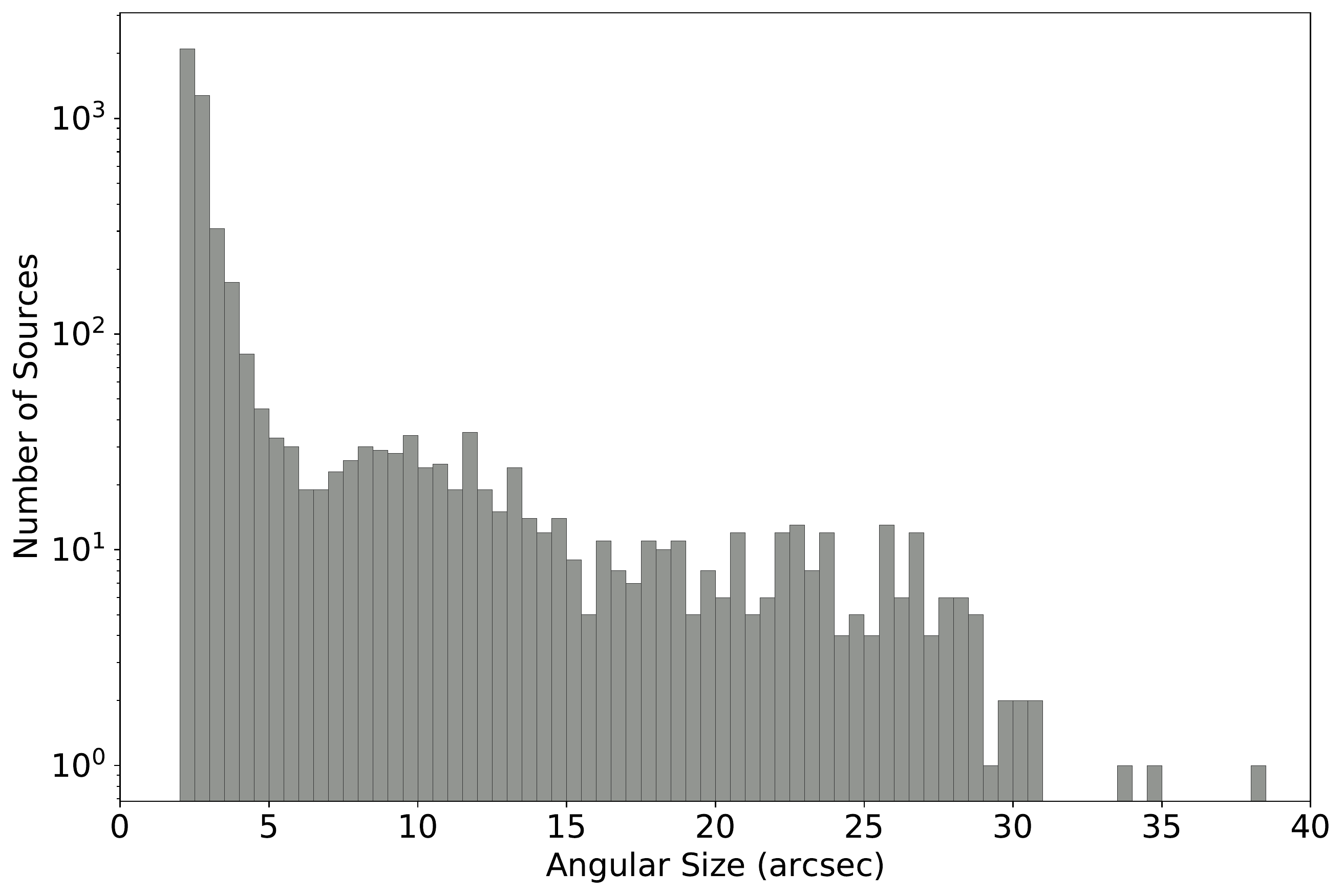}
	\caption{Angular size distribution of the 7$\mathrm{\sigma}$ CORNISH-South sources.}
\label{popper6}
\end{figure}

The catalogued angular size for both the Gaussian and non-Gaussian sources is the geometric mean (see section \ref{meas_uncern}). Based on the mean error of the angular sizes, which is $\mathrm{\sim 0.3\arcsec}$ and the size of the restoring beam ($\mathrm{2.5\arcsec}$), resolved sources are defined as sources with angular sizes $\mathrm{>2.8{''}}$ for the CORNISH-South catalogue. The angular size distribution in Figure \ref{popper6} is dominated by unresolved sources (66.3 per cent) and accounts for the obvious peak at $\mathrm{\sim 2.5\arcsec}$.  Resolved sources (1584) account for 33.6 per cent of the catalogue, of which 38 per cent are polygonal sources (608). The distribution of resolved sources is fairly flat out to 30$\mathrm{\arcsec}$ after a steep drop from 2.5$\mathrm{\arcsec}$ to 5$\mathrm{\arcsec}$. This is consistent with the maximum recoverable size (see Figure \ref{gauss_im}).

As characteristic of all interferometric observations, very extended emission will not be properly imaged due to missing information on large scale structures, limited by the shortest baseline in the array. Thus, caution should be applied in interpreting the angular sizes and flux densities of very extended sources ($>17\arcsec$). This is also demonstrated in Section \ref{ext_art}.  

\subsubsection{Galactic Latitude and Longitude Distributions}

\begin{figure}
\centering
\begin{subfigure}[l]{0.45\textwidth}
    \caption{Galactic latitude distribution.}\label{popper1}
	\includegraphics[height=6.5cm, width=\textwidth]{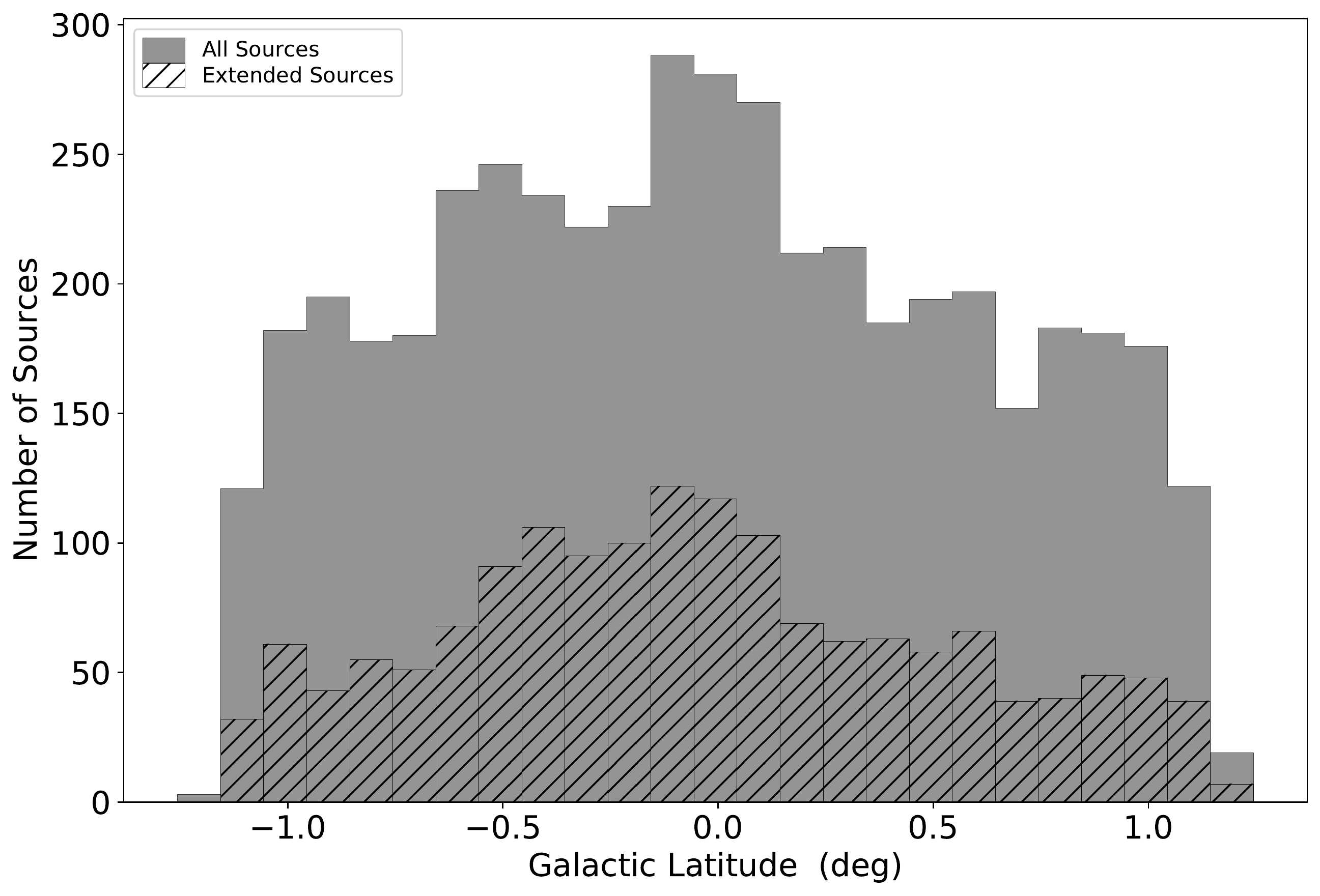}
\end{subfigure}	
\begin{subfigure}[c]{0.45\textwidth}
	\caption{Galactic longitude distribution with a bin size of $\mathrm{2^{\circ}}$.}\label{popper2}
	\includegraphics[height=6.5cm, width=\textwidth]{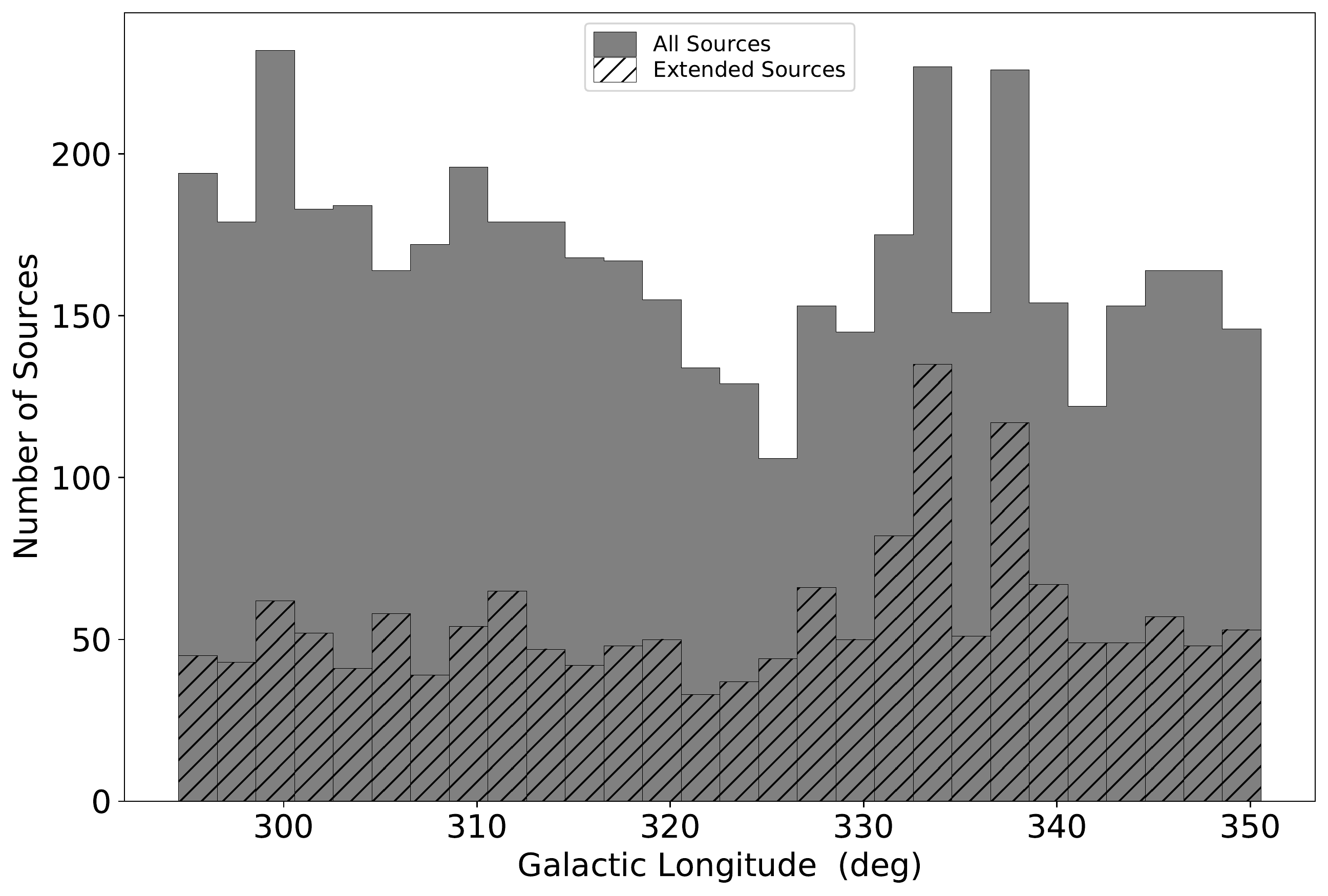}
\end{subfigure}	
	\caption{Galactic latitude (\ref{popper1}) and longitude (\ref{popper2}) distributions of the 7$\mathrm{\sigma}$ CORNISH-South sources.}
\label{suppop}
\end{figure}

Figures \ref{popper1} and \ref{popper2} show the distributions of the CORNISH-South sources in Galactic latitude and longitude, respectively. The coverage of the CORNISH-South survey is complete within the $\mathrm{|b|\leqslant 1.0}$ region as shown in Figure \ref{popper1}. The distributions are similar compared to the Galactic distributions of the CORNISH-North catalogue \citepalias{cornissh2013}.

The latitude and longitude distributions of the resolved sources correspond to the Galactic region traced by high-mass star formation \citep{urq2011,urq2009}. Known star formation complexes G333 and G338.398+00.164, can be seen at $\mathrm{l=333^{\circ}}$ and $\mathrm{l=338^{\circ}}$, respectively \citep{urq2013,urq_2013}. Based on the Galactic distribution of the resolved sources, they are expected to be dominated by HII regions that are concentrated towards the Galactic mid-plane (\citealt{urq2013}, \citetalias{kalprep}).

\subsubsection{Flux Density Distribution}
\begin{figure}
\centering	

	\includegraphics[height=6.5cm, width=7cm]{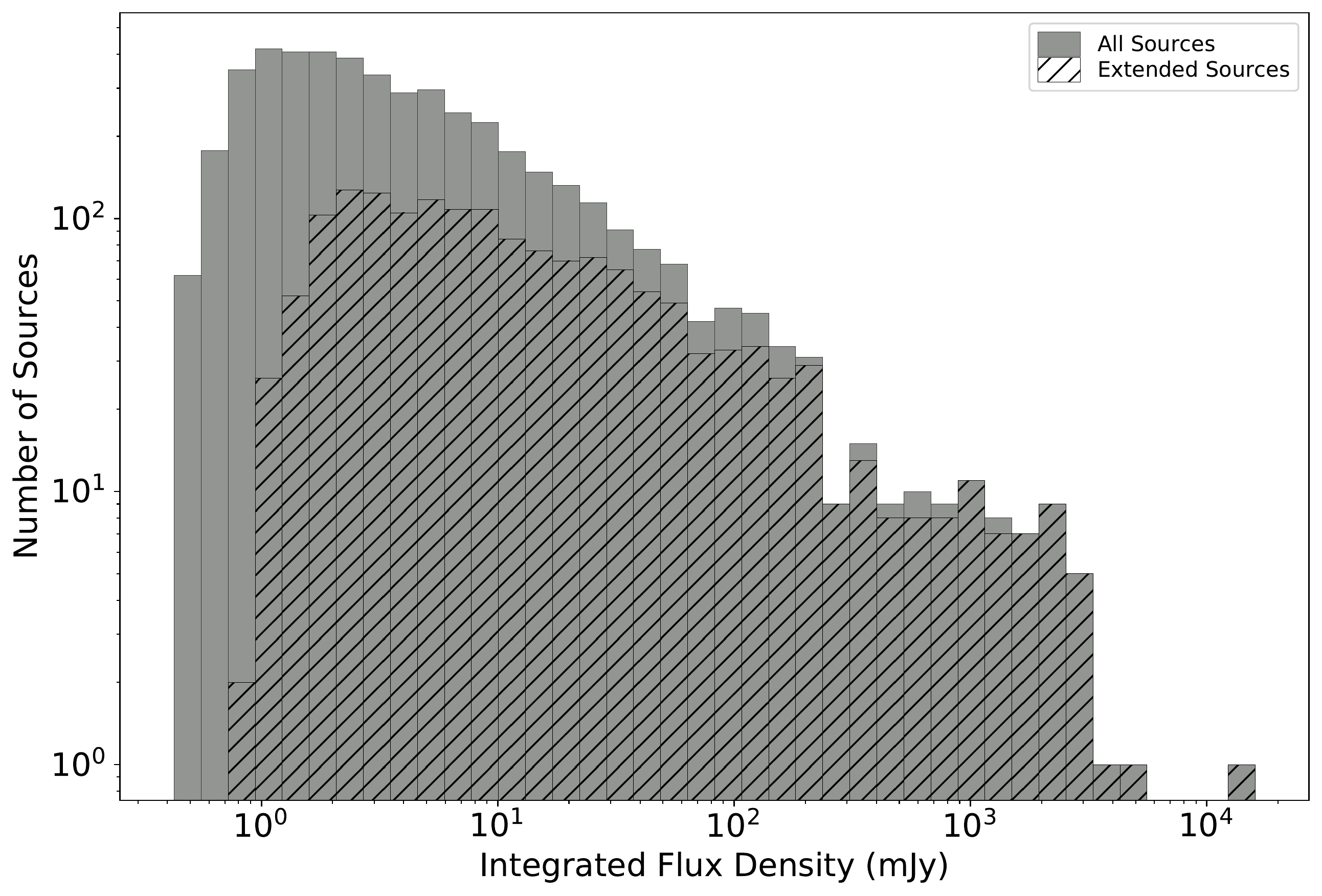}

\caption{Distributions of the integrated flux density of the 7$\mathrm{\sigma}$ CORNISH-South catalogue. Resolved/extended sources are represented by the hatched regions. Figure 19 in  \citetalias{cornissh2013} shows the distribution of the CORNISH-North.}
\label{supwlat}
\end{figure}

The integrated flux density and peak flux distribution in Figure \ref{supwlat} shows similar distributions compared to the CORNISH-North sources  \citepalias{cornissh2013}. The flux density distribution peaks at $\mathrm{\sim}$1 mJy, below which the number of sources drops off due to increasing incompleteness (see Section \ref{incomp}). For the resolved/extended sources, the flux density distribution peaks at $\mathrm{\sim 3}$ mJy and gently falls off, extending up to 10$\mathrm{^4}$ mJy. Compared to the CORNISH-North, we have picked up more faint sources as expected due to the sensitivity being two times better.

\subsection{Source classification and Example Sources}\label{example}

Initial classification of the CORNISH-South sources has utilized the availability of comparable high-resolution and high-sensitivity surveys (GLIMPSE, VVV, VPHAS, Hi-Gal and ATLASGAL) of the Galactic plane. As with the CORNISH-North survey, one of us (MGH) has visually inspected the multi-wavelength images of each source on the CORNISH-South website. Examples of multi-wavelength images of the main different types of sources are shown in Figure \ref{ex_sources1}. Using experience gained from classification of sources in the RMS (Red MSX Source) survey \citep{lum2013}, the following visual classification criteria were used.

HII regions have strong, and usually extended, mid-IR, far-IR and sub-millimetre counterpart emission. The morphology of the IR emission is usually irregular and complex, as well as often being part of a clustered environment. Their radio emission is usually fairly strong and when resolved can have a cometary, shell or irregular morphology. The mid-IR emission, dominated by polycyclic aromatic hydrocarbons (PAH) emission, arises from just outside the radio emitting region and often reflects the same morphology \citep{hoa2007}. An initial sub-classification on angular size has been used in the catalogue  , pending distance information. All H II regions less than 5$\mathrm{\arcsec}$ in size have been labelled as UCHIIs. This corresponds to the typical 0.1 pc size of UCHIIs if they were at a typical distance of $\mathrm{\sim}$ 4 kpc for the more nearby part of the population of UCHIIs (\citetalias{kalprep}). The larger ones were mostly labelled as H II regions if the radio emission was clearly identifiable as part of a single source across the radio and IR bands. If the radio emission was due to over-resolution of a much larger, multiple and complex source in the IR then it was labelled as a diffuse H II region. A few HII regions were hidden behind large amounts of dust extinction at 8$\mathrm{\mu m}$ and were labelled as IR-Dark H II regions.

PNe also have strong mid-IR counterparts due to dust and PAH emission  \citep{smith2008, guz2014,cox2016}, but are much fainter at far-IR wavelengths than H II regions, and are usually undetected in sub-millimetre plane surveys.  The SEDs of PNe generally peak at $\mathrm{\sim 24\mu m}$ but some young and dense PNe have their peaks extending up to 70$\mathrm{\mu m}$ and beyond (see \citetalias{irabor2018}; \citealt{anderson2012}; \citealt{urq2013}).  A few of the bipolar, Type I, PNe can have more far-IR and sub-millimetre emission, but are still significantly weaker than H II regions. Morphologically they are much simpler than H II regions and are isolated rather than being in clustered, complex environments.

Radio stars are point sources in every waveband they are detected in. They are also isolated sources in the field. Depending on the type of radio star they can either have blue or red colours in the optical and IR. Very red radio stars like dusty symbiotic stars are difficult to distinguish from unresolved PN without further information \citep{irabor2018}. 

Due to the sensitivity of the CORNISH-South survey the radio emission from a few known massive young stellar objects (MYSOs) was detected. They share all the IR characteristics of H II regions, but have very weak radio emission compared to HII regions in general and are unresolved or jet-like \citep{Purser2016}. It can be difficult to distinguish weak, unresolved UCHIIs powered by B3 stars from MYSOs as they have similar radio luminosities \citep{Purser2016}.

Most extra-galactic radio sources are undetected in corresponding optical, IR and sub-millimetre Galactic plane surveys used here and therefore they are straightforward to identify. In the radio, they are usually unresolved single sources and we classify them as IR quiet sources in the catalogue. A small number of these may be radio stars that are so distant or obscured as to remain undetected in the optical and IR surveys. A significant number of extra-galactic sources show the classic double radio source morphology, and these are classified as ``Radio Galaxy (Lobe)'' in the catalogue. In some cases, the core of the radio galaxy is also seen and is classified as such. If both lobes, or a lobe and a core, or both lobes and a core are visible, but part of the same extended source in the catalogue they are referred to as ``Radio Galaxy (Both)''. It is possible that some normal, star-forming galaxies, as opposed to AGN, are detected in CORNISH-South. Some radio sources had very faint, resolved counterparts in the mid-IR, where it was not clear if they are very distant PNe or galaxies. Further studies will be required to rule out a PN classification.

A few sources did not fit in to any of the above categories, either because they are known sources of unusual type, or their origin is currently unknown. These were classified as ``Other''.  

\begin{figure}
\begin{center}
	\includegraphics[height=4.4cm, width=\columnwidth]{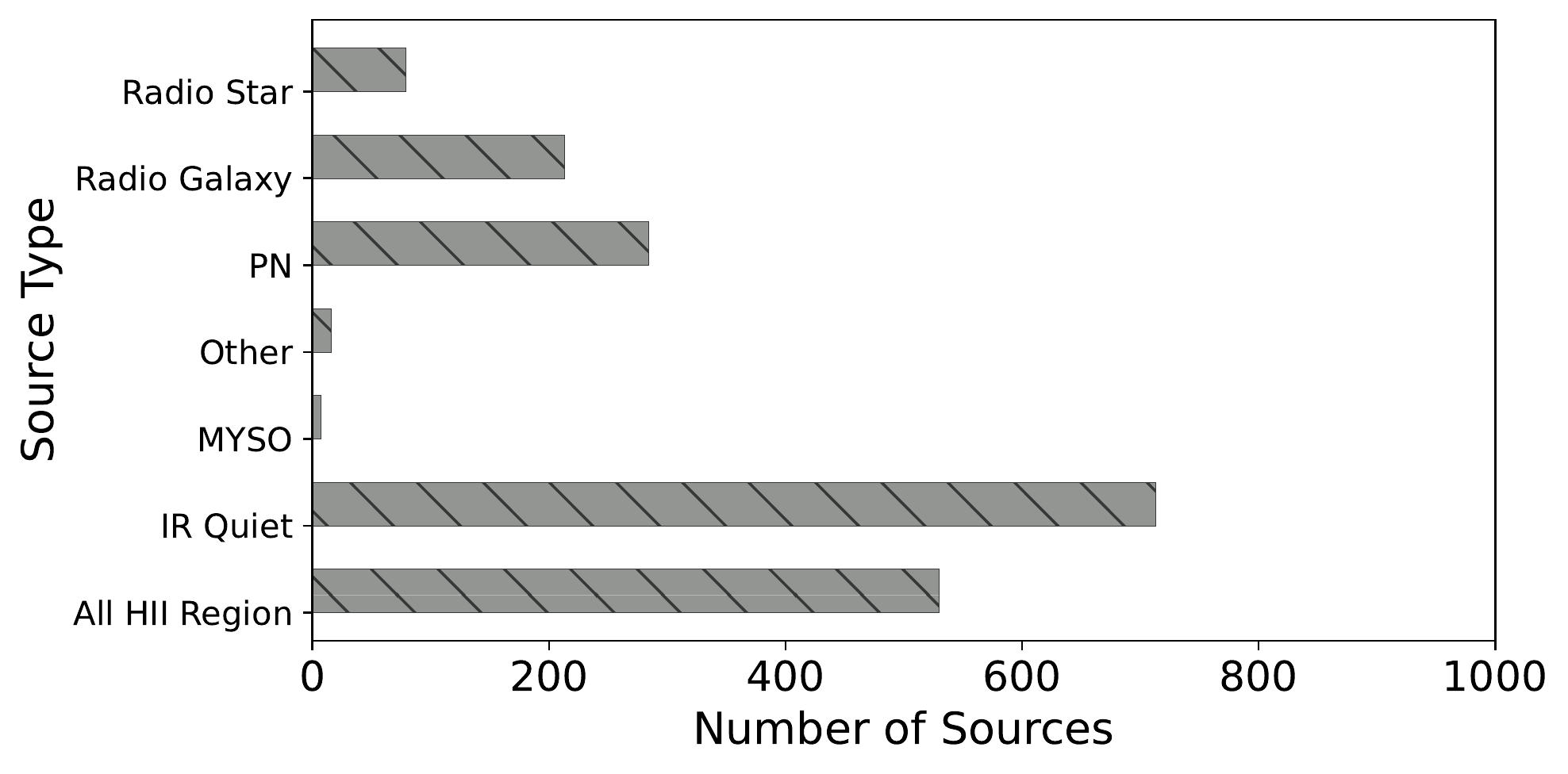}
	\caption{Distribution of classified CORNISH-South sources. Fifty-percent has been classified so far. All HII regions sum up to 530, of which 257 are UCHII regions. Unclassified sources add up to $\sim 2300$.}
\label{class_sources}
\end{center}
\end{figure}

At the time of publication, all sources with detectable flux at 8 $\mathrm{\mu m}$ seen within the same aperture used for radio fluxes have been classified. This should account for the vast majority of Galactic sources.  Figure \ref{class_sources} shows the distribution of classifions of these sources. Resolved sources with infrared counterparts are dominated by PNe and HII regions. We find more UCHII regions compared to the northern counterpart and have also identified six known MYSOs. With a sensitivity of 0.11 mJy beam$\mathrm{^{-1}}$, we have also detected PNe with lower radio flux density compared to CORNISH-North. We expect that the vast majority of the sources that remain to be visually classified will be extra-galactic and under the IR Quiet or Radio Galaxy categories. The latest classifications will be those on the CORNISH-South website.

\subsubsection{Catalogue Format}

An excerpt of the catalogue is presented in Table \ref{comp_final_tab} and the columns are arranged in the following format: Column (1) - CORNISH Source name ($\mathrm{(l+b)}$);  Columns (2) and (3) -  right ascension ($\mathrm{\alpha}$) and declination ($\mathrm{\delta}$) in (J2000) with their associated errors in brackets; Column (4)  – Peak flux and associated error in mJy beam$\mathrm{^{-1}}$;  Column (5) -  integrated flux density and associated error in mJy. Because the clean bias is close to the rms noise level and within the errors, the flux densities were not corrected for the clean bias effect. Column (6) - Angular size and associated error in arcsec; Column (7) - Gaussian FWHM major axis and error in arcsec; Column (8) - Gaussian FWHM minor axis and error in arcsec; Column (9) - position angle (E of N) of elliptical Gaussian; Column (10) - local rms noise level, ($\mathrm{\sigma_a}$), in mJy beam$\mathrm{^{-1}}$ as measured in the annulus described in Section \ref{aperturephot}; Column (11) - Signal to noise ratio of the source given by Peak flux divided by $\mathrm{\sigma_a}$; Column (12) - Type of source tells if the source is Gaussian fitted (G) or non-Gaussian, in which case a polygon (P) was drawn around the source; Column (12) - This column indicates the classification of the source. Gaussian sources have both the Gaussian fitted sizes and geometric mean sizes in the final catalogue.  The final version of the full table is made available on the CORNISH website (\url{http://cornish.leeds.ac.uk/public/index.php}).

\subsection{Astrometry and Flux Density Quality Check}\label{crossmatch}

\subsubsection{GLIMPSE}\label{malt}
\begin{figure*}
\centering
\includegraphics[ width=5.2cm]{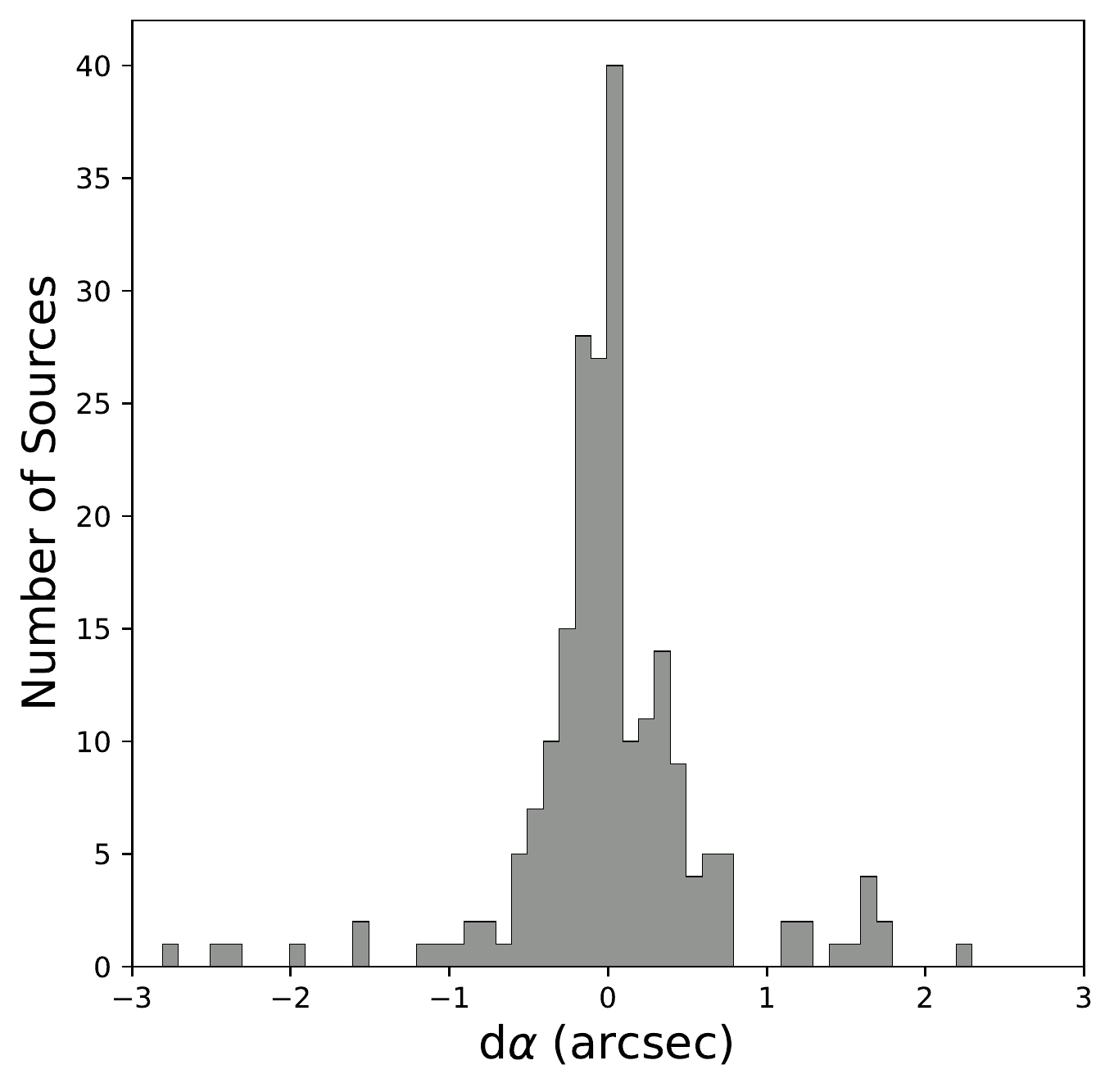}
\includegraphics[width=5.2cm]{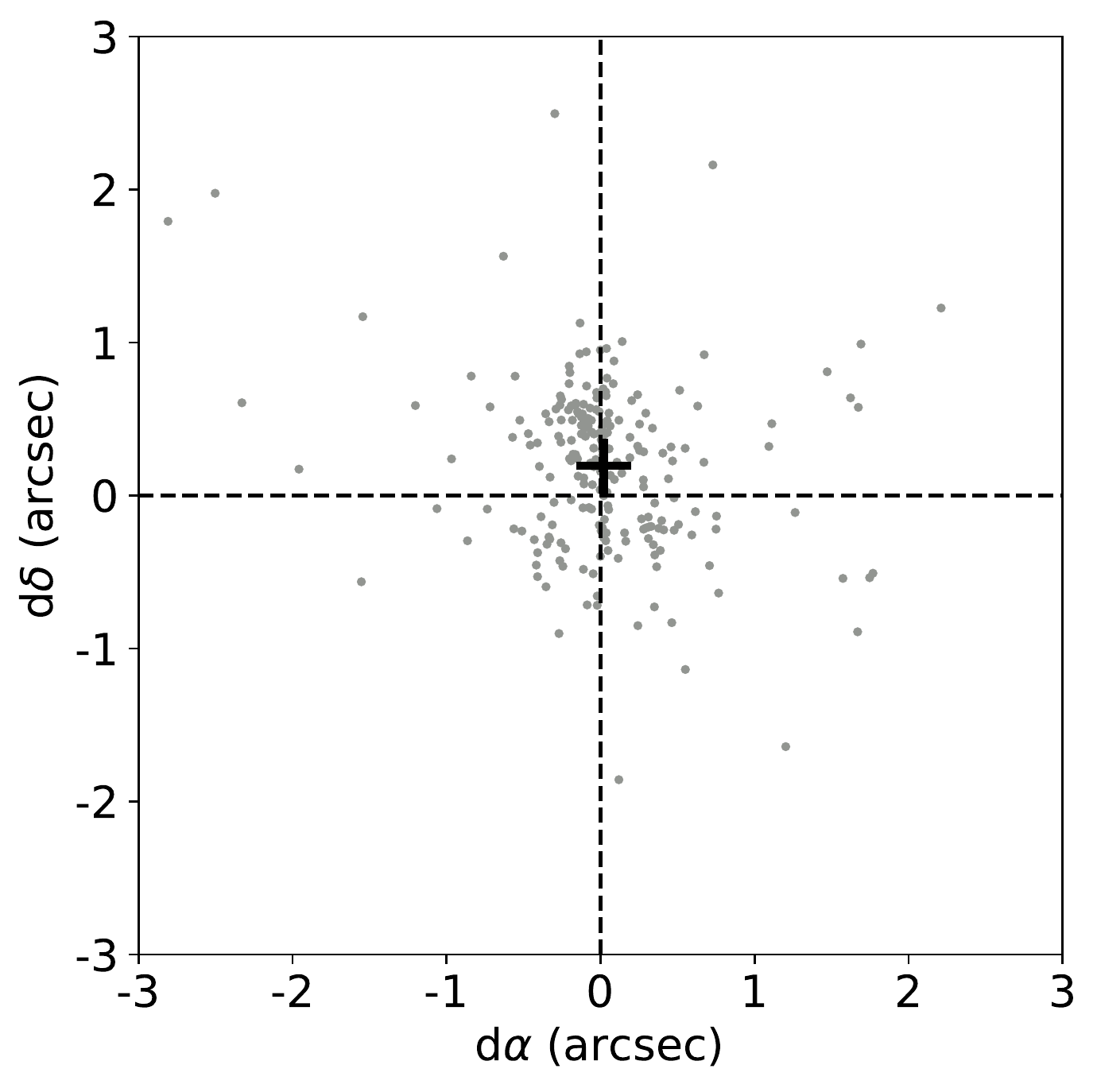}
\includegraphics[width=5.2cm]{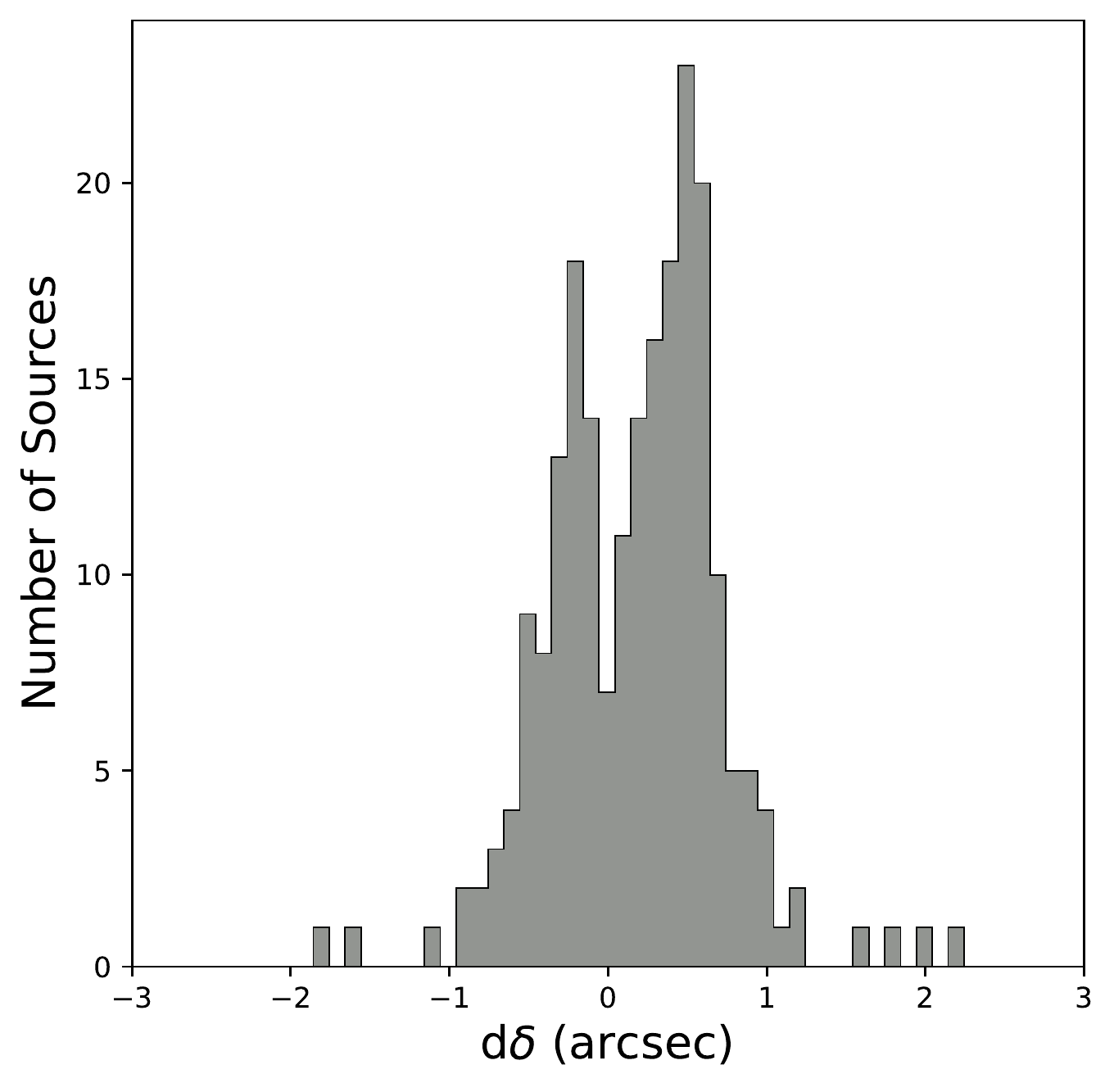}
	\caption{CORNISH-South and GLIMPSE cross-matched sources (218) within 3\arcsec. Top left: The angular offset between cross-matched sources. Top right: Offset distribution in $\mathrm{\alpha}$ (arcsec). Bottom left: Offset distribution in $\mathrm{\delta}$ (arcsec). Bottom right: Scatter plot of offsets in $\mathrm{\alpha}$ against $\mathrm{\delta}$. The cross symbol indicates the mean in $\mathrm{\alpha}$ ($\mathrm{0.02 \pm 0.04 \arcsec}$) and $\mathrm{\delta}$ ($\mathrm{0.19 \pm 0.04 \arcsec}$ ). The error is the standard error on the mean.}
\label{glimp_con}
\end{figure*}

In order to check the astrometry of the CORNISH-South catalogue, we cross-matched a sub-set of the CORNISH-South sources with the GLIMPSE point source catalogue. To avoid mis-matches and multiple matches, the CORNISH-South catalogue was limited to classified sources,  excluding extended HII regions, radio-galaxies and infrared quiet sources. Additionally, sources with angular size $\mathrm{> 3\arcsec}$ were excluded. The cross-match returned 218 sources and Figure \ref{glimp_con} shows the distributions of the offsets in right ascension ($\mathrm{\alpha}$) and declination ($\mathrm{\delta}$).

The distribution of the angular offsets in Figure \ref{glimp_con} (top left) peaks at $\mathrm{\sim}$0.4\arcsec \ and steeply falls to 1.5\arcsec \ before continuing gently out to 3\arcsec  . The offset distribution in $\mathrm{\alpha}$ is tightly peaked at about 0\arcsec \ compared to the distribution in $\mathrm{\delta}$ that has double peaks at 0.5 and -0.2\arcsec . The spread in $\mathrm{\delta}$ offset can be attributed to the distribution of the intrinsic beam along the major axis seen in the epoch I data (see Figure \ref{maj_area33}).

\subsubsection{The Red MSX Source (RMS) 6 cm ATCA survey}
Targeted radio continuum observations were conducted to identify UCHII regions and PNe as part of the Red MSX Source (RMS) survey \citep{lum2013}. The observations were carried out with the ATCA within the $\mathrm{235^\circ<l<350^\circ}$ region at 3.6 and 6 cm \citep{urqu2007rms}. Table \ref{con_rms_tab} compares the observational parameters of the RMS 4.8-GHz (6 cm) and the CORNISH-South 5.5-GHz. Both surveys have similar observational properties but the CORNISH-South observing bandwidth of 2-GHz provides better image fidelity compared to the 128 MHz of the RMS survey. 

\begin{figure*}
\includegraphics[width=5.2cm]{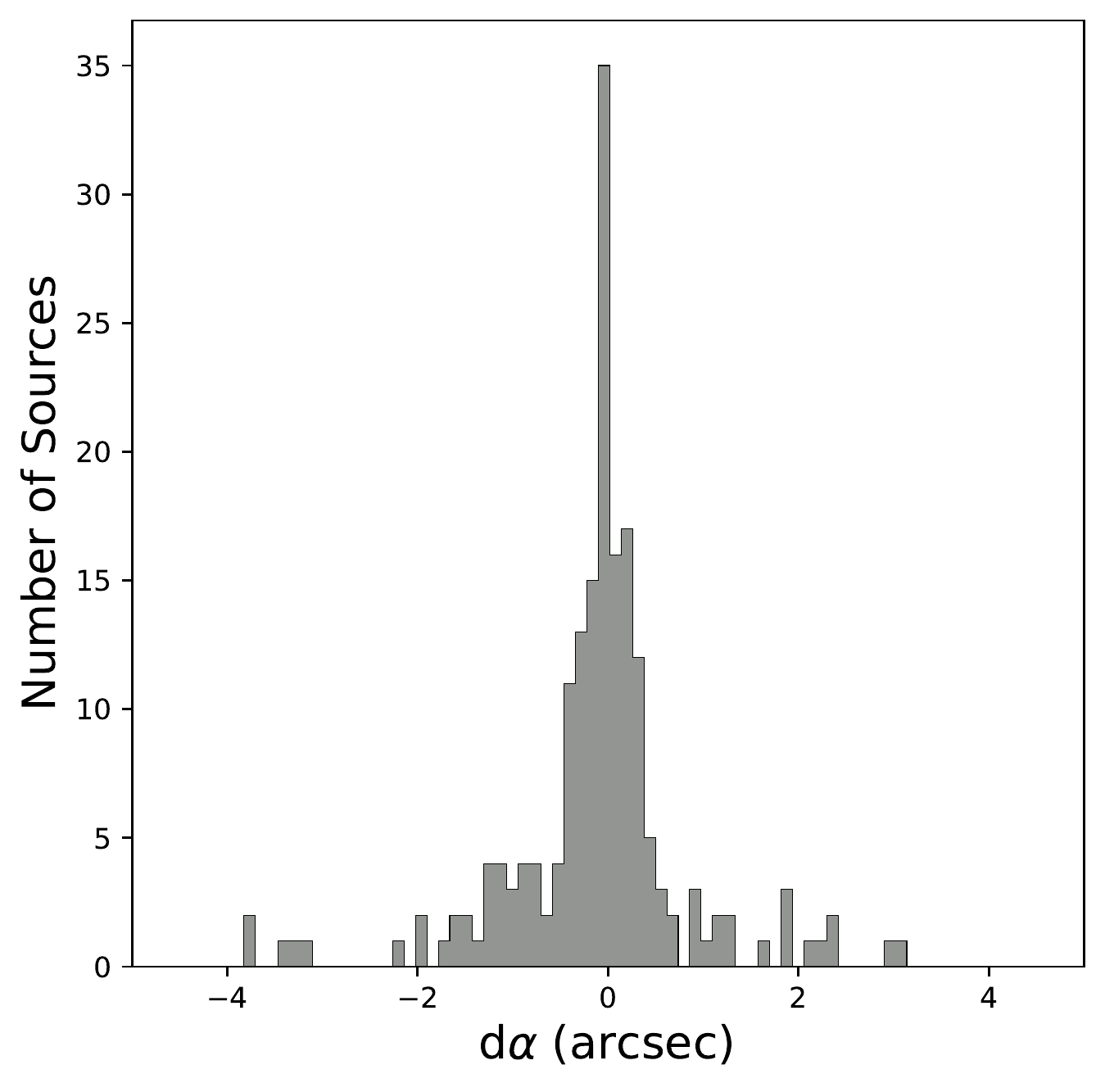}
\includegraphics[width=5.2cm]{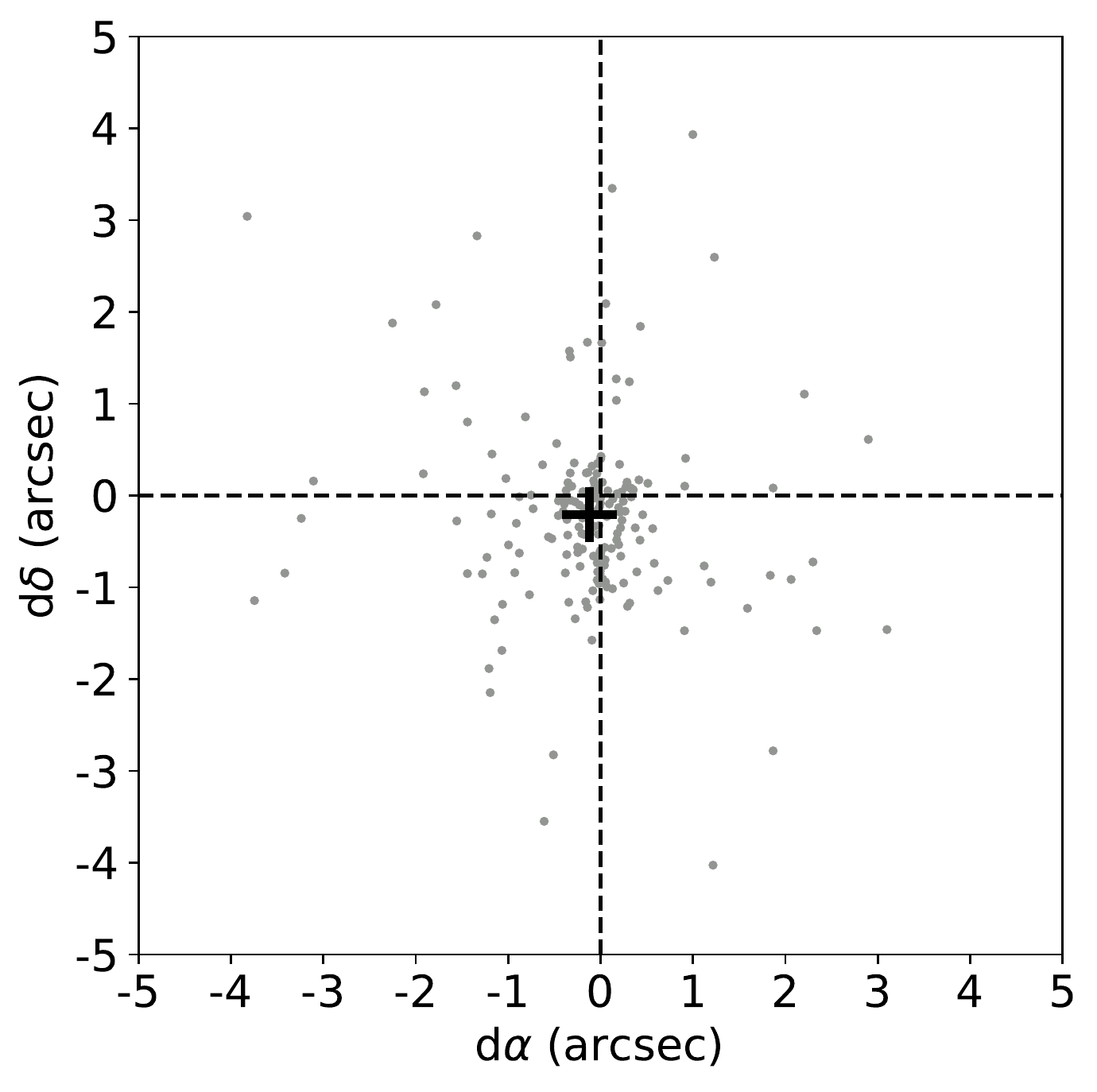}
\includegraphics[width=5.2cm]{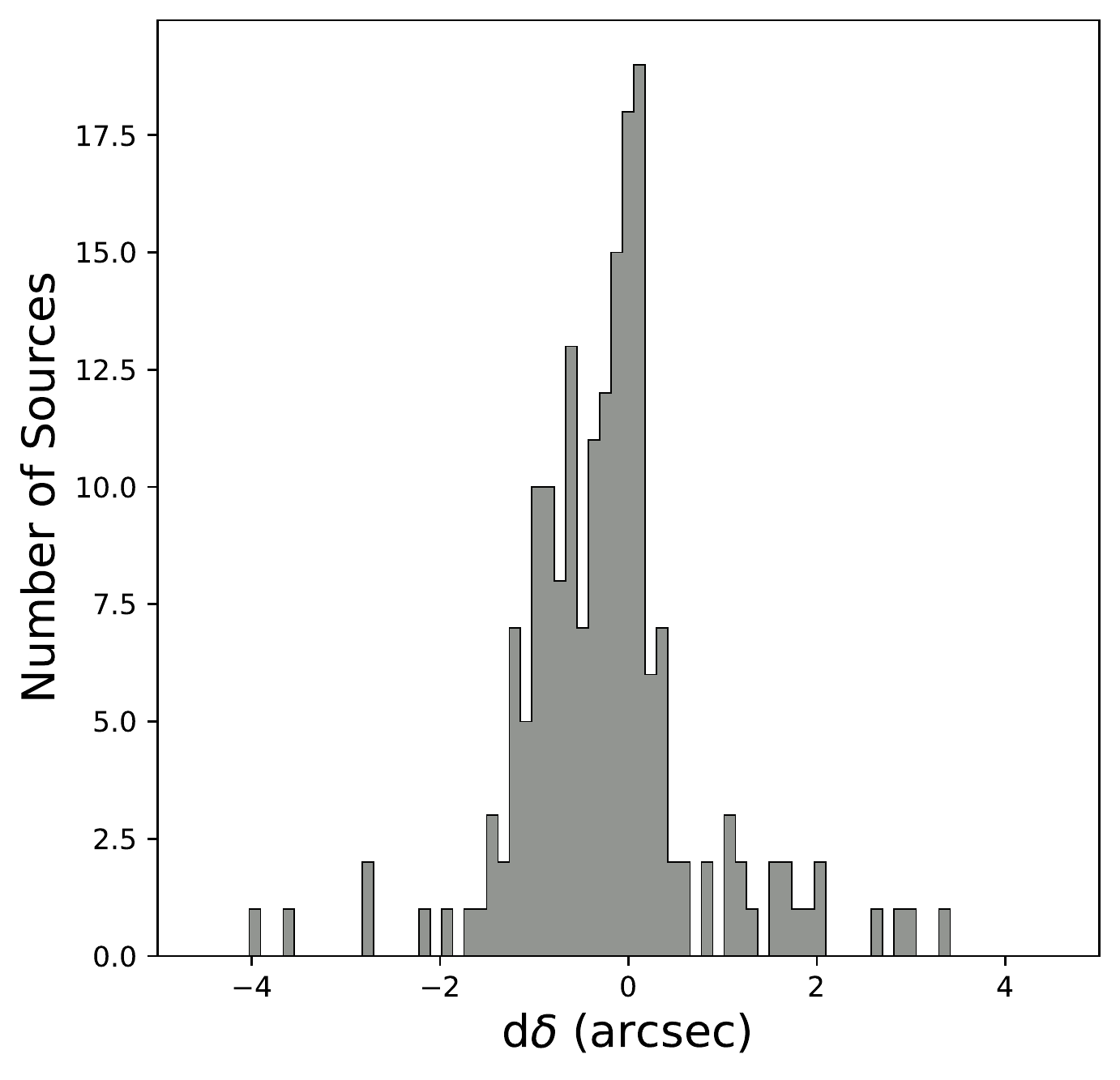}

\caption{CORNISH-South and RMS 4.8-GHz cross-matched radio sources (186) within 5\arcsec. Top left: The angular offset between cross-matched sources. Top right: Offset distribution in $\mathrm{\alpha}$ (arcsec). Bottom left: Offset distribution in $\mathrm{\delta}$ (arcsec). Bottom right: Scatter plot of offsets in $\mathrm{\alpha}$ against $\mathrm{\delta}$. The cross symbol indicates the mean in d$\mathrm{\alpha}$ ($\mathrm{0.12 \pm 0.07\arcsec}$) and d$\mathrm{\delta}$ ($\mathrm{0.21 \pm 0.08 \arcsec}$). The mean error for the CORNISH-South is 8 mJy.}
\label{con_rms_figs}
\end{figure*}

For further checks on the CORNISH-South astrometry and flux densities, the CORNISH-South 5.5-GHz catalogue was cross-matched with the RMS 4.8-GHz catalogue \citep{urqu2007cat} within a 5$\mathrm{\arcsec}$ radius. For a one-to-one match, the CORNISH-South catalogue was limited to only sources that have been visually classified, excluding diffuse HII regions and radio-galaxies. Figure \ref{con_rms_figs} (top left) shows the distribution of the angular separation between the CORNISH-South and RMS for 186 radio sources. The distribution shows a tight correlation with a sharp peak at 0.3$\mathrm{\arcsec}$ and then a steep fall to $\mathrm{\sim 1.5\arcsec}$  before gently falling off to $\mathrm{\sim 4.3\arcsec}$.

Seventy-five per cent of the cross-matched sources fall within 1.5$\mathrm{\arcsec}$ and 94 per cent fall within 3\arcsec . Compared to the distribution of the offset in $\mathrm{\delta}$, the offset in $\mathrm{\alpha}$ shows a narrower distribution that is strongly peaked at 0\arcsec . Similar offset distribution in $\mathrm{\delta}$ is seen in the CORNISH-South-GLIMPSE cross-match that is attributed to the spread in the intrinsic major axis distribution. The mean offset in $\mathrm{\alpha}$ and $\mathrm{\delta}$ is 0.1\arcsec\ and 0.2\arcsec , respectively. Based on this, and the CORNISH-South-GLIMPSE cross-match, we adopt a positional accuracy of 0.22 $\mathrm{\pm}$ 0.11$\mathrm{\arcsec}$ for the CORNISH-South catalogue. This is also in line with the mean positional accuracy of the secondary calibrators (see Section \ref{calib_posit}).

\begin{table}
\begin{center}
\caption{Comparison of the RMS 6 cm (4.8-GHz) and the CORNISH-South observation parameters.}

\begin{tabular}{l|c|c}\hline 
\multicolumn{1}{c}{Parameters}
& \multicolumn{1}{c}{CORNISH-South}
& \multicolumn{1}{c}{RMS}
\\  

\hline

Rest frequency (GHz)&5.5&4.8\\
Array&6A&6C/6D\\
Bandwidth (GHz)&2&0.128\\
Synthesised beam&2.5$\mathrm{\arcsec}$&2.5$\mathrm{\arcsec}$\\
Typical image rms (mJy beam$\mathrm{^{-1}}$)&0.11&0.27\\
Image pixel size&0.6$\mathrm{\arcsec}$&0.6$\mathrm{\arcsec}$\\
\hline
\end{tabular}
\label{con_rms_tab}
\end{center}
\end{table}

\begin{figure}
\includegraphics[height=7.5cm, width=\columnwidth]{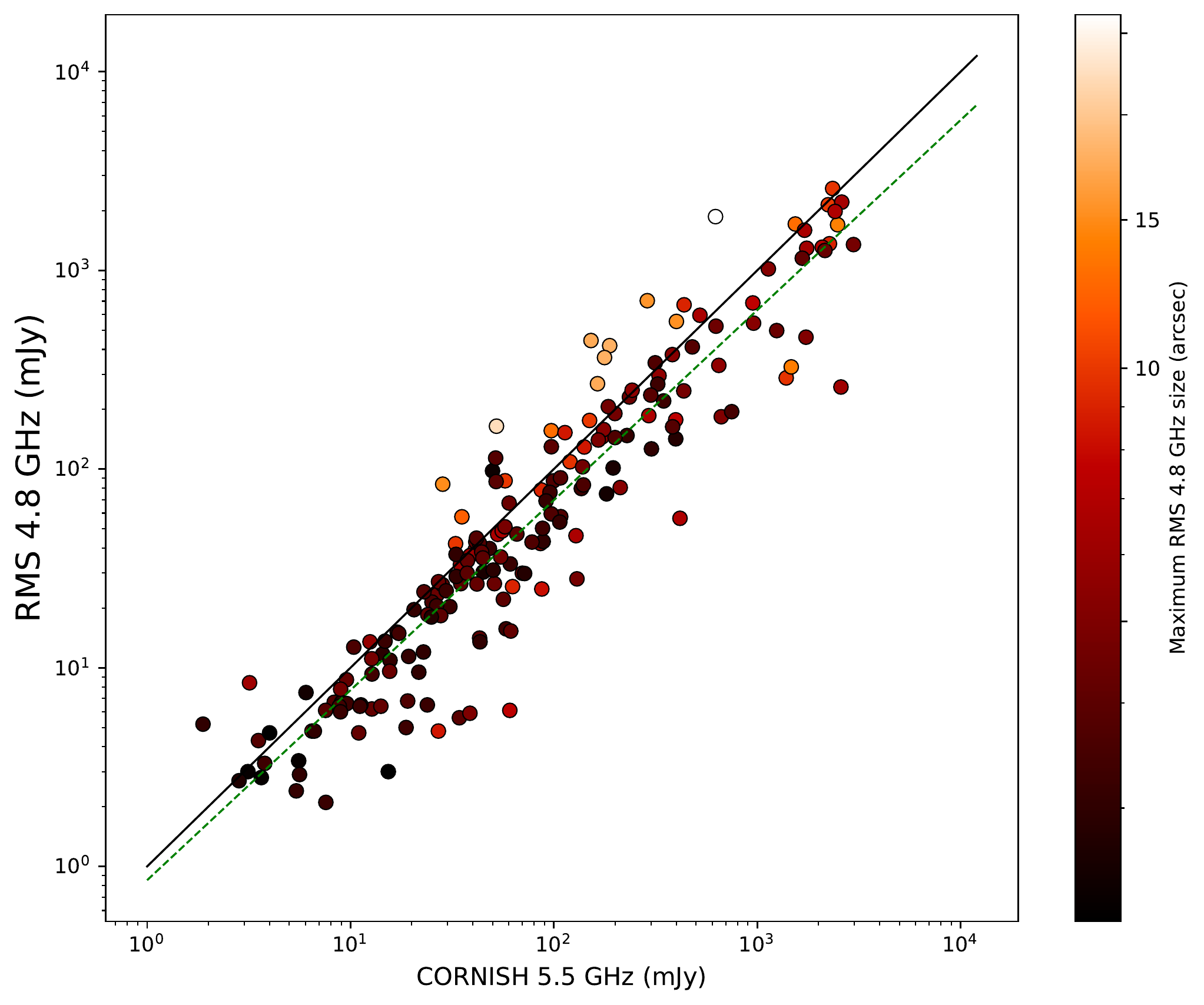}
\caption{A plot of the 5.5-GHz CORNISH-South flux densities against the RMS 4.8-GHz \citep{urqu2007rms} flux densities. The black line is a one-to-one line ($\mathrm{y=x}$) and the dotted green line is the best fit line defined by $\mathrm{log(y)=(0.99\pm 0.03)log(x)-0.11\pm 0.05}$. The error on the fitted line is the standard deviation.}
\label{con_rms_figs2}
\end{figure}

In Figure \ref{con_rms_figs2}, the flux densities of the cross-matched sources are compared. A one-to-one line (black line) shows that the CORNISH-South 5.5-GHz flux densities are higher than the RMS 4.8-GHz flux densities on average. The few sources where the RMS flux densities are higher are found to be HII regions with larger angular sizes. The two surveys used different ATCA configurations (see Table \ref{con_rms_tab}), which are sensitive to different angular sizes. Compared to the shortest baselines of the 6C (153 m) and 6D (77 m) configurations used for the RMS survey \citep{urqu2007rms}, the shortest baseline of the CORNISH-South 6A configuration (337 m) makes it less sensitive to extended structures. Hence, for optically thin and angularly large HII regions, the RMS flux densities are expected to be higher, as seen in Figure \ref{con_rms_figs2}. However, the better \textit{uv}-coverage of the CORNISH-South recovers more extended emission on some sources.

A best-fit line (dotted green line) predicts $\mathrm{\sim}$ 24 per cent flux density increase for the CORNISH-South counterparts that is more than the typical calibration error of 10 per cent. According to \cite{urqu2007rms},  the distribution of spectral indices, ($\mathrm{\alpha}$, where $\mathrm{S_\nu \propto \nu^{\alpha}}$) between the 3.6 and 6 cm data is slightly skewed towards positive indices, which suggests optically thick sources in their sample. This provides a possible explanation for the compact sources where the 5.5 GHz CORNISH-South flux densities are higher than the 4.8 GHz RMS ones.

\section{Conclusion and Future Work}
The CORNISH program has successfully mapped the southern Galactic plane at radio wavelengths with unprecedented resolution and sensitivity. We have presented radio continuum ATCA data at 5.5-GHz, covering the $\mathrm{295\degr < l < 350\degr}$; $\mathrm{|b|\leq 1\degr}$ region of the southern Galactic plane. The resolution of 2.5$\mathrm{\arcsec}$ and noise level of 0.11 mJy beam$\mathrm{^{-1}}$ deliver radio data matched to the existing high resolution, multi-wavelength surveys GLIMPSE, VVV and VPHAS+ of the southern Galactic plane. 

Utilizing the MIRIAD program for data reduction and AEGEAN source finding algorithm, we have identified 4701 sources above 7$\mathrm{\sigma}$. Data arising from fields with poor uv-coverage make up to only 2 per cent of the data set. In addition to several measures undertaken to ensure data quality, visual inspection has also been used to exclude artefacts and hence the data are highly reliable. The survey has a 90\% completeness level at a flux density of 1.1 mJy. The measured properties show distributions that are similar to that of the CORNISH-North catalogue presented in \citetalias{cornissh2013}. 

All sources with potential counterparts at 8$\mathrm{\mu m}$ (GLIMPSE) have been visually classified. This corresponds to 43 per cent of the sources. We have identified 530 HII regions, of which 257 are UCHIIs. Additionally, we identified 287 PNe and 79 radio stars. The vast majority of the remaining unclassified sources without infrared counterparts are expected to be extra-galactic. With the sensitivity of the CORNISH-South survey being two times better than the CORNISH-North counterpart, sources with lower flux densities have been detected, including a few MYSOs. A detailed analysis of the statistical properties of individual catalogues will be presented in future papers. The CORNISH-South survey also carried out simultaneous observations of the entire field at 9-GHz, which will be presented in a separate paper. The 9-GHz data provide higher spatial resolution at reduced sensitivity and will enable an examination of spectral indices of the sources.

The CORNISH-South data are particularly important in the characterization of compact ionized sources towards the Galactic mid-plane, as no radio survey has previously covered the southern Galactic plane in such resolution and sensitivity. Previous surveys of the southern Galactic plane were limited in resolution and  coverage, especially within the $\mathrm{|b|<1^\circ}$ region,  hence not suitable for studies of compact ionized regions.  New radio facilities such as the, ASKAP\footnote{\url{https://www.atnf.csiro.au/projects/askap/index.html}}\citep{hotan2014}, MeerKAT\footnote{\url{https://www.sarao.ac.za/gallery/meerkat/}}\citep{jonas2016} and ultimately the SKA\footnote{\url{https://www.skatelescope.org/}}\citep{braun2019} are exploring the southern Galactic plane at greater depth. The CORNISH-South catalogue is well positioned as a very useful resource to characterise the population of radio sources seen in the MEERKAT L-band survey (8\arcsec \ resolution and 10$\mathrm{\mu}$Jy/beam noise level; Goedhart et al., In preparation.) and also for follow-up observations with ALMA\footnote{\url{https://www.almaobservatory.org/en/home/}}. It will also be useful in combination with existing multi-wavelength surveys of the southern Galactic plane to address key questions in stellar formation and evolution. This will allow multi-wavelength exploration and statistical studies of compact ionized regions, which can then be compared with population synthesis models.  To date, the CORNISH program has delivered the most sensitive and unbiased compact radio source catalogue towards the southern Galactic mid-plane. Data products in the form of FITS images of individual sources and catalogues are available on the CORNISH-South Website \footnote{\url{https://cornish-south.leeds.ac.uk/}}.

With deeper and larger radio surveys on the way, it is also important that we look at building and improving existing machine-learning models for image classification (e.g. Adhiambo et al., in preparation). Having used multi-wavelength visual classification, the CORNISH sources provide a good training and validation set for machine-learning models that will allow automated classification of radio sources in these upcoming larger and deeper surveys.

\begin{landscape}

\begin{table}
\begin{center}
\caption{An excerpt of the 5.5-GHz CORNISH-South catalogue. The full version of the catalogue is available online. Measurement errors are in parentheses. The flux densities have not been corrected for the clean bias effect.}\label{complete_cal}

\begin{tabular}{p{2cm}|p{2.17cm}|p{2.17cm}|p{1.5cm}|p{1.5cm}|l|c|c|c|c|l|c|c}\hline
{Source Name}
& {$\mathrm{\alpha}$ ($\mathrm{^{h}\ ^{m} \ ^{s}}$)} 
&{$\mathrm{\delta}$ ($\mathrm{^{\circ}\ \arcmin\  \arcsec}$)}
&{Peak}
& {$\mathrm{S_{5.5-GHz}}$}
& {$\mathrm{\theta_s}$}
&{$\mathrm{\theta_{maj}}$}
&{$\mathrm{\theta_{min}}$}
& {PA}
&{rms}
& {Sigma}
& {Type}
& {Class}\\ 

(l+b)&(J2000)&(J2000)&(mJy beam$\mathrm{^{-1}}$)&mJy&(\arcsec)&(\arcsec)&(\arcsec)&(deg)&mJy beam$\mathrm{^{-1}}$&&& \\

\hline
G295.1757$-$0.5744	&11:43:38.84	(0.65)&	-62:25:17.5	(0.44)&	4.12	 (0.21)&	58.13 (7.20)&	20.84 (0.14)&	0.0&	0.0&	0.0	&0.12&23&	P&	HII Region\\
G296.5033$-$0.2695&11:55:25.27	(0.02)&	-62:26:24.6	(0.02)&	10.92	(1.22)&	15.06 (1.87)&8.43 (0.05)&0.0&	0.0&	0.0&	0.10&72&	P&PN\\

G297.3943$-$0.6347&12:02:23.29	(0.96)&	-62:58:42.8	(1.28)&	0.94	(0.02)&	7.09	(2.61)&	16.67(0.34)&	0.0	&0.0&	0.0&0.10&	9.9&P&HII Region\\

G298.3434+0.1466&	12:11:42.23	(0.21)&-62:22:14.9 (0.35)&	0.74	(0.01)&	1.30	(0.81)&	7.18(0.22)&0.0&	0.0&	0.0&	0.08&	9.8	&P&	Radio Galaxy\\

G298.8382$-$0.3388	&12:15:20.73	(0.03)&	-62:55:25.4	(0.03)&	11.29	(1.30)&	35.34	(2.89)&	11.29 (0.06)&	0.0&	0.0&	0.0&	0.17&	68&	P&HII Region\\
...&...&...&...&...&...&...&...&...&...&...&...&...\\
...&...&...&...&...&...&...&...&...&...&...&...&...\\
...&...&...&...&...&...&...&...&...&...&...&...&...\\
G349.9207+0.6682&	17:16:31.30	(0.02)&	-36:59:40.29	(0.02)&	13.16	(0.26)&	13.92	(0.61)&	2.57	(0.06)&	2.62	(0.05)&	2.52(0.05)&	0.73&	0.17&	75&	G&PN\\

G349.9260+0.0811&	17:18:56.62	(0.07)&	-37:19:44.7	(0.11)&	2.61	(0.08)&	17.17	(0.54)&	6.42	(0.01)&	7.75	(0.01)&	5.31	(0.01)	&-0.71&	0.14&	21&	G& HII Region\\

G350.0290$-$0.4950&	17:21:37.33	(0.16)&	-37:34:25.77	(0.18)&	2.64	(0.42)&	2.58	(0.92)&	2.48	(0.44)&	2.72	(0.41)&	2.25	(0.37)&	-7.17&	0.35&	7.4&	G&	Radio Galaxy\\

G350.0920+0.2309&	17:18:48.44	(0.01)&	-37:06:25.28	(0.01)&	89.33	(0.46)&	96.32	(1.08)&	2.60	(0.02)&	2.68	(0.01)&	2.52	(0.01)&	-8.31&	0.27&	330	&G	&PN\\

G350.1237$-$0.5563&	17:22:08.88	(0.06)&	-37:31:50.61	(0.07)&	3.68	(0.2)&	3.27	(0.49)&	2.36	(0.18)&	2.74	(0.17)&	2.03	(0.15)&	-0.78&	0.12&	29&	G&	IR Quiet\\

\hline
\end{tabular}
\label{comp_final_tab}
\end{center}

\end{table}
\clearpage
\end{landscape}

\section*{Acknowledgements}

TI acknowledges the support of the Science and Technology Facilities Council of the United Kingdom (STFC) through grant ST/P00041X/1 and the OSAPND scholarship (Nigeria). The work done by PFG was carried out in part at the Jet Propulsion Laboratory, which is operated by the California Institute of Technology under a contract with the National Aeronautics and Space Administration (80NM0018D0004). JM acknowledges financial support from the State Agency for Research of the Spanish Ministry of Science and Innovation under grant PID2019-105510GB-C32. JMP acknowledge financial support from the State Agency for Research of the Spanish Ministry of Science and Innovation under grant PID2019-105510GB-C31 and through the Unit of Excellence María de Maeztu 2020-2023 award to the Institute of Cosmos Sciences (CEX2019-000918-M). G.A.F  acknowledges support from the Collaborative Research Centre 956, funded by the Deutsche Forschungsgemeinschaft (DFG) project ID 184018867.

The Australia Telescope Compact Array (ATCA) is part of the Australia Telescope National Facility which is funded by the Australian Government for operation as a National Facility managed by CSIRO. We acknowledge the Gomeroi people as the traditional owners of the Observatory site. This work made use of Montage, aplpy and astropy python libraries in the batch processing of FITS files.

%%%%%%%%%%%%%%%%%%%%%%%%%%%%%%%%%%%%%%%%%%%%%%%%%%
\section*{Data Availability}

All data (images and catalogues) are described in the text and available on the cornish website at \url{https://cornish-south.leeds.ac.uk/} to download in csv (catalogues), uv files and FITS (images) formats.

\bibliographystyle{mnras}
\bibliography{Accepted_TIRABOR} % if your bibtex file is called example.bib

\newcommand{\noop}[1]{}
\begin{thebibliography}{}
\makeatletter
\relax
\def\mn@urlcharsother{\let\do\@makeother \do\$\do\&\do\#\do\^\do\_\do\%\do\~}
\def\mn@doi{\begingroup\mn@urlcharsother \@ifnextchar [ {\mn@doi@}
  {\mn@doi@[]}}
\def\mn@doi@[#1]#2{\def\@tempa{#1}\ifx\@tempa\@empty \href
  {http://dx.doi.org/#2} {doi:#2}\else \href {http://dx.doi.org/#2} {#1}\fi
  \endgroup}
\def\mn@eprint#1#2{\mn@eprint@#1:#2::\@nil}
\def\mn@eprint@arXiv#1{\href {http://arxiv.org/abs/#1} {{\tt arXiv:#1}}}
\def\mn@eprint@dblp#1{\href {http://dblp.uni-trier.de/rec/bibtex/#1.xml}
  {dblp:#1}}
\def\mn@eprint@#1:#2:#3:#4\@nil{\def\@tempa {#1}\def\@tempb {#2}\def\@tempc
  {#3}\ifx \@tempc \@empty \let \@tempc \@tempb \let \@tempb \@tempa \fi \ifx
  \@tempb \@empty \def\@tempb {arXiv}\fi \@ifundefined
  {mn@eprint@\@tempb}{\@tempb:\@tempc}{\expandafter \expandafter \csname
  mn@eprint@\@tempb\endcsname \expandafter{\@tempc}}}

\bibitem[\protect\citeauthoryear{{Anderson}, {Zavagno}, {Barlow},
  {Garc{\'{\i}}a-Lario}  \& {Noriega-Crespo}}{{Anderson}
  et~al.}{2012}]{anderson2012}
{Anderson} L.~D.,  {Zavagno} A.,  {Barlow} M.~J.,  {Garc{\'{\i}}a-Lario} P.,
  {Noriega-Crespo} A.,  2012, \mn@doi [\aap] {10.1051/0004-6361/201117640},
  \href {http://adsabs.harvard.edu/abs/2012A%26A...537A...1A} {537, A1}

\bibitem[\protect\citeauthoryear{{Bertin} \& {Arnouts}}{{Bertin} \&
  {Arnouts}}{1996}]{bertin1996}
{Bertin} E.,  {Arnouts} S.,  1996, \mn@doi [\aaps] {10.1051/aas:1996164}, \href
  {https://ui.adsabs.harvard.edu/abs/1996A&AS..117..393B} {117, 393}

\bibitem[\protect\citeauthoryear{{Braun}, {Bonaldi}, {Bourke}, {Keane}  \&
  {Wagg}}{{Braun} et~al.}{2019}]{braun2019}
{Braun} R.,  {Bonaldi} A.,  {Bourke} T.,  {Keane} E.,   {Wagg} J.,  2019, arXiv
  e-prints, \href {https://ui.adsabs.harvard.edu/abs/2019arXiv191212699B} {p.
  arXiv:1912.12699}

\bibitem[\protect\citeauthoryear{{Briggs}}{{Briggs}}{1995}]{briggs1995}
{Briggs} D.~S.,  1995, PhD thesis

\bibitem[\protect\citeauthoryear{{Brunthaler} et~al.}{{Brunthaler}
  et~al.}{2021}]{Brunthaler2021}
{Brunthaler} A.,  et~al., 2021, \mn@doi [\aap] {10.1051/0004-6361/202039856},
  \href {https://ui.adsabs.harvard.edu/abs/2021A&A...651A..85B} {651, A85}

\bibitem[\protect\citeauthoryear{{Carey} et~al.,}{{Carey}
  et~al.}{2009}]{carey2009}
{Carey} S.~J.,  et~al., 2009, \mn@doi [\pasp] {10.1086/596581}, \href
  {http://adsabs.harvard.edu/abs/2009PASP..121...76C} {121, 76}

\bibitem[\protect\citeauthoryear{{Cesaroni} et~al.,}{{Cesaroni}
  et~al.}{2015}]{cesa2015}
{Cesaroni} R.,  et~al., 2015, \mn@doi [\aap] {10.1051/0004-6361/201525953},
  \href {https://ui.adsabs.harvard.edu/abs/2015A&A...579A..71C} {579, A71}

\bibitem[\protect\citeauthoryear{Churchwell}{Churchwell}{1990}]{ChurchwellEd1990UHrt}
Churchwell E.,  1990, The Astronomy and astrophysics review, 2, 79

\bibitem[\protect\citeauthoryear{Churchwell}{Churchwell}{2002}]{church_2002_new}
Churchwell E.,  2002, \mn@doi [Annual Review of Astronomy and Astrophysics]
  {10.1146/annurev.astro.40.060401.093845}, 40, 27

\bibitem[\protect\citeauthoryear{{Churchwell} et~al.}{{Churchwell}
  et~al.}{2009}]{churchwell2009}
{Churchwell} E.,  et~al., 2009, \mn@doi [\pasp] {10.1086/597811}, \href
  {http://adsabs.harvard.edu/abs/2009PASP..121..213C} {121, 213}

\bibitem[\protect\citeauthoryear{{Comeron} \& {Torra}}{{Comeron} \&
  {Torra}}{1996}]{come1996}
{Comeron} F.,  {Torra} J.,  1996, \aap, \href
  {https://ui.adsabs.harvard.edu/abs/1996A&A...314..776C} {314, 776}

\bibitem[\protect\citeauthoryear{{Condon}}{{Condon}}{1997}]{con1997}
{Condon} J.~J.,  1997, \mn@doi [\pasp] {10.1086/133871}, \href
  {http://adsabs.harvard.edu/abs/1997PASP..109..166C} {109, 166}

\bibitem[\protect\citeauthoryear{{Condon}, {Cotton}, {Greisen}, {Yin},
  {Perley}, {Taylor}  \& {Broderick}}{{Condon} et~al.}{1998}]{condon21998}
{Condon} J.~J.,  {Cotton} W.~D.,  {Greisen} E.~W.,  {Yin} Q.~F.,  {Perley}
  R.~A.,  {Taylor} G.~B.,   {Broderick} J.~J.,  1998, \mn@doi [\aj]
  {10.1086/300337}, \href
  {https://ui.adsabs.harvard.edu/abs/1998AJ....115.1693C} {115, 1693}

\bibitem[\protect\citeauthoryear{{Cox}, {Pilleri}, {Bern{\'e}}, {Cernicharo}
  \& {Joblin}}{{Cox} et~al.}{2016}]{cox2016}
{Cox} N.~L.~J.,  {Pilleri} P.,  {Bern{\'e}} O.,  {Cernicharo} J.,   {Joblin}
  C.,  2016, \mn@doi [\mnras] {10.1093/mnrasl/slv184}, \href
  {http://adsabs.harvard.edu/abs/2016MNRAS.456L..89C} {456, L89}

\bibitem[\protect\citeauthoryear{{Davies}, {Hoare}, {Lumsden}, {Hosokawa},
  {Oudmaijer}, {Urquhart}, {Mottram}  \& {Stead}}{{Davies}
  et~al.}{2011}]{davies2011}
{Davies} B.,  {Hoare} M.~G.,  {Lumsden} S.~L.,  {Hosokawa} T.,  {Oudmaijer}
  R.~D.,  {Urquhart} J.~S.,  {Mottram} J.~C.,   {Stead} J.,  2011, \mn@doi
  [\mnras] {10.1111/j.1365-2966.2011.19095.x}, \href
  {http://adsabs.harvard.edu/abs/2011MNRAS.416..972D} {416, 972}

\bibitem[\protect\citeauthoryear{{Djordjevic}, {Thompson}, {Urquhart}  \&
  {Forbrich}}{{Djordjevic} et~al.}{2019}]{djor2019}
{Djordjevic} J.~O.,  {Thompson} M.~A.,  {Urquhart} J.~S.,   {Forbrich} J.,
  2019, \mn@doi [\mnras] {10.1093/mnras/stz1262}, \href
  {https://ui.adsabs.harvard.edu/abs/2019MNRAS.487.1057D} {487, 1057}

\bibitem[\protect\citeauthoryear{{Drew} et~al.,}{{Drew}
  et~al.}{2005}]{drew2005}
{Drew} J.~E.,  et~al., 2005, \mn@doi [\mnras]
  {10.1111/j.1365-2966.2005.09330.x}, \href
  {http://adsabs.harvard.edu/abs/2005MNRAS.362..753D} {362, 753}

\bibitem[\protect\citeauthoryear{{Drew} et~al.,}{{Drew}
  et~al.}{2014}]{drew2014}
{Drew} J.~E.,  et~al., 2014, \mn@doi [\mnras] {10.1093/mnras/stu394}, \href
  {http://adsabs.harvard.edu/abs/2014MNRAS.440.2036D} {440, 2036}

\bibitem[\protect\citeauthoryear{{Fragkou}, {Parker}, {Boji{\v{c}}i{\'c}}  \&
  {Aksaker}}{{Fragkou} et~al.}{2018}]{frag2016}
{Fragkou} V.,  {Parker} Q.~A.,  {Boji{\v{c}}i{\'c}} I.~S.,   {Aksaker} N.,
  2018, \mn@doi [\mnras] {10.1093/mnras/sty1977}, \href
  {https://ui.adsabs.harvard.edu/abs/2018MNRAS.480.2916F} {480, 2916}

\bibitem[\protect\citeauthoryear{{Franzen} et~al.,}{{Franzen}
  et~al.}{2015}]{fran2015}
{Franzen} T.~M.~O.,  et~al., 2015, \mn@doi [\mnras] {10.1093/mnras/stv1866},
  \href {https://ui.adsabs.harvard.edu/abs/2015MNRAS.453.4020F} {453, 4020}

\bibitem[\protect\citeauthoryear{{Green}}{{Green}}{1999}]{green1999}
{Green} A.~J.,  1999, in {Taylor} A.~R.,  {Landecker} T.~L.,   {Joncas} G.,
  eds,  Astronomical Society of the Pacific Conference Series Vol. 168, New
  Perspectives on the Interstellar Medium. p.~43

\bibitem[\protect\citeauthoryear{{Green} et~al.,}{{Green}
  et~al.}{2012}]{green_MMB_2012}
{Green} J.~A.,  et~al., 2012, \mn@doi [\mnras]
  {10.1111/j.1365-2966.2011.20229.x}, \href
  {https://ui.adsabs.harvard.edu/abs/2012MNRAS.420.3108G} {420, 3108}

\bibitem[\protect\citeauthoryear{{Green} et~al.,}{{Green}
  et~al.}{2017}]{green_MMB_2017}
{Green} J.~A.,  et~al., 2017, \mn@doi [\mnras] {10.1093/mnras/stx887}, \href
  {https://ui.adsabs.harvard.edu/abs/2017MNRAS.469.1383G} {469, 1383}

\bibitem[\protect\citeauthoryear{{Guzman-Ramirez}, {Lagadec}, {Jones},
  {Zijlstra}  \& {Gesicki}}{{Guzman-Ramirez} et~al.}{2014}]{guz2014}
{Guzman-Ramirez} L.,  {Lagadec} E.,  {Jones} D.,  {Zijlstra} A.~A.,   {Gesicki}
  K.,  2014, \mn@doi [\mnras] {10.1093/mnras/stu454}, \href
  {http://adsabs.harvard.edu/abs/2014MNRAS.441..364G} {441, 364}

\bibitem[\protect\citeauthoryear{{Hancock}, {Murphy}, {Gaensler}, {Hopkins}  \&
  {Curran}}{{Hancock} et~al.}{2012}]{Hancock_Aegean_2012}
{Hancock} P.~J.,  {Murphy} T.,  {Gaensler} B.~M.,  {Hopkins} A.,   {Curran}
  J.~R.,  2012, \mn@doi [\mnras] {10.1111/j.1365-2966.2012.20768.x}, \href
  {http://adsabs.harvard.edu/abs/2012MNRAS.422.1812H} {422, 1812}

\bibitem[\protect\citeauthoryear{{Hancock}, {Trott}  \&
  {Hurley-Walker}}{{Hancock} et~al.}{2018}]{Hancock_Aegean2_2018}
{Hancock} P.~J.,  {Trott} C.~M.,   {Hurley-Walker} N.,  2018, \mn@doi [\pasa]
  {10.1017/pasa.2018.3}, \href
  {http://adsabs.harvard.edu/abs/2018PASA...35...11H} {35, e011}

\bibitem[\protect\citeauthoryear{{Haverkorn}, {Gaensler}, {McClure-Griffiths},
  {Dickey}  \& {Green}}{{Haverkorn} et~al.}{2006}]{hav2006}
{Haverkorn} M.,  {Gaensler} B.~M.,  {McClure-Griffiths} N.~M.,  {Dickey} J.~M.,
    {Green} A.~J.,  2006, \mn@doi [\apjs] {10.1086/508467}, \href
  {http://adsabs.harvard.edu/abs/2006ApJS..167..230H} {167, 230}

\bibitem[\protect\citeauthoryear{Hoare, Kurtz, Lizano, Keto  \& Hofner}{Hoare
  et~al.}{2007}]{hoa2007}
Hoare M.,  Kurtz S.,  Lizano S.,  Keto E.,   Hofner P.,  2007, Protostars and
  Planets V

\bibitem[\protect\citeauthoryear{{Hoare} et~al.,}{{Hoare}
  et~al.}{2012}]{cornish2012}
{Hoare} M.~G.,  et~al., 2012, \mn@doi [\pasp] {10.1086/668058}, \href
  {http://cdsads.u-strasbg.fr/abs/2012PASP..124..939H} {124, 939}

\bibitem[\protect\citeauthoryear{{Homan} \& {Lister}}{{Homan} \&
  {Lister}}{2006}]{ho2006}
{Homan} D.~C.,  {Lister} M.~L.,  2006, \mn@doi [\aj] {10.1086/500256}, \href
  {http://adsabs.harvard.edu/abs/2006AJ....131.1262H} {131, 1262}

\bibitem[\protect\citeauthoryear{{Hopkins}, {Miller}, {Connolly}, {Genovese},
  {Nichol}  \& {Wasserman}}{{Hopkins} et~al.}{2002}]{hopkins2002}
{Hopkins} A.~M.,  {Miller} C.~J.,  {Connolly} A.~J.,  {Genovese} C.,  {Nichol}
  R.~C.,   {Wasserman} L.,  2002, \mn@doi [\aj] {10.1086/338316}, \href
  {https://ui.adsabs.harvard.edu/abs/2002AJ....123.1086H} {123, 1086}

\bibitem[\protect\citeauthoryear{{Hotan} et~al.,}{{Hotan}
  et~al.}{2014}]{hotan2014}
{Hotan} A.~W.,  et~al., 2014, \mn@doi [\pasa] {10.1017/pasa.2014.36}, \href
  {https://ui.adsabs.harvard.edu/abs/2014PASA...31...41H} {31, e041}

\bibitem[\protect\citeauthoryear{{Hurley-Walker} et~al.,}{{Hurley-Walker}
  et~al.}{2019}]{hurley-walker2019}
{Hurley-Walker} N.,  et~al., 2019, \mn@doi [\pasa] {10.1017/pasa.2019.37},
  \href {https://ui.adsabs.harvard.edu/abs/2019PASA...36...47H} {36, e047}

\bibitem[\protect\citeauthoryear{{Huynh}, {Hopkins}, {Norris}, {Hancock},
  {Murphy}, {Jurek}  \& {Whiting}}{{Huynh} et~al.}{2012}]{huy2012}
{Huynh} M.~T.,  {Hopkins} A.,  {Norris} R.,  {Hancock} P.,  {Murphy} T.,
  {Jurek} R.,   {Whiting} M.,  2012, \mn@doi [\pasa] {10.1071/AS11026}, \href
  {https://ui.adsabs.harvard.edu/abs/2012PASA...29..229H} {29, 229}

\bibitem[\protect\citeauthoryear{{Intema}, {Jagannathan}, {Mooley}  \&
  {Frail}}{{Intema} et~al.}{2017}]{intema2017}
{Intema} H.~T.,  {Jagannathan} P.,  {Mooley} K.~P.,   {Frail} D.~A.,  2017,
  \mn@doi [\aap] {10.1051/0004-6361/201628536}, \href
  {https://ui.adsabs.harvard.edu/abs/2017A&A...598A..78I} {598, A78}

\bibitem[\protect\citeauthoryear{{Irabor} et~al.,}{{Irabor}
  et~al.}{2018}]{irabor2018}
{Irabor} T.,  et~al., 2018, \mn@doi [\mnras] {10.1093/mnras/sty1929}, \href
  {https://ui.adsabs.harvard.edu/abs/2018MNRAS.480.2423I} {480, 2423}

\bibitem[\protect\citeauthoryear{{Jonas} \& {MeerKAT Team}}{{Jonas} \& {MeerKAT
  Team}}{2016}]{jonas2016}
{Jonas} J.,  {MeerKAT Team} 2016, in MeerKAT Science: On the Pathway to the
  SKA. p.~1, \mn@doi{10.22323/1.277.0001}

\bibitem[\protect\citeauthoryear{{Kalcheva}}{{Kalcheva}}{2018}]{kal2018PhD}
{Kalcheva} I.~E.,  2018, PhD thesis, University of Leeds

\bibitem[\protect\citeauthoryear{{Kalcheva} et~al.}{{Kalcheva}
  et~al.}{2018}]{kalprep}
{Kalcheva} I.~E.,  et~al., 2018, \mn@doi [\aap] {10.1051/0004-6361/201832734},
  \href {https://ui.adsabs.harvard.edu/abs/2018A&A...615A.103K} {615, A103}

\bibitem[\protect\citeauthoryear{{Kurtz}}{{Kurtz}}{2000}]{kurt2000}
{Kurtz} S.~E.,  2000, in {Arthur} S.~J.,  {Brickhouse} N.~S.,   {Franco} J.,
  eds,  Revista Mexicana de Astronomia y Astrofisica Conference Series Vol. 9,
  {Ultracompact H II Regions: New Challenges}. pp 169--176

\bibitem[\protect\citeauthoryear{{Kurtz} \& {Franco}}{{Kurtz} \&
  {Franco}}{2002}]{kurtz2002}
{Kurtz} S.,  {Franco} J.,  2002, in {Henney} W.~J.,  {Franco} J.,   {Martos}
  M.,  eds,  Revista Mexicana de Astronomia y Astrofisica Conference Series
  Vol. 12, {Ultracompact H II Regions}. pp 16--21

\bibitem[\protect\citeauthoryear{{Lucas} et~al.,}{{Lucas}
  et~al.}{2008}]{lucas2008}
{Lucas} P.~W.,  et~al., 2008, \mn@doi [\mnras]
  {10.1111/j.1365-2966.2008.13924.x}, \href
  {http://adsabs.harvard.edu/abs/2008MNRAS.391..136L} {391, 136}

\bibitem[\protect\citeauthoryear{{Lumsden}, {Hoare}, {Urquhart}, {Oudmaijer},
  {Davies}, {Mottram}, {Cooper}  \& {Moore}}{{Lumsden} et~al.}{2013}]{lum2013}
{Lumsden} S.~L.,  {Hoare} M.~G.,  {Urquhart} J.~S.,  {Oudmaijer} R.~D.,
  {Davies} B.,  {Mottram} J.~C.,  {Cooper} H.~D.~B.,   {Moore} T.~J.~T.,  2013,
  \mn@doi [\apjs] {10.1088/0067-0049/208/1/11}, \href
  {http://adsabs.harvard.edu/abs/2013ApJS..208...11L} {208, 11}

\bibitem[\protect\citeauthoryear{{McClure-Griffiths}, {Dickey}, {Gaensler},
  {Green}, {Haverkorn}  \& {Strasser}}{{McClure-Griffiths}
  et~al.}{2005}]{mcclure2005}
{McClure-Griffiths} N.~M.,  {Dickey} J.~M.,  {Gaensler} B.~M.,  {Green} A.~J.,
  {Haverkorn} M.,   {Strasser} S.,  2005, \mn@doi [\apjs] {10.1086/430114},
  \href {http://adsabs.harvard.edu/abs/2005ApJS..158..178M} {158, 178}

\bibitem[\protect\citeauthoryear{{Minniti} et~al.,}{{Minniti}
  et~al.}{2011}]{min2011}
{Minniti} D.,  et~al., 2011, Boletin de la Asociacion Argentina de Astronomia
  La Plata Argentina, \href {http://adsabs.harvard.edu/abs/2011BAAA...54..265M}
  {54, 265}

\bibitem[\protect\citeauthoryear{{Molinari} et~al.}{{Molinari}
  et~al.}{2010}]{molinari2010}
{Molinari} S.,  et~al., 2010, \mn@doi [\pasp] {10.1086/651314}, \href
  {http://adsabs.harvard.edu/abs/2010PASP..122..314M} {122, 314}

\bibitem[\protect\citeauthoryear{{Murphy}, {Mauch}, {Green}, {Hunstead},
  {Piestrzynska}, {Kels}  \& {Sztajer}}{{Murphy} et~al.}{2007}]{murphy2007}
{Murphy} T.,  {Mauch} T.,  {Green} A.,  {Hunstead} R.~W.,  {Piestrzynska} B.,
  {Kels} A.~P.,   {Sztajer} P.,  2007, \mn@doi [\mnras]
  {10.1111/j.1365-2966.2007.12379.x}, \href
  {http://adsabs.harvard.edu/abs/2007MNRAS.382..382M} {382, 382}

\bibitem[\protect\citeauthoryear{{Prandoni}, {Gregorini}, {Parma}, {de Ruiter},
  {Vettolani}, {Wieringa}  \& {Ekers}}{{Prandoni} et~al.}{2000}]{pran2000}
{Prandoni} I.,  {Gregorini} L.,  {Parma} P.,  {de Ruiter} H.~R.,  {Vettolani}
  G.,  {Wieringa} M.~H.,   {Ekers} R.~D.,  2000, \mn@doi [\aaps]
  {10.1051/aas:2000360}, \href
  {https://ui.adsabs.harvard.edu/abs/2000A&AS..146...31P} {146, 31}

\bibitem[\protect\citeauthoryear{{Purcell} et~al.,}{{Purcell}
  et~al.}{2013}]{cornissh2013}
{Purcell} C.~R.,  et~al., 2013, \mn@doi [\apjs] {10.1088/0067-0049/205/1/1},
  \href {http://cdsads.u-strasbg.fr/abs/2013ApJS..205....1P} {205, 1}

\bibitem[\protect\citeauthoryear{{Purser} et~al.,}{{Purser}
  et~al.}{2016}]{Purser2016}
{Purser} S.~J.~D.,  et~al., 2016, \mn@doi [\mnras] {10.1093/mnras/stw1027},
  \href {https://ui.adsabs.harvard.edu/abs/2016MNRAS.460.1039P} {460, 1039}

\bibitem[\protect\citeauthoryear{{Roberts}, {Roger}, {Ribes}, {Cooke},
  {Murray}, {Cooper}  \& {Biraud}}{{Roberts} et~al.}{1975}]{rob1975}
{Roberts} J.~A.,  {Roger} R.~S.,  {Ribes} J.-C.,  {Cooke} D.~J.,  {Murray}
  J.~D.,  {Cooper} B.~F.~C.,   {Biraud} F.,  1975, \mn@doi [Australian Journal
  of Physics] {10.1071/PH750325}, \href
  {http://adsabs.harvard.edu/abs/1975AuJPh..28..325R} {28, 325}

\bibitem[\protect\citeauthoryear{{Sault} \& {Conway}}{{Sault} \&
  {Conway}}{1999}]{sault1999}
{Sault} R.~J.,  {Conway} J.~E.,  1999, in {Taylor} G.~B.,  {Carilli} C.~L.,
  {Perley} R.~A.,  eds,  Astronomical Society of the Pacific Conference Series
  Vol. 180, Synthesis Imaging in Radio Astronomy II. p.~419

\bibitem[\protect\citeauthoryear{{Sault}, {Teuben}  \& {Wright}}{{Sault}
  et~al.}{1995}]{sault1995}
{Sault} R.~J.,  {Teuben} P.~J.,   {Wright} M.~C.~H.,  1995, in {Shaw} R.~A.,
  {Payne} H.~E.,   {Hayes} J.~J.~E.,  eds,  Astronomical Society of the Pacific
  Conference Series Vol. 77, Astronomical Data Analysis Software and Systems
  IV. p.~433 (\mn@eprint {arXiv} {astro-ph/0612759})

\bibitem[\protect\citeauthoryear{{Sault}, {Staveley-Smith}  \& {Brouw}}{{Sault}
  et~al.}{1996}]{salt1996}
{Sault} R.~J.,  {Staveley-Smith} L.,   {Brouw} W.~N.,  1996, \aaps, \href
  {https://ui.adsabs.harvard.edu/abs/1996A&AS..120..375S} {120, 375}

\bibitem[\protect\citeauthoryear{{Schuller} et~al.,}{{Schuller}
  et~al.}{2009}]{schu2009}
{Schuller} F.,  et~al., 2009, \mn@doi [\aap] {10.1051/0004-6361/200811568},
  \href {http://adsabs.harvard.edu/abs/2009A%26A...504..415S} {504, 415}

\bibitem[\protect\citeauthoryear{{Smith} \& {McLean}}{{Smith} \&
  {McLean}}{2008}]{smith2008}
{Smith} E.~C.,  {McLean} I.~S.,  2008, in {Kwok} S.,  {Sanford} S.,  eds,  Vol.
  251, Organic Matter in Space. pp 219--220, \mn@doi{10.1017/S1743921308021613}

\bibitem[\protect\citeauthoryear{{Steggles}}{{Steggles}}{2016}]{harry}
{Steggles} H.~G.,  2016, PhD thesis, University of Leeds

\bibitem[\protect\citeauthoryear{{Tremblay}, {Walsh}, {Longmore}, {Urquhart}
  \& {K{\"o}nig}}{{Tremblay} et~al.}{2015}]{trem2015}
{Tremblay} C.~D.,  {Walsh} A.~J.,  {Longmore} S.~N.,  {Urquhart} J.~S.,
  {K{\"o}nig} C.,  2015, \mn@doi [\pasa] {10.1017/pasa.2015.48}, \href
  {https://ui.adsabs.harvard.edu/abs/2015PASA...32...47T} {32, e047}

\bibitem[\protect\citeauthoryear{{Urquhart}, {Busfield}, {Hoare}, {Lumsden},
  {Clarke}, {Moore}, {Mottram}  \& {Oudmaijer}}{{Urquhart}
  et~al.}{2007a}]{urqu2007cat}
{Urquhart} J.~S.,  {Busfield} A.~L.,  {Hoare} M.~G.,  {Lumsden} S.~L.,
  {Clarke} A.~J.,  {Moore} T.~J.~T.,  {Mottram} J.~C.,   {Oudmaijer} R.~D.,
  2007a, VizieR Online Data Catalog, \href
  {https://ui.adsabs.harvard.edu/abs/2007yCat..34610011U} {pp J/A+A/461/11}

\bibitem[\protect\citeauthoryear{{Urquhart}, {Busfield}, {Hoare}, {Lumsden},
  {Clarke}, {Moore}, {Mottram}  \& {Oudmaijer}}{{Urquhart}
  et~al.}{2007b}]{urqu2007rms}
{Urquhart} J.~S.,  {Busfield} A.~L.,  {Hoare} M.~G.,  {Lumsden} S.~L.,
  {Clarke} A.~J.,  {Moore} T.~J.~T.,  {Mottram} J.~C.,   {Oudmaijer} R.~D.,
  2007b, \mn@doi [\aap] {10.1051/0004-6361:20065837}, \href
  {https://ui.adsabs.harvard.edu/abs/2007A&A...461...11U} {461, 11}

\bibitem[\protect\citeauthoryear{{Urquhart} et~al.}{{Urquhart}
  et~al.}{2009}]{urq2009}
{Urquhart} J.~S.,  et~al., 2009, \mn@doi [\aap] {10.1051/0004-6361/200912108},
  \href {http://cdsads.u-strasbg.fr/abs/2009A%26A...501..539U} {501, 539}

\bibitem[\protect\citeauthoryear{{Urquhart} et~al.}{{Urquhart}
  et~al.}{2011}]{urq2011}
{Urquhart} J.~S.,  et~al., 2011, \mn@doi [\mnras]
  {10.1111/j.1365-2966.2011.19594.x}, \href
  {http://adsabs.harvard.edu/abs/2011MNRAS.418.1689U} {418, 1689}

\bibitem[\protect\citeauthoryear{{Urquhart} et~al.,}{{Urquhart}
  et~al.}{2013a}]{urq2013}
{Urquhart} J.~S.,  et~al., 2013a, \mn@doi [\mnras] {10.1093/mnras/stt1310},
  \href {https://ui.adsabs.harvard.edu/abs/2013MNRAS.435..400U} {435, 400}

\bibitem[\protect\citeauthoryear{Urquhart, Figura, Moore, Hoare, Lumsden,
  Mottram, Thompson  \& Oudmaijer}{Urquhart et~al.}{2013b}]{urq_2013}
Urquhart J.~S.,  Figura C.~C.,  Moore T. J.~T.,  Hoare M.~G.,  Lumsden S.~L.,
  Mottram J.~C.,  Thompson M.~A.,   Oudmaijer R.~D.,  2013b, \mn@doi [\mnras]
  {10.1093/mnras/stt2006}, 437, 1791

\bibitem[\protect\citeauthoryear{{Walsh} et~al.,}{{Walsh}
  et~al.}{2011}]{walsh2011}
{Walsh} A.~J.,  et~al., 2011, \mn@doi [\mnras]
  {10.1111/j.1365-2966.2011.19115.x}, \href
  {https://ui.adsabs.harvard.edu/abs/2011MNRAS.416.1764W} {416, 1764}

\bibitem[\protect\citeauthoryear{{White}, {Becker}, {Helfand}  \&
  {Gregg}}{{White} et~al.}{1997}]{white_1997}
{White} R.~L.,  {Becker} R.~H.,  {Helfand} D.~J.,   {Gregg} M.~D.,  1997,
  \mn@doi [\apj] {10.1086/303564}, \href
  {https://ui.adsabs.harvard.edu/abs/1997ApJ...475..479W} {475, 479}

\bibitem[\protect\citeauthoryear{{Wilson} et~al.,}{{Wilson}
  et~al.}{2011}]{wilson2011}
{Wilson} W.~E.,  et~al., 2011, \mn@doi [\mnras]
  {10.1111/j.1365-2966.2011.19054.x}, \href
  {http://adsabs.harvard.edu/abs/2011MNRAS.416..832W} {416, 832}

\bibitem[\protect\citeauthoryear{{Wood} \& {Churchwell}}{{Wood} \&
  {Churchwell}}{1989}]{wood21989}
{Wood} D.~O.~S.,  {Churchwell} E.,  1989, \mn@doi [\apjs] {10.1086/191329},
  \href {http://adsabs.harvard.edu/abs/1989ApJS...69..831W} {69, 831}

\bibitem[\protect\citeauthoryear{{Yang}, {Thompson}, {Tian}, {Bihr}, {Beuther}
  \& {Hindson}}{{Yang} et~al.}{2019}]{yang2019}
{Yang} A.~Y.,  {Thompson} M.~A.,  {Tian} W.~W.,  {Bihr} S.,  {Beuther} H.,
  {Hindson} L.,  2019, \mn@doi [\mnras] {10.1093/mnras/sty2811}, \href
  {https://ui.adsabs.harvard.edu/abs/2019MNRAS.482.2681Y} {482, 2681}

\makeatother
\end{thebibliography}

%%%%%%%%%%%%%%%%%%%%%%%%%%%%%%%%%%%%%%%%%%%%%%%%%%

%%%%%%%%%%%%%%%%% APPENDICES %%%%%%%%%%%%%%%%%%%%%

\appendix

\section{Example sources in the CORNISH-South catalogue that have been classified.}

\begin{landscape}
\begin{figure}

\begin{center}

	\includegraphics[height=5cm, width=6cm]{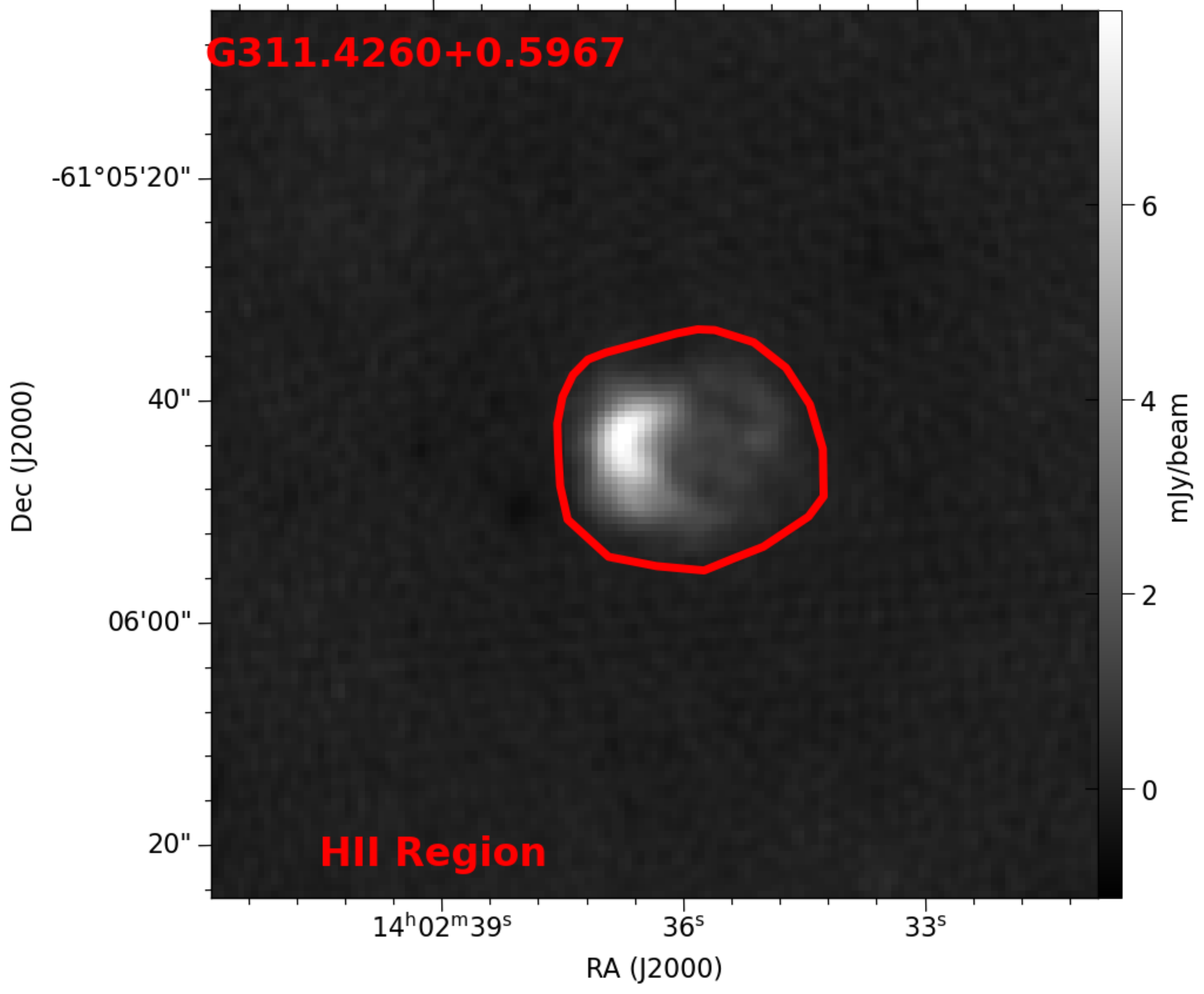}
    \includegraphics[height=5cm, width=5.2cm]{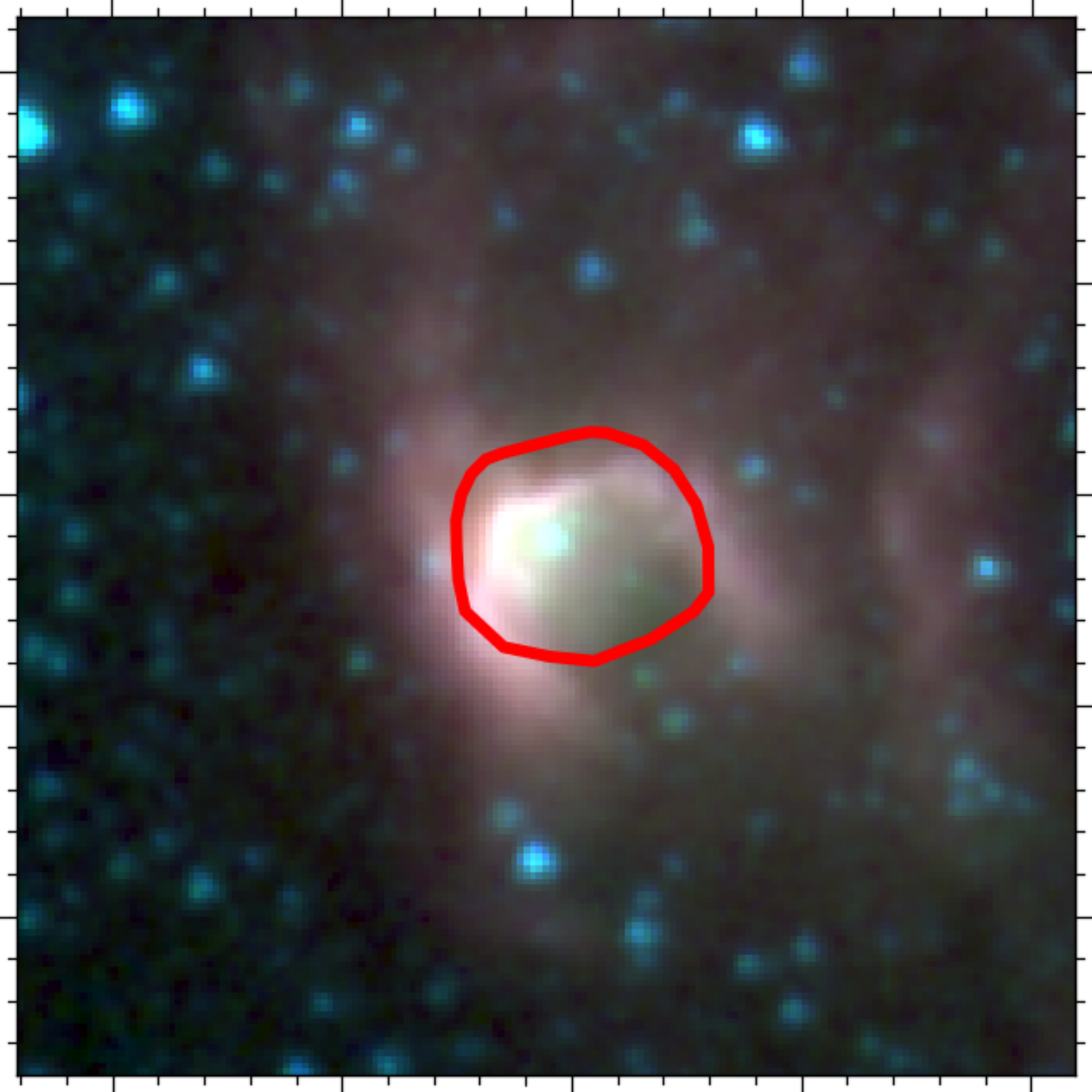}	
    \includegraphics[height=5cm, width=5.2cm]{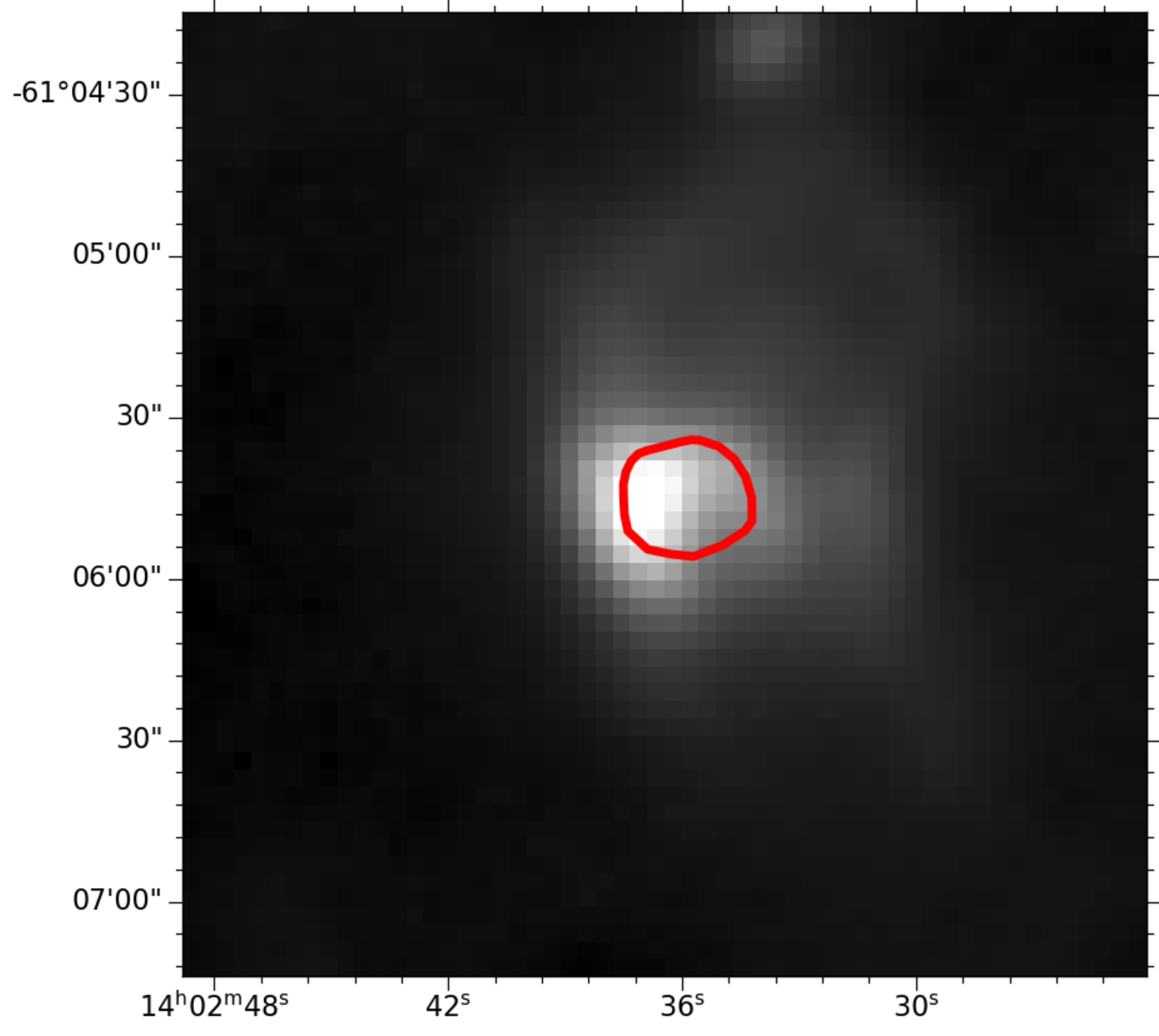}	
     \includegraphics[height=5cm, width=5.2cm]{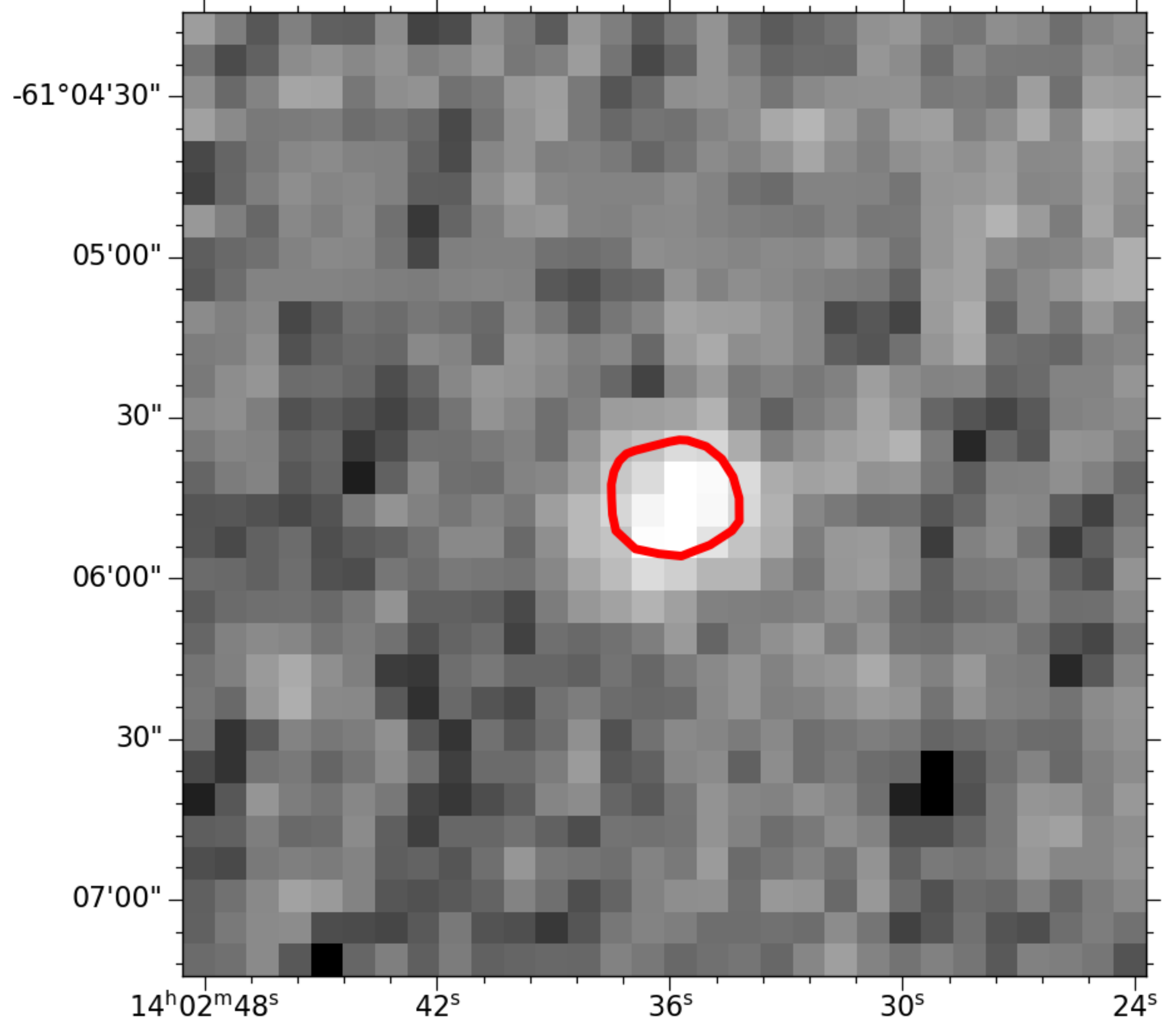}

	\includegraphics[height=5cm, width=6cm]{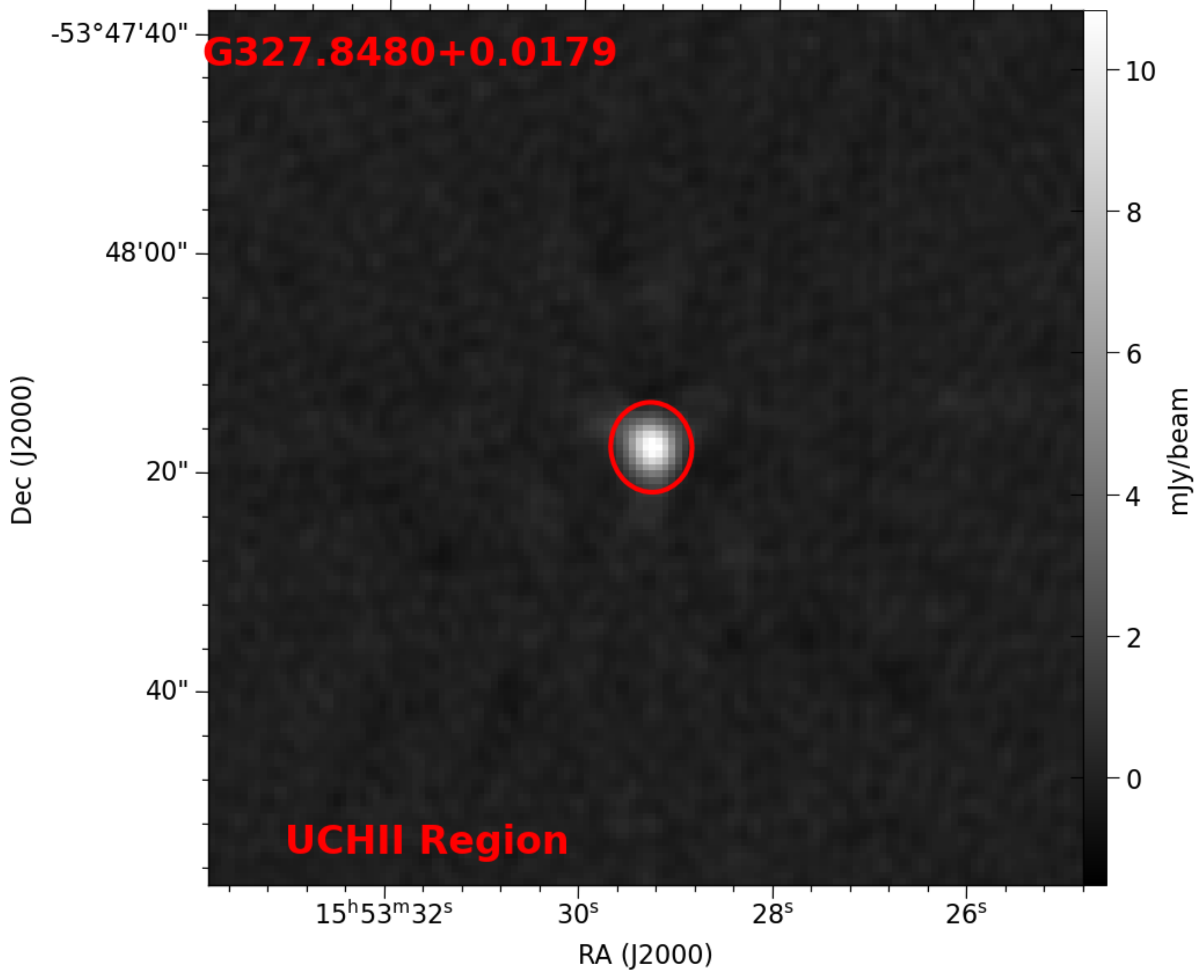}
    \includegraphics[height=5cm, width=5.2cm]{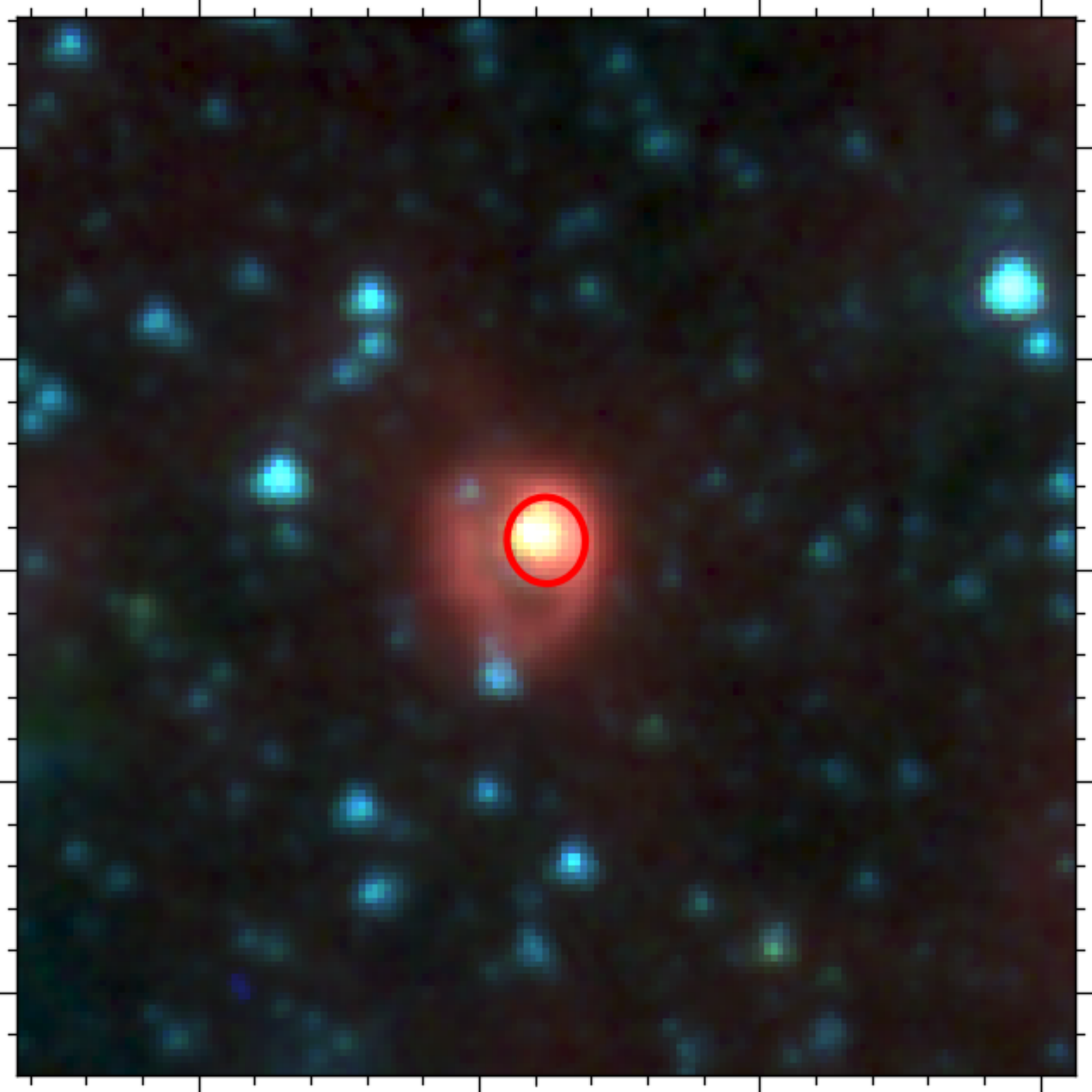}	
    \includegraphics[height=5cm, width=5.2cm]{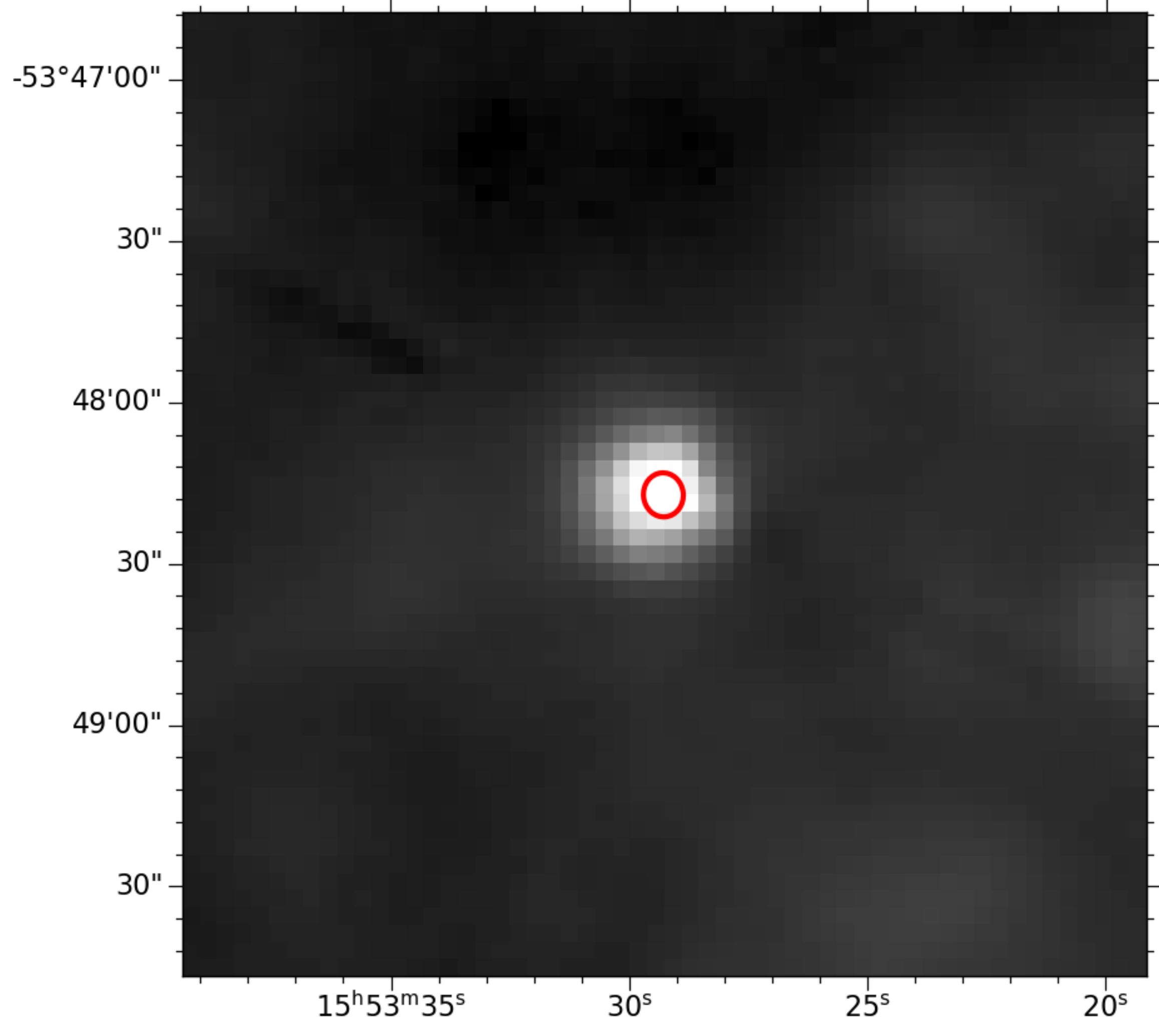}	
     \includegraphics[height=5cm, width=5.2cm]{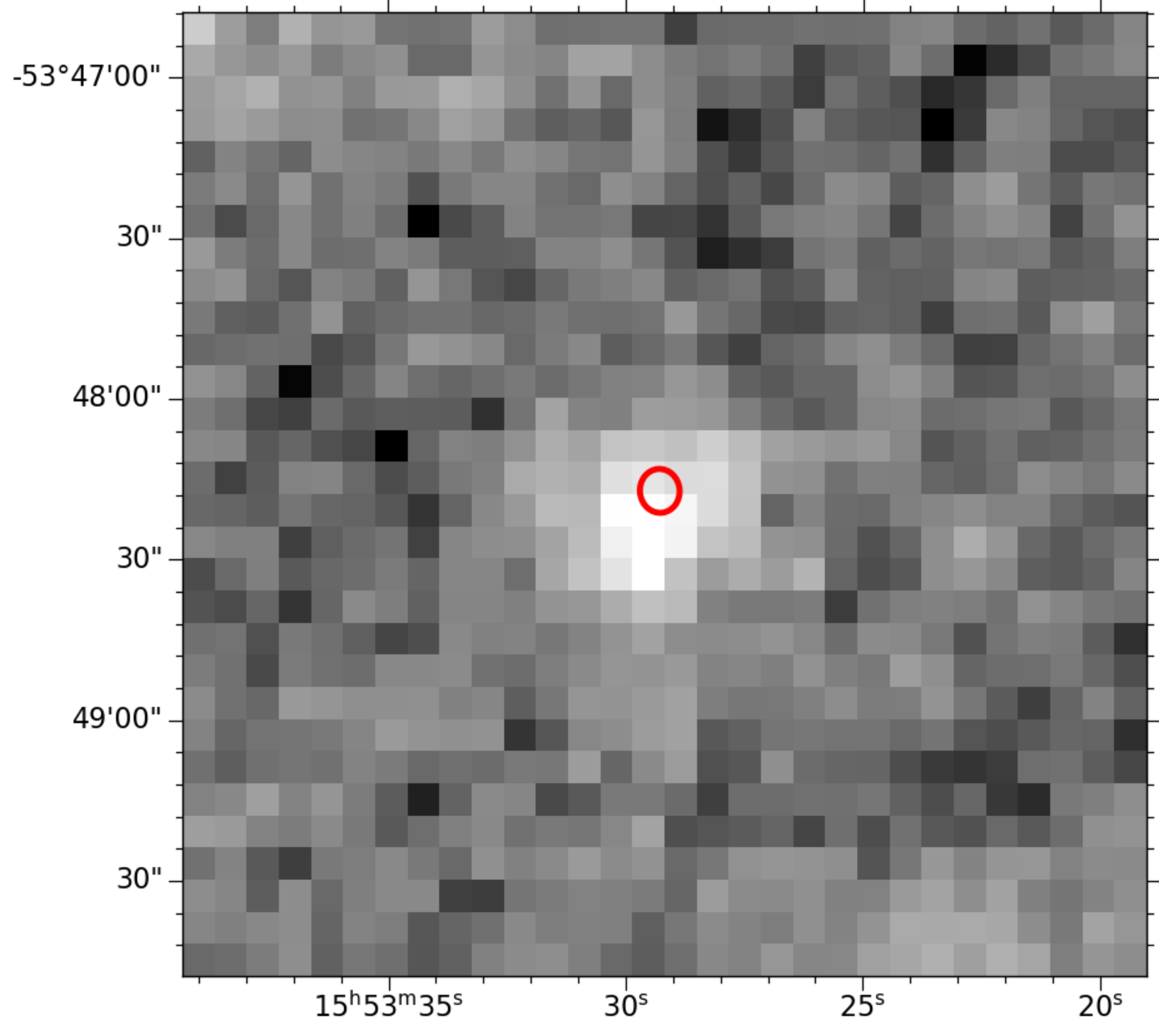}

	\includegraphics[height=5.2cm, width=6.2cm]{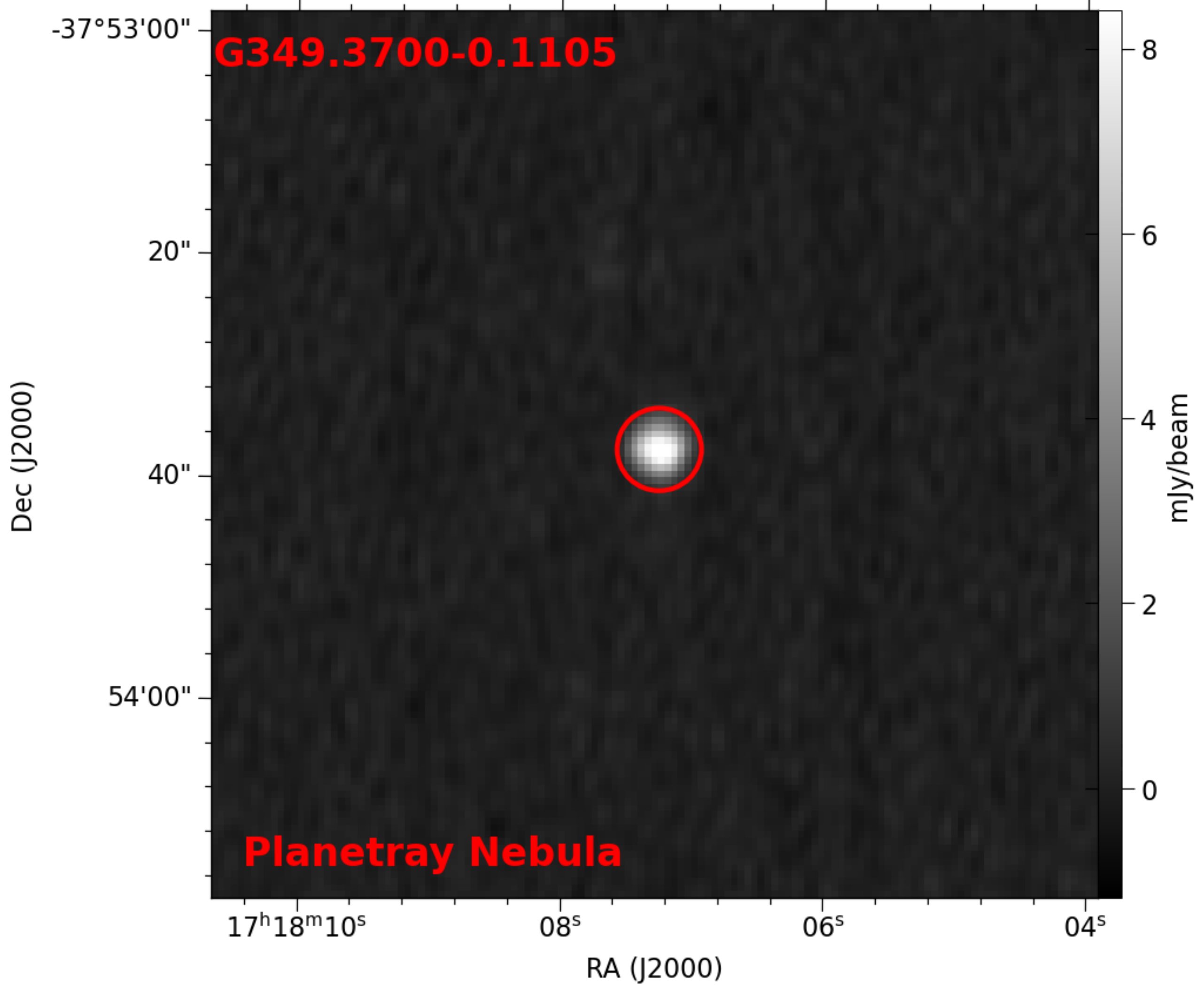}
    \includegraphics[height=5cm, width=5.2cm]{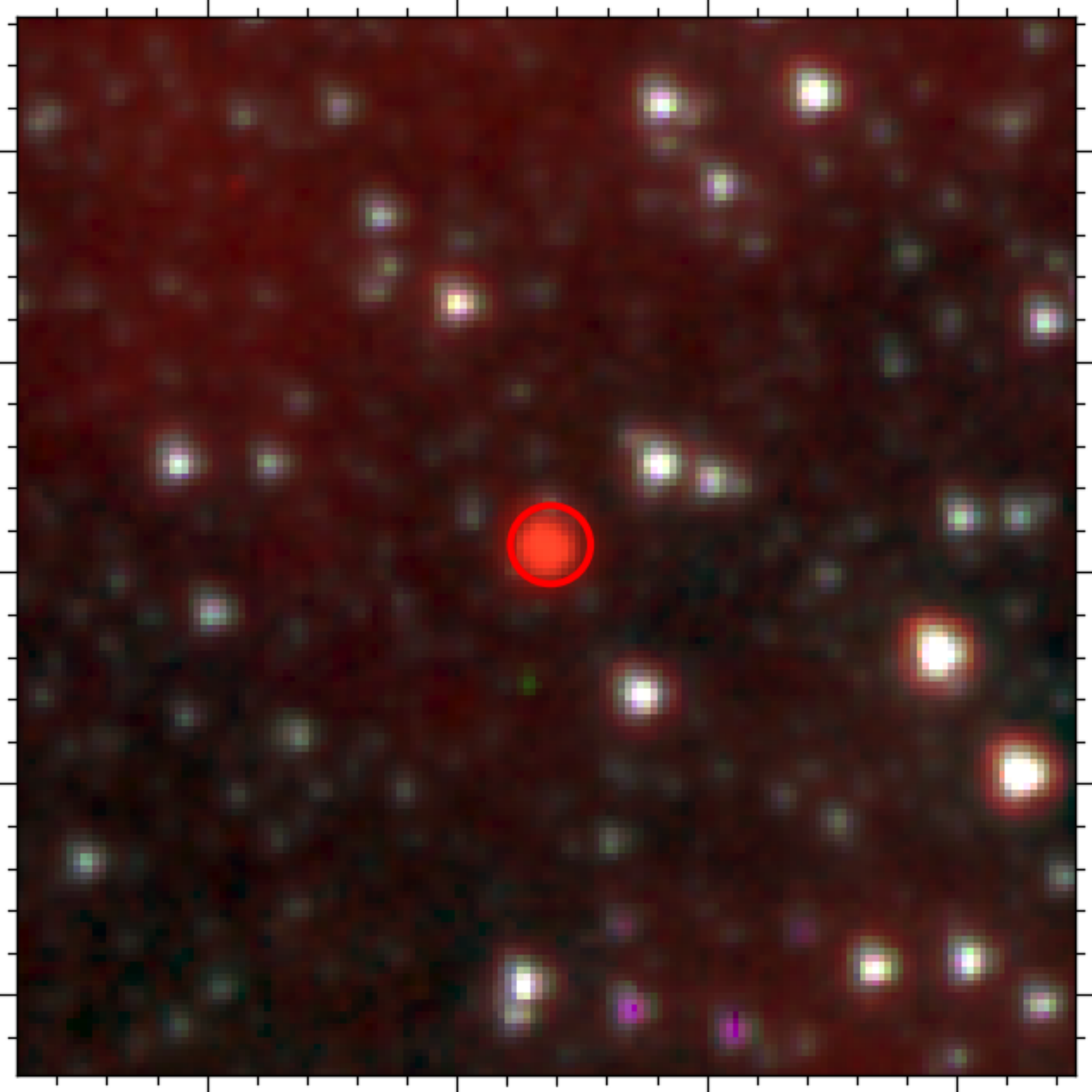}	
    \includegraphics[height=5cm, width=5.2cm]{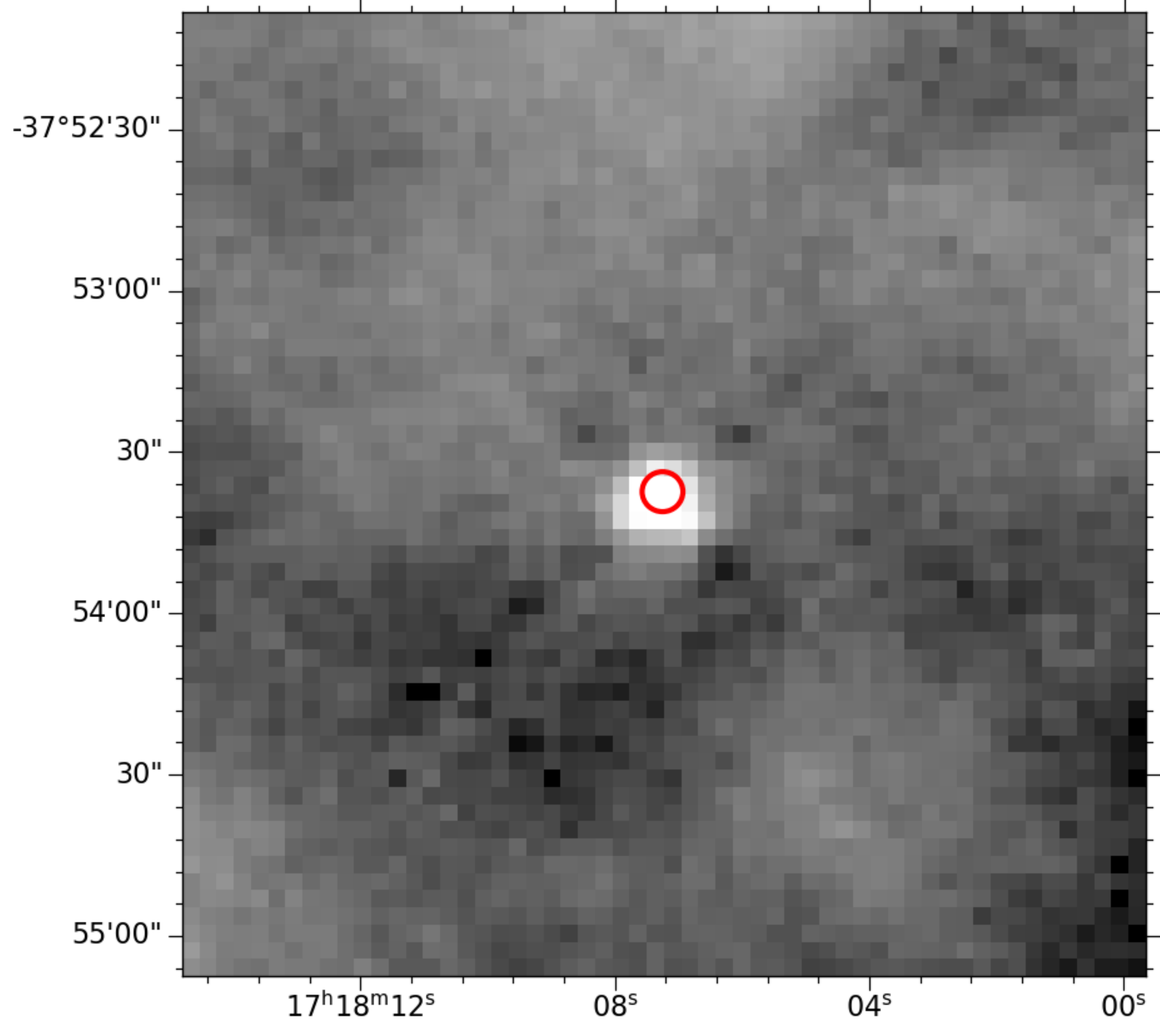}	
     \includegraphics[height=5cm, width=5.2cm]{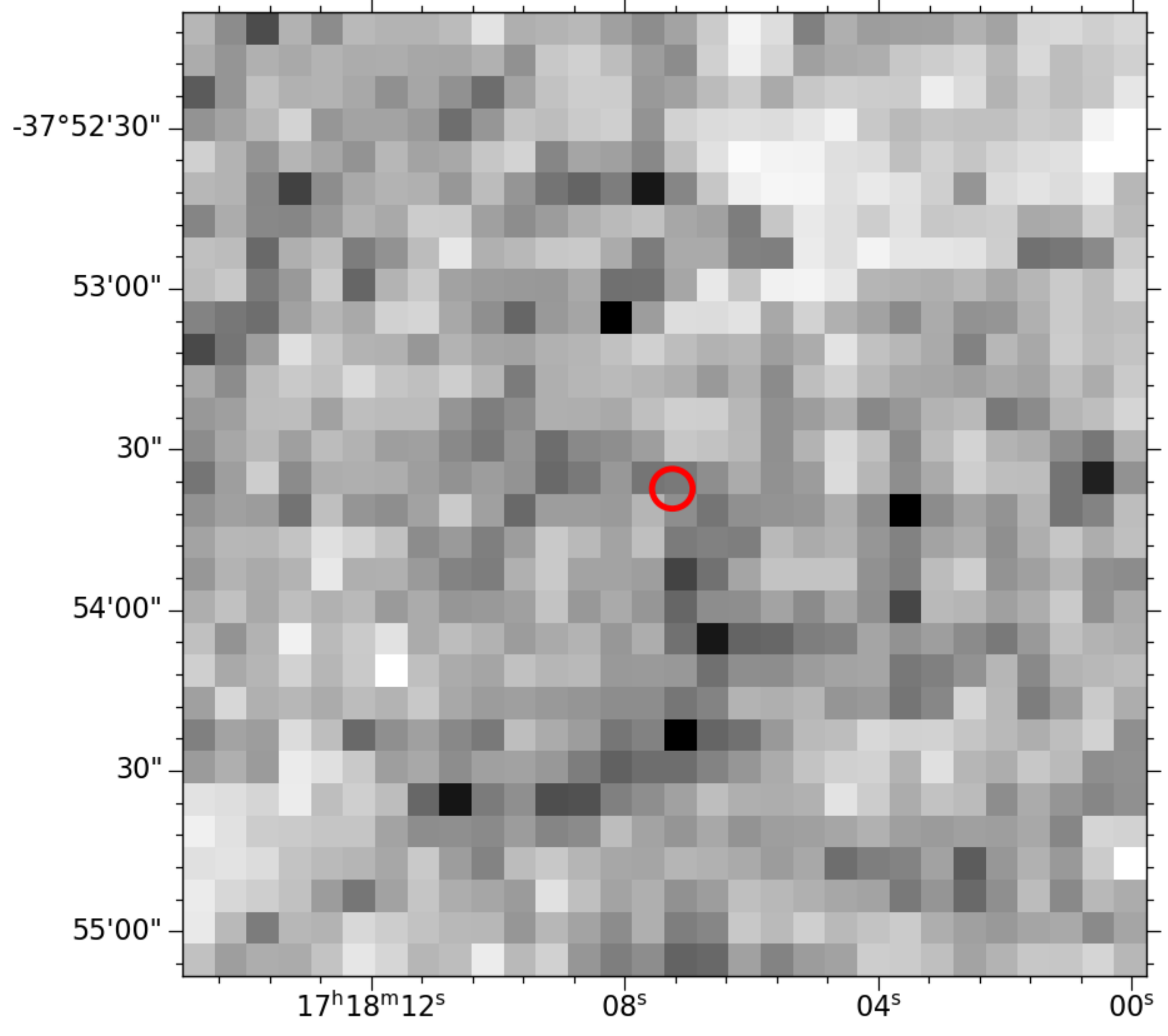}

	\caption{From the left: 5.5-GHz radio (CORNISH), 3-colour GLIMPSE, 70$\mathrm{\mu m}$ image (Hi-Gal) and ATLASGAL 850$\mathrm{\mu m}$ image . From top to bottom: Example HII region, UCHII region, PN, radio-star, radio-galaxy and infrared-quiet sources. CORNISH radio images are 80$\mathrm{\arcsec}$ by 80$\mathrm{\arcsec}$; GLIMPSE images are 100$\mathrm{\arcsec}$ by 100$\mathrm{\arcsec}$; HI-GAL and ATLASGAL images are 180$\mathrm{\arcsec}$ by 180$\mathrm{\arcsec}$. The red polygons/gaussian overlays are that of the 5.5 GHz CORNISH sizes. Images are available online.}
\label{ex_sources1}
\end{center}

\end{figure}
\end{landscape}

\begin{landscape}
\begin{figure}
\begin{center}

	\includegraphics[height=5cm, width=6cm]{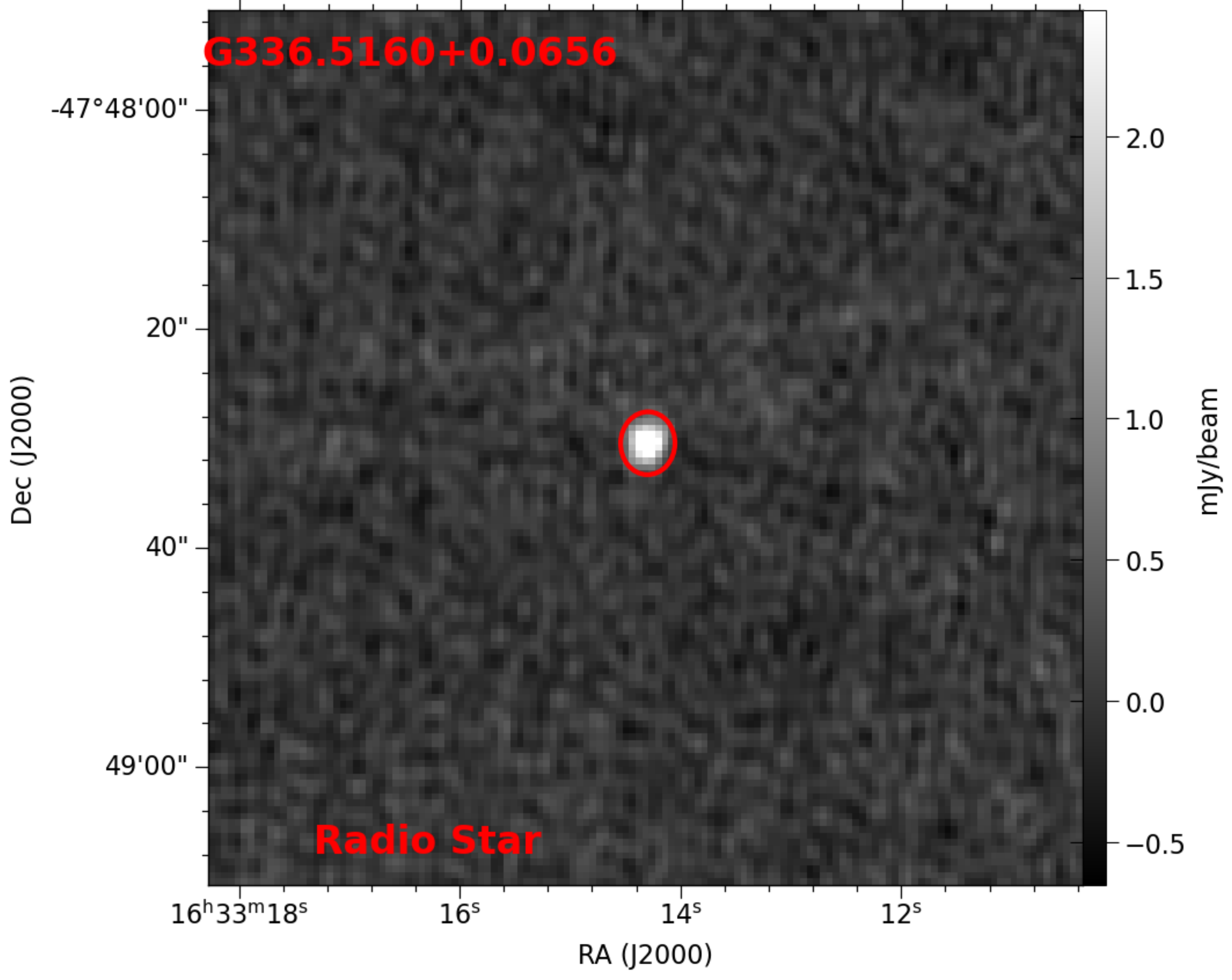}
    \includegraphics[height=5cm, width=5.2cm]{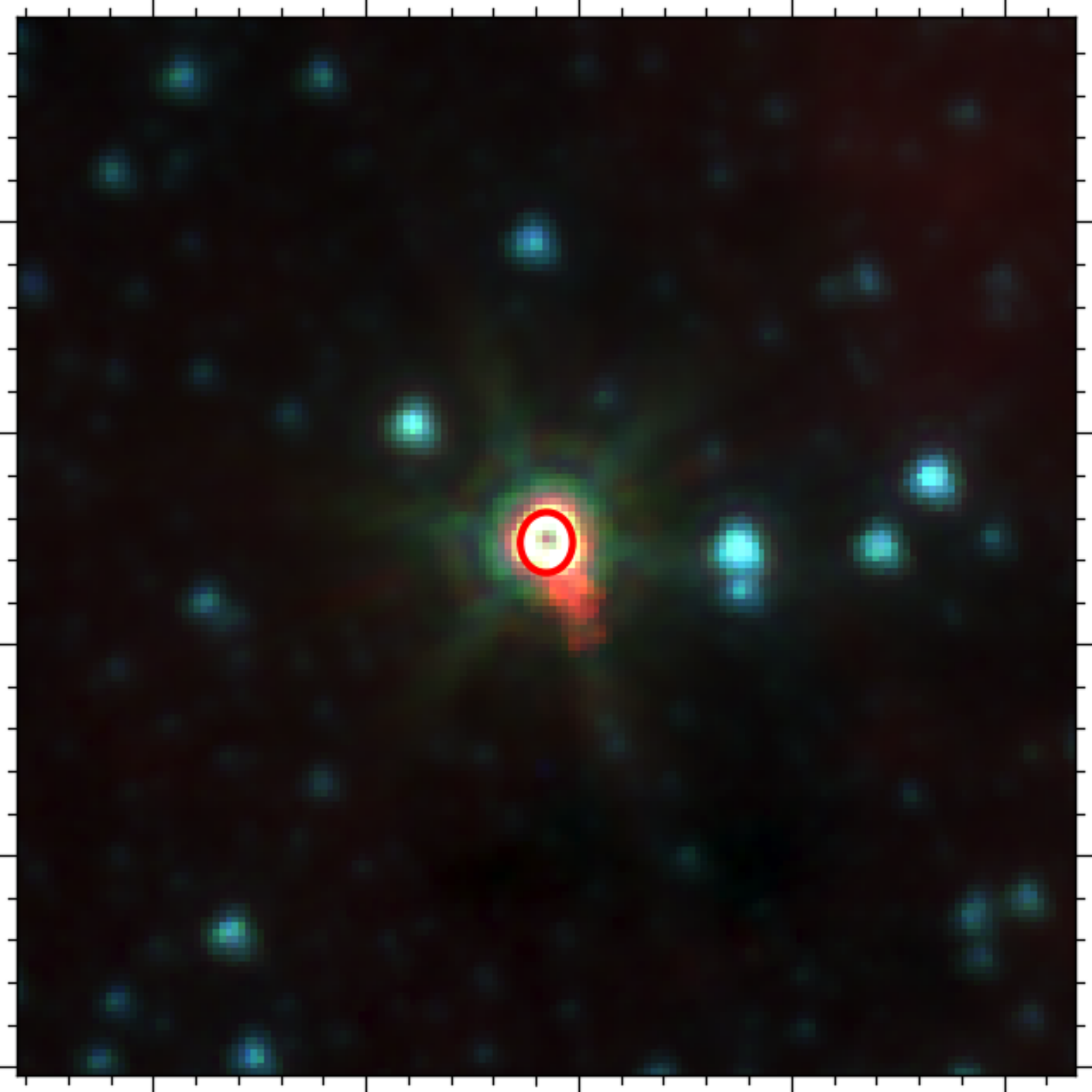}	
    \includegraphics[height=5cm, width=5.2cm]{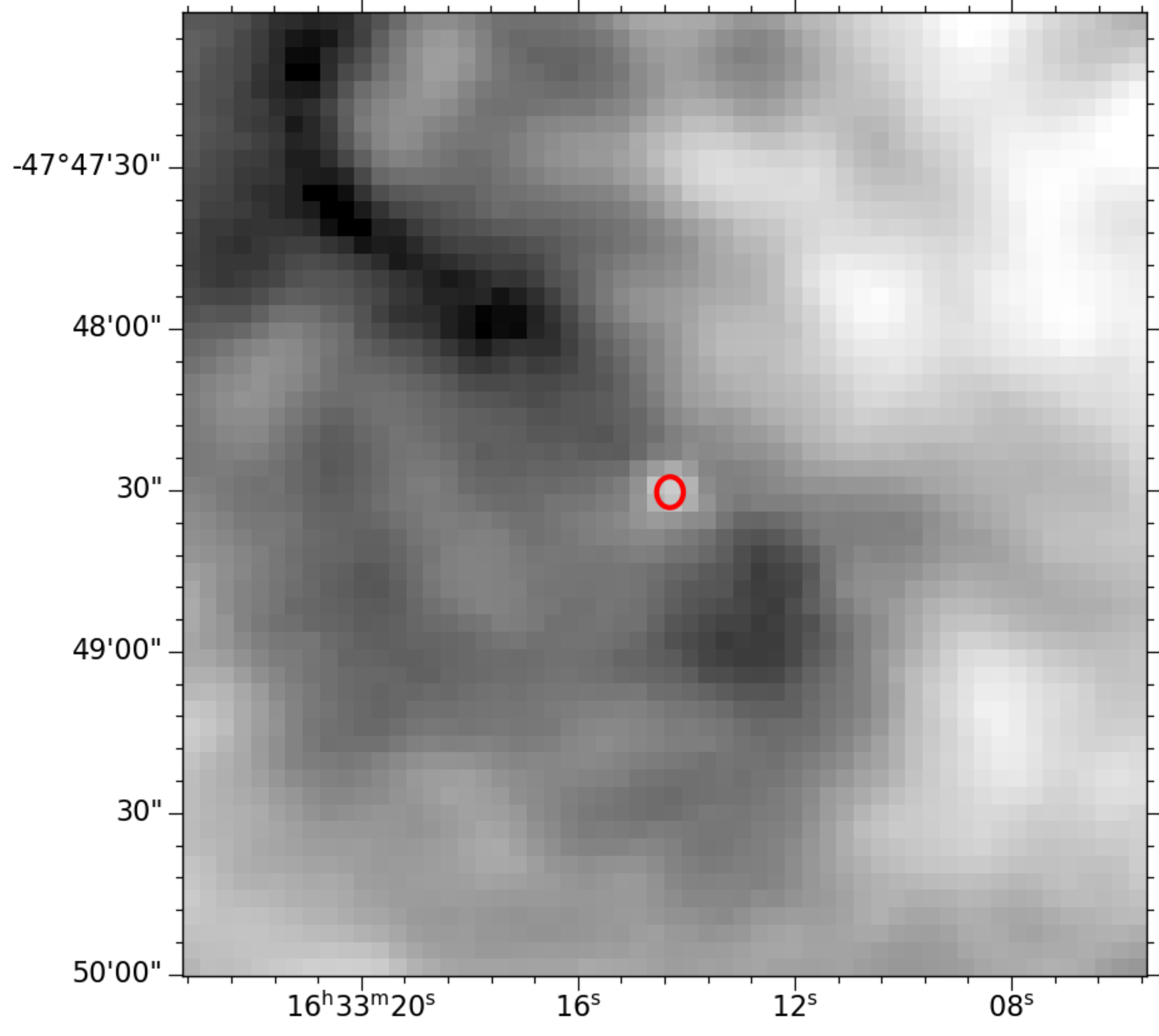}	
     \includegraphics[height=5cm, width=5.2cm]{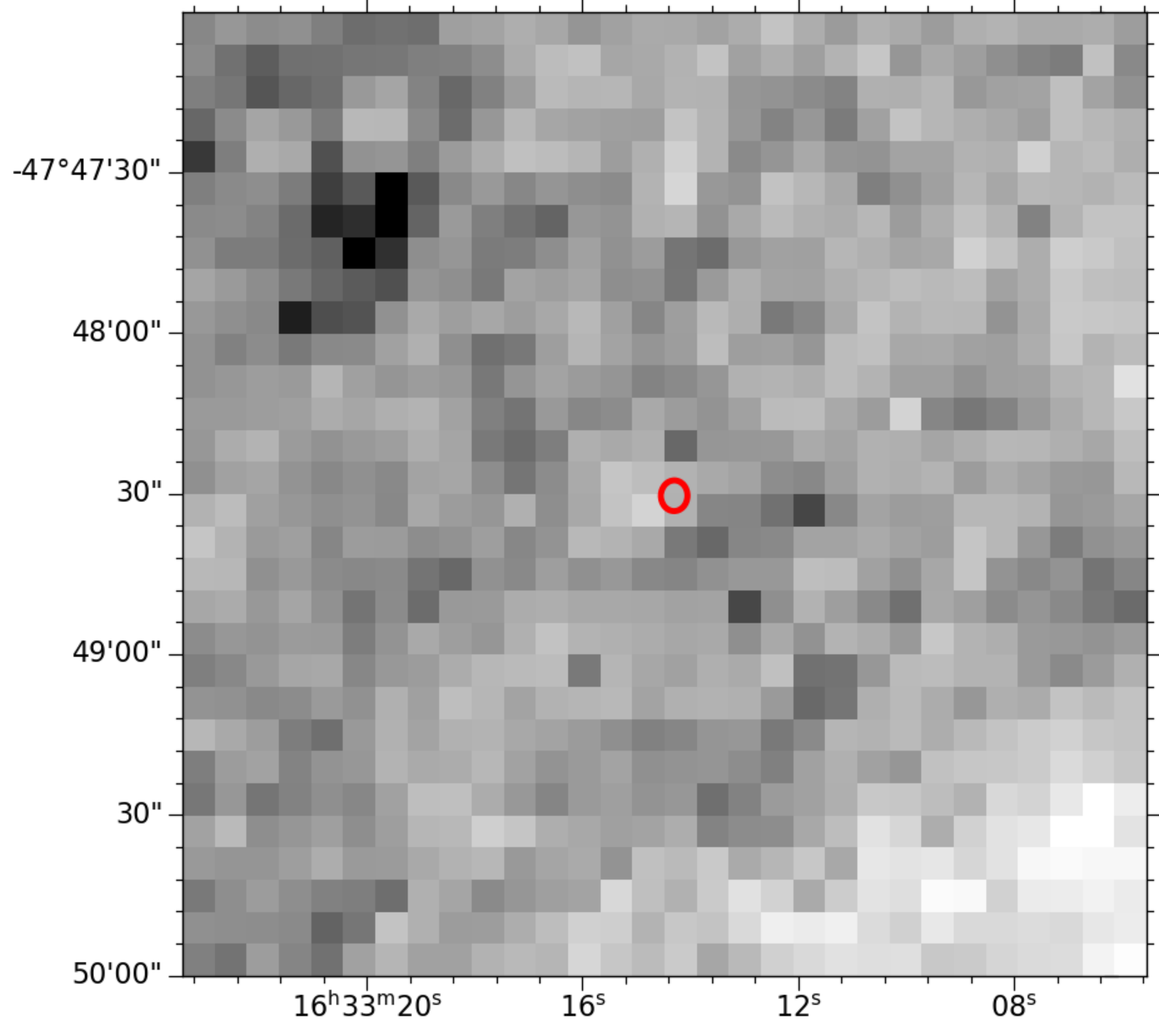}

	\includegraphics[height=5cm, width=6cm]{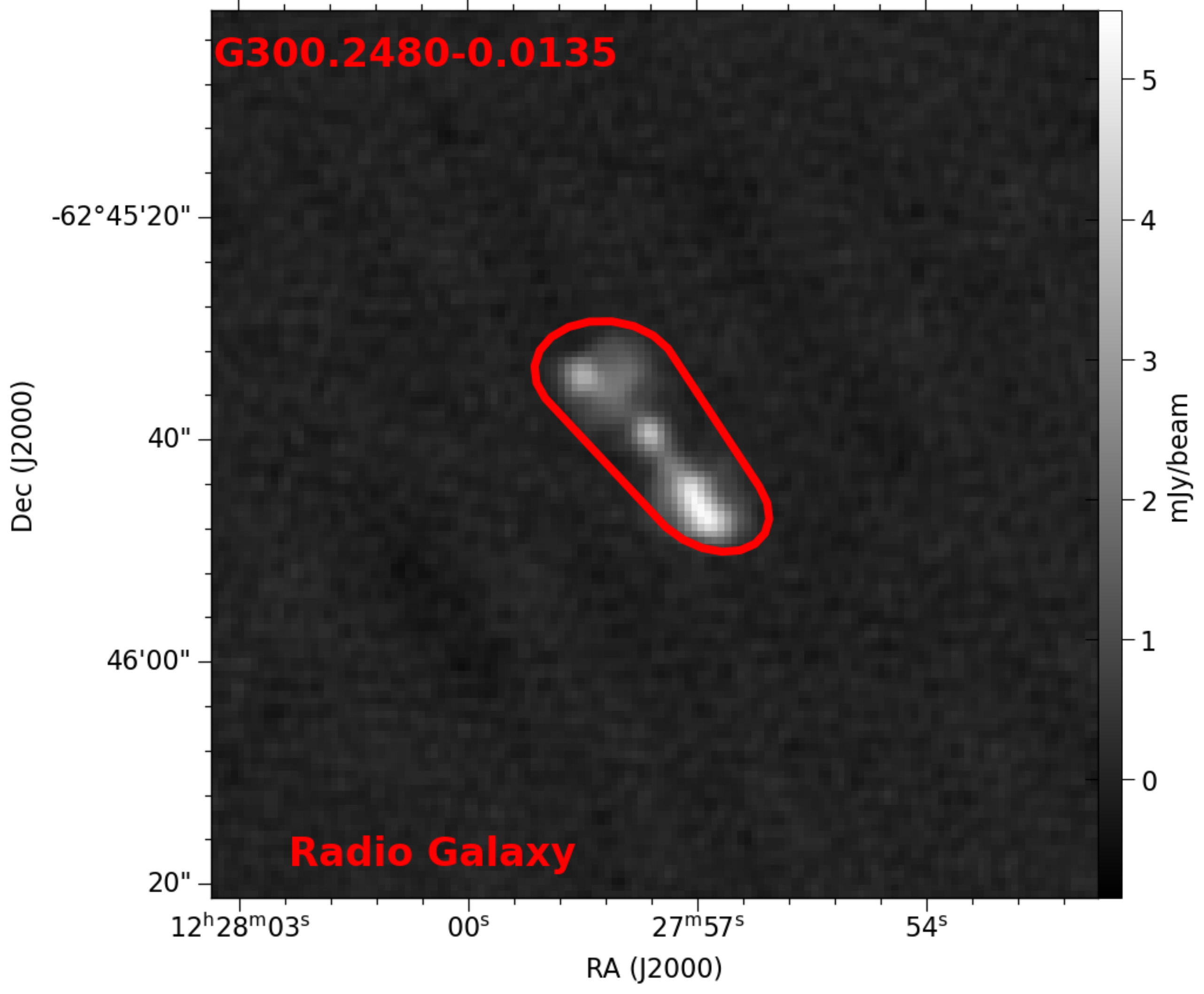}
    \includegraphics[height=5cm, width=5.2cm]{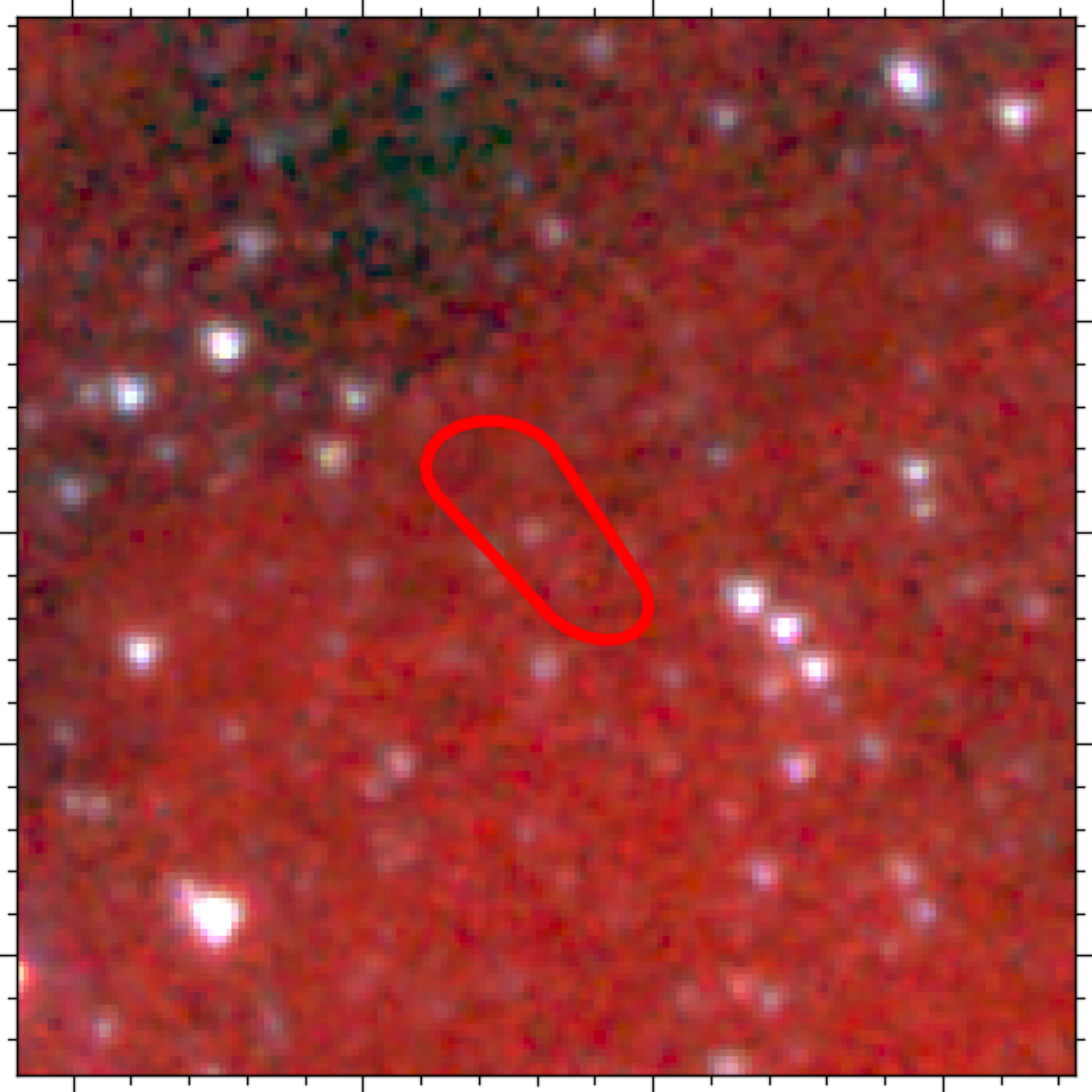}	
    \includegraphics[height=5cm, width=5.2cm]{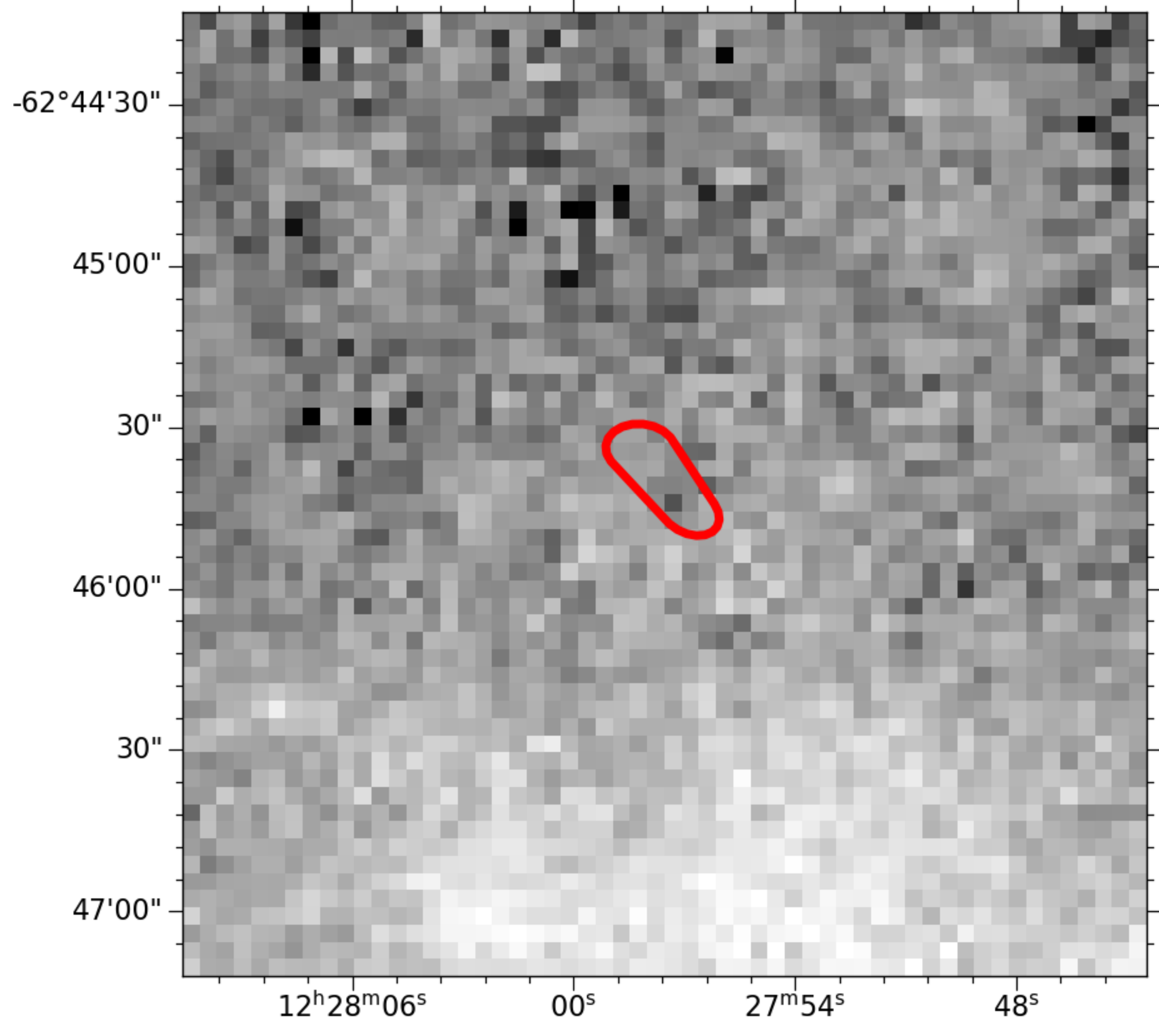}	
     \includegraphics[height=5cm, width=5.2cm]{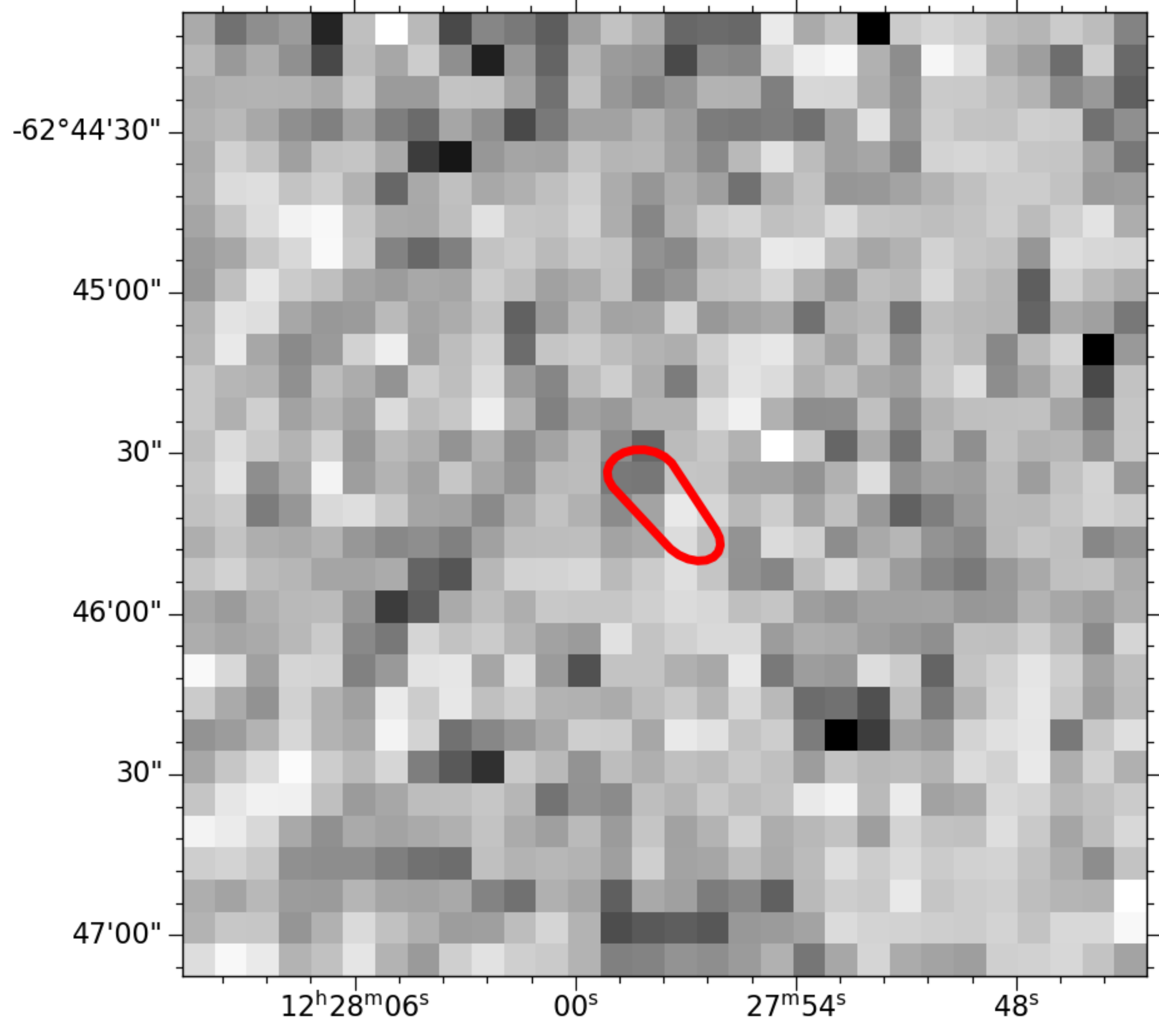}

	\includegraphics[height=5.2cm, width=6.2cm]{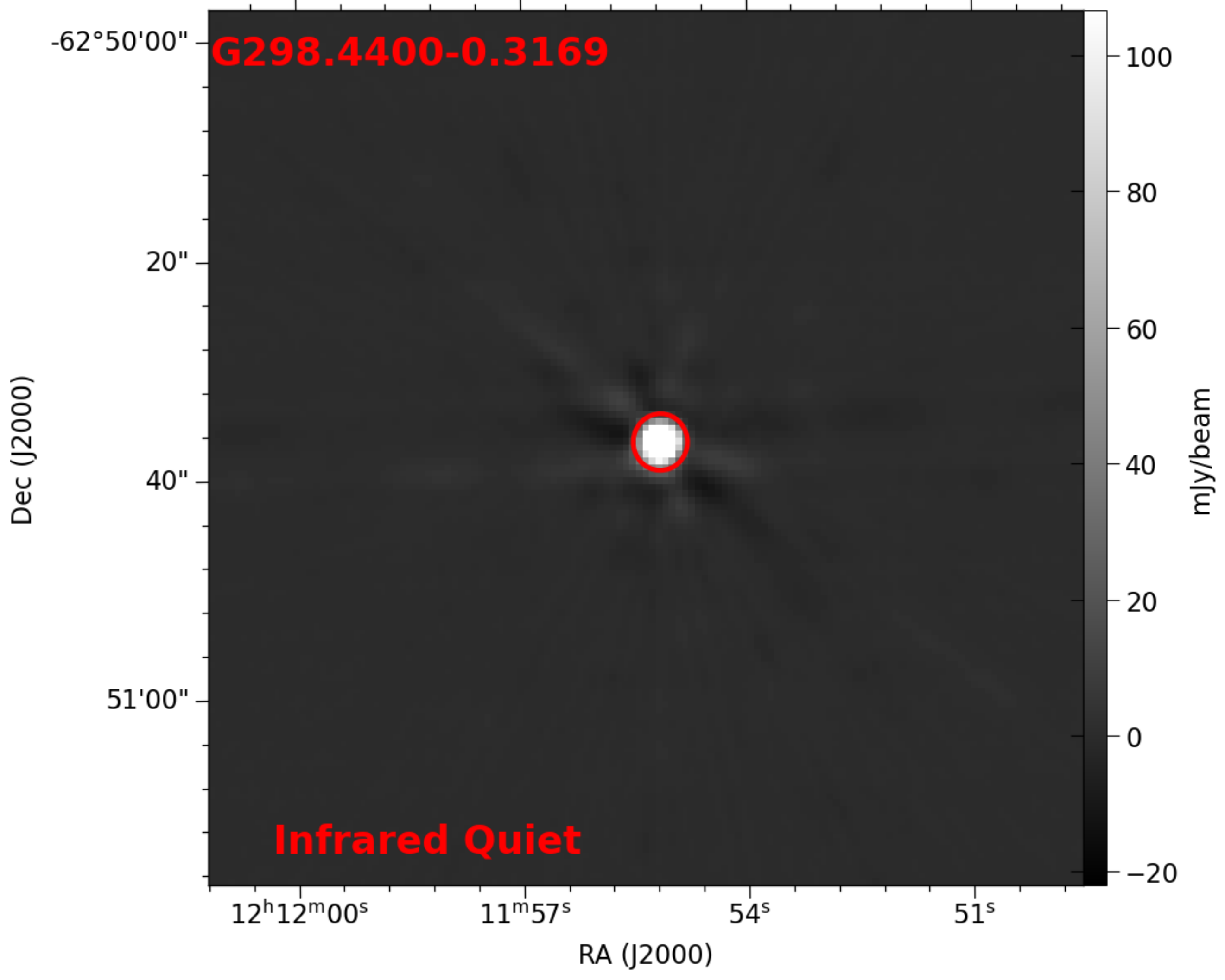}
    \includegraphics[height=5cm, width=5.2cm]{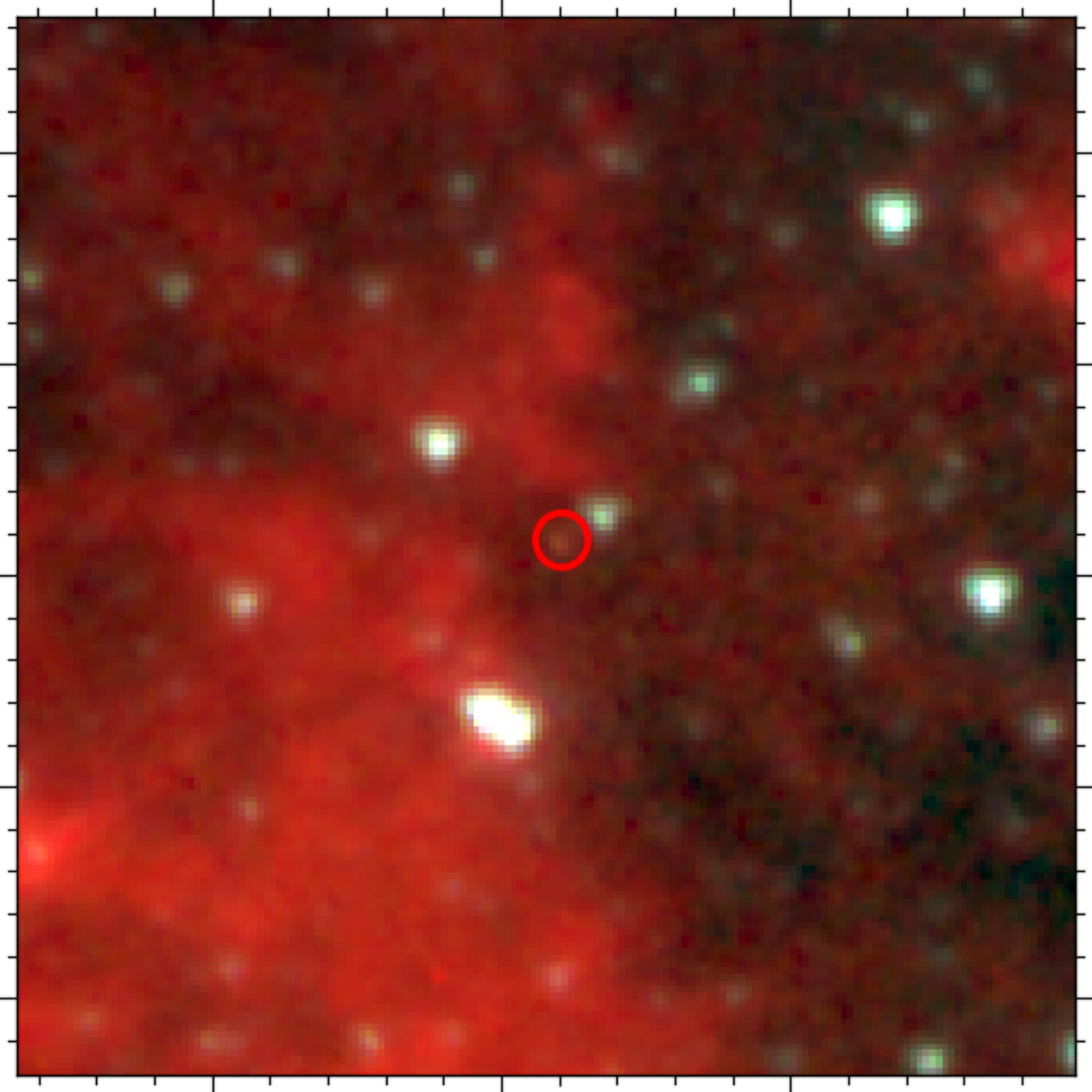}	
    \includegraphics[height=5cm, width=5.2cm]{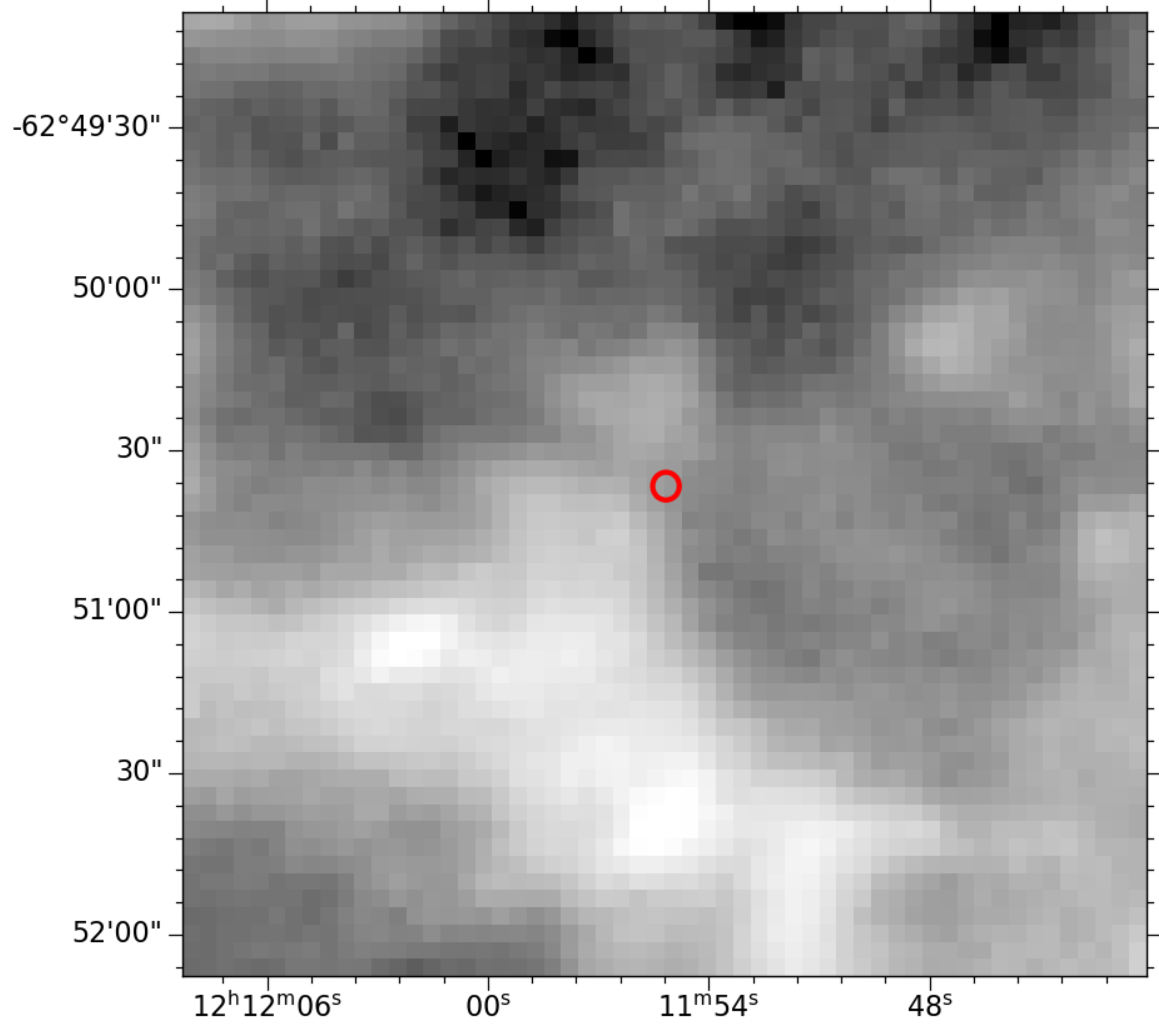}	
     \includegraphics[height=5cm, width=5.2cm]{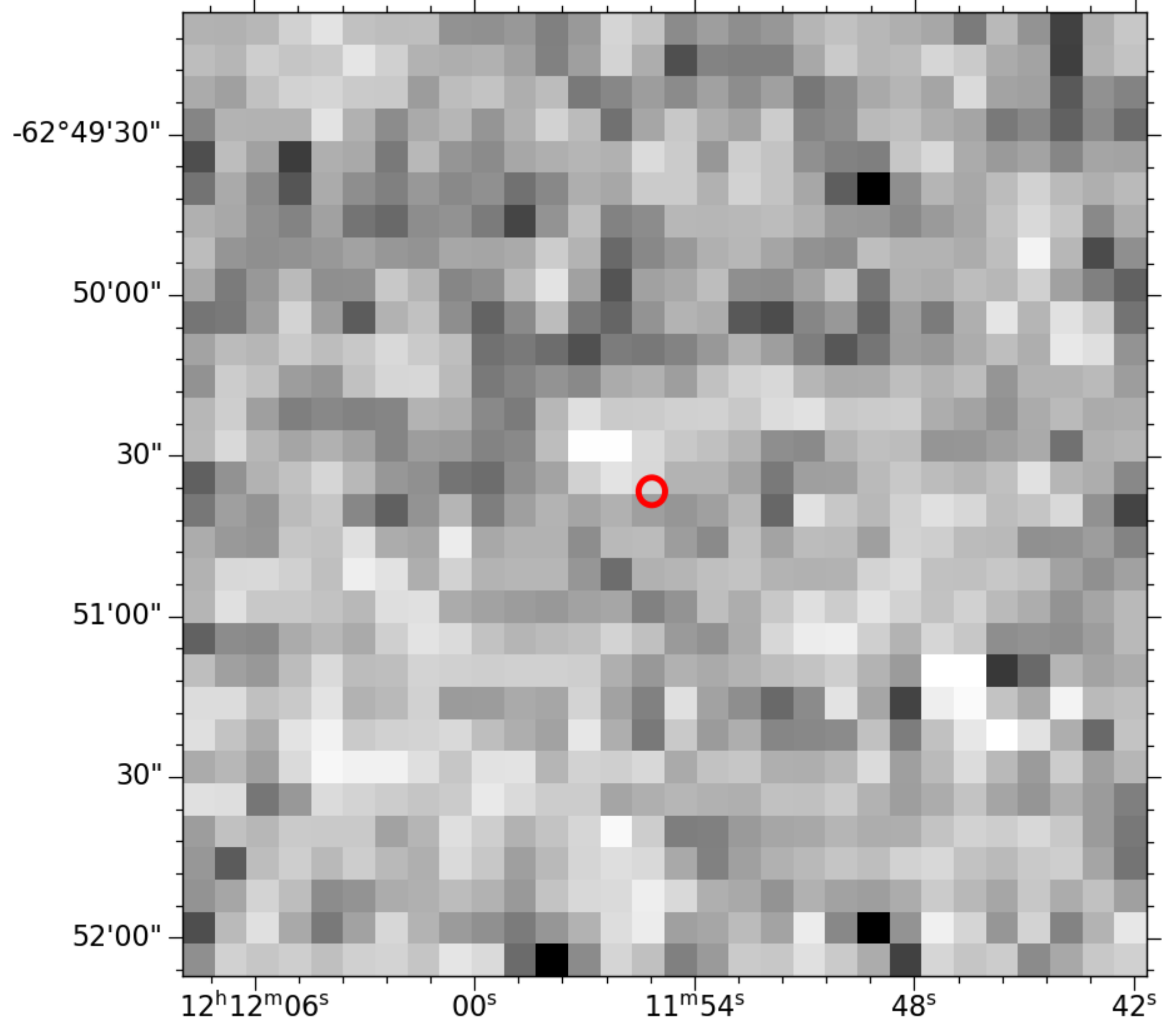}

	\caption{Continuation from Figure \ref{ex_sources1}. Images are available online.}
\label{ex_sources2}
\end{center}
\end{figure}
\end{landscape}

%%%%%%%%%%%%%%%%%%%%%%%%%%%%%%%%%%%%%%%%%%%%%%%%%%

% Don't change these lines
\bsp	% typesetting comment
\label{lastpage}
\end{document}